\documentclass[a4paper,11pt]{article}

\usepackage{jheppub} 

\usepackage{autobreak}

\usepackage{graphicx}
\usepackage{subfigure}
\usepackage{color}
\graphicspath{{figures/}{fig/}}
\usepackage{amsmath}
\usepackage{bm}
\usepackage{slashed}
\usepackage{epsfig}
\usepackage{amsfonts}
\usepackage{epstopdf}
\usepackage{hyperref}
\usepackage{booktabs}
\usepackage{multirow}
\usepackage{float}
\usepackage{makecell}
\usepackage{graphicx}
\allowdisplaybreaks
\preprint{MPP-2023-292}
\begin{document}

\title{$Z_c$ and $Z_{cs}$ systems with operator mixing at NLO in QCD sum rules}

\author[a]{Ren-Hua Wu}

\author[a,b]{Chen-Yu Wang}

\author[a]{Ce Meng}

\author[a,c]{Yan-Qing Ma}

\author[a,c]{Kuang-Ta Chao}

\affiliation[a]{School of Physics and State Key Laboratory of Nuclear Physics and Technology, Peking University,\\Beijing 100871, China}
\affiliation[b]{Max-Planck-Institut f\"ur Physik, Boltzmannstr. 8,  85748, Garching, Germany}
\affiliation[c]{Center for High Energy Physics, Peking University, Beijing 100871, China}

\emailAdd{renhuawu@pku.edu.cn}

\emailAdd{cywang@mpp.mpg.de}

\emailAdd{mengce75@pku.edu.cn}

\emailAdd{yqma@pku.edu.cn}

\emailAdd{ktchao@pku.edu.cn}

\date{\today}

\abstract{We study the mass spectra of hidden-charm tetraquark systems with quantum numbers $(I^G)J^P=(1^+)1^+$ (and their $I=\frac{1}{2}$ partners) using QCD sum rules. The analysis incorporates the complete next-to-leading order (NLO) contribution to the perturbative QCD part of the operator product expansions, with particular attention to operator mixing effects due to renormalization group evolution. We find that both the parametric dependence and the perturbative convergence are significantly improved for the two mixed operators $J_{1,5}^{\text{Mixed}}$ and $J_{2,6}^{\text{Mixed}}$, compared with those for the unmixed meson-meson or diquark-antidiquark type ones.  For the $\bar{d}c\bar{c}u$ system, the masses of $J_{1,5}^{\text{Mixed}}$ and $J_{2,6}^{\text{Mixed}}$ are determined to be $3.89^{+0.18}_{-0.12}$ GeV and $4.03^{+0.06}_{-0.07}$ GeV, respectively, closely matching those of $Z_c$(3900) and $Z_c(4020)$. Similarly, for the $\bar{s}c\bar{c}u$ states, the masses of $J_{1,5}^{\text{Mixed}}$ and $J_{2,6}^{\text{Mixed}}$ are found to be $4.02^{+0.17}_{-0.09}$ GeV and $4.21^{+0.08}_{-0.07}$ GeV, respectively, in close proximity to $Z_{cs}$(3985)/$Z_{cs}$(4000) and $Z_{cs}$(4220), consistent with the expectation that they are the partners of $Z_c$(3900) and $Z_c$(4020). Our results highlight the crucial role of operator mixing, an inevitable effect in a complete NLO calculation, in achieving a robust phenomenological description for the tetraquark system.
}

\maketitle
\flushbottom

\section{Introduction}

In recent years, a large number of new hadronic states containing heavy quarks (the charm quark $c$ or bottom quark $b$) have been observed at hadron colliders and $e^+e^-$ colliders~\cite{Zyla:2020zbs}. They are expected to be candidates of tetraquark states, pentaquark states, and other exotic states which contain two heavy quarks~\cite{Chen:2016spr,Chen:2016qju,Karliner:2017qhf,Guo:2017jvc,Olsen:2017bmm,Liu:2019zoy,Brambilla:2019esw}. These findings have opened up a new stage for the study of hadron physics and QCD. Among these exotic states, the $Z_c$(3900) is a good candidate of tetraquark state, and this state can help us to better and deeper understand tetraquark structure and properties.

In 2013, the first charged new hadronic state $Z_c$(3900), was discovered in the $J/\psi\pi^\pm$ invariant mass spectrum by $\text{BES\uppercase\expandafter{\romannumeral3}}$~\cite{BESIII:2013ris} and Belle~\cite{Belle:2013yex}.
At present, according to the particle data group~\cite{Zyla:2020zbs}, the average mass and decay width of $Z_c$(3900) are
\begin{align}
	\bar{m}(Z_c(3900)) &=3887.1\pm2.6 \, \text{MeV}\,  ,\\
	\bar{\Gamma}(Z_c(3900)) &=28.6\pm 2.6 \, \text{MeV}\, .
\end{align}
The $Z_c$ family may have other members, like the $Z_c$(4020)/$Z_c$(4025)~\cite{BESIII:2013ouc,BESIII:2013mhi}, which was found by BESIII. The other members could include $Z_c$(4200)~\cite{Belle:2014nuw},  $Z_c$(4430)~\cite{Belle:2007hrb,Belle:2013shl,LHCb:2014zfx}. Those states may have the same quantum numbers ($I^G(J^P)$) as $Z_c$(3900). Moreover, some partners of $Z_c$ family are also found, i.e. the possible $\bar{s}c\bar{c}u$ states $Z_{cs}$(3985)/$Z_{cs}$(4000)~\cite{BESIII:2020qkh,LHCb:2021uow}, and $Z_{cs}$(4220)~\cite{LHCb:2021uow}. Among those states, the $Z_c$(3900) is the most plausible candidate of the tetraquark ground state.

 Since $Z_c$(3900) is a charged state and can decay to $J/\psi \pi^{\pm}$ by strong interaction, this state could be assigned  as the $uc\bar{d}\bar{c}$ compact tetraquark\cite{Braaten:2013boa,Qiao:2013raa,Dias:2013xfa,Deng:2014gqa} or, more probably, the $D\bar{D}^{*}$ molecular state\cite{Wang:2013cya,Wilbring:2013cha,Guo:2013sya,Dong:2013iqa,Zhang:2013aoa,Albaladejo:2015lob,Aceti:2014uea,Goerke:2016hxf,Albaladejo:2016jsg,Du:2020vwb,Wang:2020dgr,Chen:2023def,Cheng:2023vyv}.   Up to now, $Z_c$(3900) has been studied by various models and tools, like QCD Sum Rules\cite{Dias:2013xfa,Qiao:2013raa,Zhang:2013aoa,Cui:2013yva,Wang:2013daa,Chen:2015ata,Wang:2020dgr,Wang:2019tlw}, Lattice QCD\cite{Prelovsek:2013xba,Prelovsek:2014swa,Albaladejo:2016jsg,Chen:2014afa,CLQCD:2019npr,Liu:2019gmh}, potential models\cite{Zhao:2014gqa,He:2015mja} and other methods \cite{Maiani:2013nmn,Chen:2013coa,Deng:2014gqa,He:2014nya,Swanson:2014tra,Guo:2014iya,Du:2022jjv,Ozdem:2017jqh,Ozdem:2021hka}. Although many works give some results which can describe $Z_c$(3900), they may suffer from large theoretical errors. Therefore, further study of $Z_c$(3900) is still needed. Moreover, the $Z_{cs}$ system was also studied as the hadronic molecular state\cite{Lee:2008uy,Baru:2021ddn,Meng:2022xdf,Wang:2023vtx,Ozdem:2021hka,Ozdem:2021yvo}, the compact tetraquark state\cite{Ebert:2005nc,Dias:2013qga,Ferretti:2020ewe,Ozdem:2021yvo}, or with other explanations\cite{Chen:2013wca,Voloshin:2019ilw}, and  it is still worth studying further (For more discussions, see, e.g.,  a recent review\cite{Chen:2022asf}).

The QCD Sum Rules~\cite{Shifman:1978bx,Shifman:1978by} approach is a powerful tool to study hadronic properties~\cite{Colangelo:2000dp,Narison:2010wb,Narison:2014wqa,Albuquerque:2018jkn}. Currently, there have been many leading order (LO) in $\alpha_s$ calculations for the $Z_{c(s)}$ system~\cite{Dias:2013xfa,Qiao:2013raa,Zhang:2013aoa,Cui:2013yva,Wang:2013daa,Khemchandani:2013iwa,Chen:2015ata,Wang:2020dgr,Albuquerque:2021tqd,Albuquerque:2022weq,Wang:2019tlw,Wang:2020iqt}. But it is well known that the LO results have large theoretical uncertainties, e.g. the \emph{ad hoc} choice of quark masses and QCD strong coupling constant, and that the next-to-leading order (NLO) QCD corrections to the perturbative part $C_1$ may significantly reduce the theoretical errors and bring about sizable corrections to the results. This has been emphasized in many works, e.g.\ for the proton~\cite{Ovchinnikov:1991mu,Groote:2008hz}, the singly heavy baryons~\cite{Groote:2008dx}, the doubly heavy baryon $\Xi_{cc}^{++}$~\cite{Wang:2017qvg}, the fully heavy baryons $\Omega_{ccc}^{++}$ and $\Omega_{bbb}^{-}$~\cite{Wu:2021tzo}, the fully heavy tetraquarks $\bar{Q}Q\bar{Q}Q\,  (Q=c,b)$~\cite{Wu:2022qwd}, and other systems (based on partial NLO contributions)  \cite{Albuquerque:2017vfq,Albuquerque:2020ugi,Albuquerque:2021tqd,Albuquerque:2022weq}.

In general, the NLO contribution of $C_1$ may lead to large corrections to the results and substantially reduce the dependence of renormalization scale and scheme as well as other parameters. This has been shown
in our studies at NLO in $\alpha_s$ for the doubly heavy baryon $\Xi_{cc}^{++}$~\cite{Wang:2017qvg}, and fully heavy baryons $\Omega_{ccc}^{++}$ and $\Omega_{bbb}^{-}$~\cite{Wu:2021tzo}. Moreover, according to the NLO study for the fully heavy tetraquark ($\bar{Q}Q\bar{Q}Q$)~\cite{Wu:2022qwd}, we find that the NLO contribution can lead to the operator-mixing (color-configuration mixing) and its effect can be vitally important in understanding the color-structure of the states and phenomenological results. In contrast, at LO in $\alpha_s$, there exist infinite operator-mixing schemes, so in the previous LO work~\cite{Dias:2013xfa,Qiao:2013raa,Zhang:2013aoa,Cui:2013yva,Wang:2013daa,Khemchandani:2013iwa,Chen:2015ata,Wang:2020dgr,Wang:2019tlw,Wang:2020iqt} it is hard to deal with this problem. Whereas at NLO, we can naturally determine the operator-mixing scheme by diagonalizing the anomalous dimension matrix of those operators. Based on our previous studies, we expect that the NLO contribution and operator-mixing effect are also important for the $Z_c$ system. So far in the literature there exist no complete calculations of the NLO contribution of $C_1$,  and in Ref.\cite{Albuquerque:2021tqd,Albuquerque:2022weq} only partial NLO contribution originating from the so-called factorized diagrams, was considered.
In contrast, we will perform a complete NLO calculation for $C_1$, and study the induced operator-mixing or color configuration mixing effects, and do phenomenological analysis. This will be the aim of our present paper.

The rest of the paper is organized as the following. In Sec.~\ref{sec:QCD sum rules},  QCD sum rules for calculation of the mass of $uc\bar{d}\bar{c}$ are given. In Sec.~\ref{sec:Calculation of Ci}, we present our methods to calculate perturbative coefficients. In Sec.~\ref{sec:Current operators}, we introduce the current operators and the operator-mixing scheme. Phenomenological results and discussions are given in Sec.~\ref{sec:Phenomenology}. Some details of our calculations and results are given in Appendices~\ref{sec:matrix}, \ref{sec:details}.

\section{QCD Sum Rules}\label{sec:QCD sum rules}

In this section, we briefly review the framework of the QCD sum rules used to calculate the mass of the tetraquark ground state. See Ref.~\cite{Colangelo:2000dp} for more details. We start with a two-point correlation function
\begin{align}\label{eq:corrFun}
\Pi(q^2) &= i \int {\mathrm{d}^D x e^{iq \cdot x} \langle \Omega|T[J(x) J^\dagger(0)] |\Omega\rangle} ,
\end{align}
where $D$ denotes the space-time dimension, $\Omega$ denotes the QCD vacuum and $J$ is a tetraquark current operator to be defined later.

On the one hand, the correlation function $\Pi(q^2)$ can be related to  the phenomenological spectrum by the  K\"{a}ll\'{e}n-Lehmann representation~\cite{Colangelo:2000dp},
\begin{align}\label{eq:K-LSR}
\Pi(q^2) &=  \int {\mathrm{d} s \frac{\rho(s)}{s-q^2-i\epsilon}}\,,
\end{align}
where $\rho(s)$ denotes the physical spectrum density. Taking the narrow resonance approximation for the physical ground state, one can parametrize the spectrum density as a pole plus a continuum part
\begin{align}\label{eq:SP}
\rho(s) = \lambda_H \delta(s-M_H^2)+ \rho_{\text{cont}}(s) \theta(s-s_{h})\,,
\end{align}
where $M_H$ and $\lambda_H$ denote the mass of the ground state and pole residue, respectively. $\rho_{\text{cont}}(s)$ denotes the continuum spectrum density, which could also contain information of higher resonances.
$s_{h}$ is the threshold of the continuum spectrum.

On the other hand, in the region where $-q^2=Q^2\gg\Lambda^2_{\text{QCD}}$, one can calculate the correlation function $\Pi(q^2)$ using the operator product expansion (OPE), which reads

\begin{align}\label{eq:OPE}
\begin{split}
\Pi(q^2) &=C_1(q^2) +\sum_i C_i(q^2) \langle O_i \rangle \, ,
\end{split}
\end{align}
where $C_{1}$ and $C_{i}$ are perturbatively calculable Wilson coefficients, and $\langle O_i  \rangle$ is a shorthand of the vacuum condensate $\langle \Omega|O_i |\Omega \rangle $, which is a nonperturbative but universal quantity. The relative importance of the vacuum condensate is power suppressed by the dimension of the operator $O_i$. In our calculations, we will only keep the relevant vacuum condensates up to dimension five,
which gives the approximated expression of OPE as
\begin{align}\label{eq:DisInt}
\begin{split}
\Pi(q^2) & = C_1(q^2)+C_{\bar{q}q}(q^2) \langle \bar{q} q \rangle+ C_{GG}(q^2) \langle g_s^2 \hat{G}\hat{G} \rangle +C_{\bar{q}qG}(q^2) \langle g_s \bar{q} q G \rangle \, ,
\end{split}
\end{align}
where $\langle g_s \bar{q} q G \rangle $ denotes $\langle g_s \bar{q} \sigma \cdot G q  \rangle $.

According to Eq.~(\ref{eq:K-LSR}), one can relate the physical spectrum density to the imaginary part of $\Pi(q^2)$ in Eq.~(\ref{eq:DisInt}) using the dispersion relation, which gives
\begin{align}\label{eq:KL-DisInt}
\begin{split}
\Pi(q^2) & =  \int {\mathrm{d} s \frac{\rho(s)}{s-q^2-i\epsilon}} \, \\
&= \frac{1}{\pi} \int_{s_{\text{th}}}^\infty \mathrm{d} s \frac{\text{Im} C_1(s)+  \sum_i \text{Im}C_i(s) \langle O_i \rangle  }{s-q^2-i\epsilon} \,,
\end{split}
\end{align}
where $s_{\text{th}}=4 m_c^2$ is the QCD threshold for the $\bar{d}c\bar{c}u$ system (light quarks are considered massless in Wilson coefficients), and the integral in the second line has been assumed to be convergent.
Then by employing the quark-hadron duality and Borel transformation~\cite{Colangelo:2000dp}, we obtain a sum rule for $\Pi(q^2)$,
\begin{align}\label{eq:Borel-QHD}
\begin{split}
\lambda_H  e^{-\frac{M_H^2}{M_B^2}} =& \frac{1}{\pi} \int_{s_{\text{th}}}^{s_0} \mathrm{d} s\,  {\text{Im}} C_1(s)\, e^{-\frac{s}{M_B^2}}  \,+ \sum_i \frac{1}{\pi} \int_{s_{\text{th}}}^\infty \mathrm{d} s\,  {\text{Im}} C_{i}(s) e^{-\frac{s}{M_B^2}}\langle O_i \rangle\,,
\end{split}
\end{align}
where $s_0$ is the threshold parameter and $M_B^2$ is the Borel parameter. They are introduced into the formula due to the qurak-hadron duality and Borel transformation, respectively.
By differentiating both sides of Eq.~(\ref{eq:Borel-QHD}) with respect to $-\frac{1}{M_B^2}$, one can get
\begin{align}\label{eq:diff}
\begin{split}
\lambda_H \ M_H^2 \ e^{-\frac{M_H^2}{M_B^2}} =& \frac{1}{\pi} \int_{s_{\text{th}}}^{s_0} \mathrm{d} s\, s\,  {\text{Im}} C_1(s)\, e^{-\frac{s}{M_B^2}}\,  +\,  \sum_i \frac{1}{\pi}\int_{s_{\text{th}}}^\infty \mathrm{d} s\,  s\,  {\text{Im}} C_{i}(s) e^{-\frac{s}{M_B^2}} \langle O_i \rangle
\,.
\end{split}
\end{align}
Finally, one can solve $M_H$ according to Eq.~(\ref{eq:Borel-QHD}) and (\ref{eq:diff}),
\begin{align}\label{eq:MH}
\begin{split}
M_H^2 &= \frac{\int_{s_{\text{th}}}^{s_0} \mathrm{d} s\, \, s \ \rho_{1}(s)\  e^{-\frac{s}{M_B^2}} + \sum_i \int_{s_{\text{th}}}^\infty \mathrm{d} s\  s\  \rho_{i}(s)\  e^{-\frac{s}{M_B^2}} \langle O_i \rangle}{\int_{s_{\text{th}}}^{s_0} \mathrm{d} s\, \, \rho_{1}(s)\ e^{-\frac{s}{M_B^2}}+ \sum_i \int_{s_{\text{th}}}^\infty \mathrm{d} s \ \rho_{i}(s)\  e^{-\frac{s}{M_B^2}}\langle O_i \rangle}\,,
\end{split}
\end{align}
where $\rho_1=\frac{1}{\pi}{\text{Im}} C_1$ and $\rho_{i}=\frac{1}{\pi}{\text{Im}} C_{i}$.

Similar to Eq.~(\ref{eq:corrFun}), for the axial vector tetraquark currents $J_\mu$ (to be defined later), one can introduce two-point correlation functions as
\begin{align}\label{eq:corrFun2}
\Pi^{A}_{\mu \nu}(q^2) &= i \int {\mathrm{d}^{D} x \, e^{iq \cdot x} \langle \Omega|T[J_\mu (x) J^\dagger_\nu(0)] |\Omega\rangle} \,,
\end{align}
For $ J^P=1^+ $ axial vector particle, the correlation function $\Pi_{\mu \nu}^A$ can be decomposed as
\begin{align}\label{eq:VAcorrFun}
		\Pi_{\mu \nu}^A(q^2) &= \left( -g_{\mu \nu}+\frac{q_\mu q_\nu}{q^2} \right) \Pi^A_{1}(q^2)+\frac{q_\mu q_\nu}{q^2} \Pi^A_2(q^2)\,.
\end{align}

In this paper we use $\Pi^{A}_{1}$ to construct sum rules, as it contains the genuine spin-1 degrees-of-freedom we are interested in.
The calculation of the corresponding ground state masses is similar to that in Eq.~(\ref{eq:MH}).

\begin{figure}[h]
	\centering
	\subfigure[$C_1$-LO]{
		\includegraphics[scale=0.37]{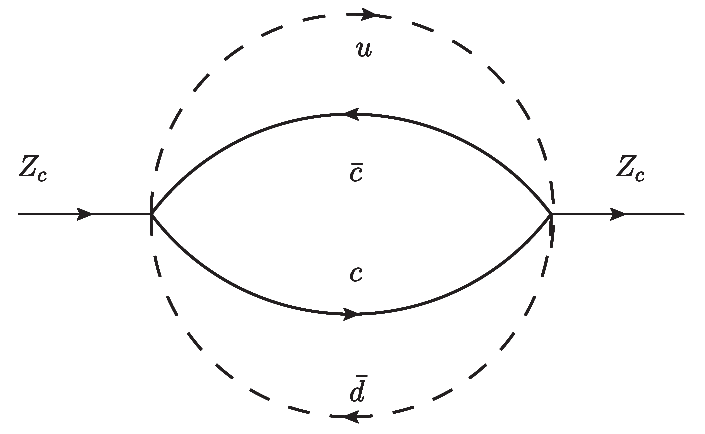}
	}
	\subfigure[$C_{\overline{q}q}$-LO]{
		\includegraphics[scale=0.52]{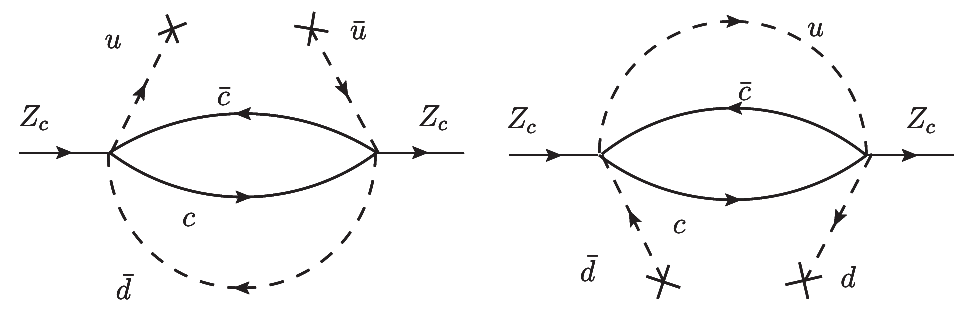}
	}
\vspace{+0.6cm}
	\subfigure[$C_{GG}$-LO]{
		\includegraphics[scale=0.92]{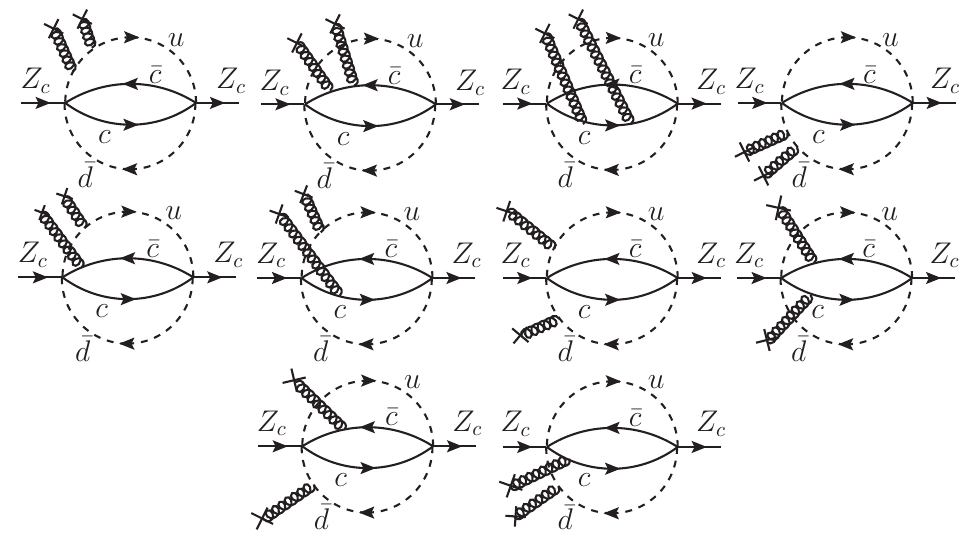}
	}
\vspace{+0.3cm}
	\subfigure[$C_{\bar{q}qG}$-LO]{
		\includegraphics[scale=0.82]{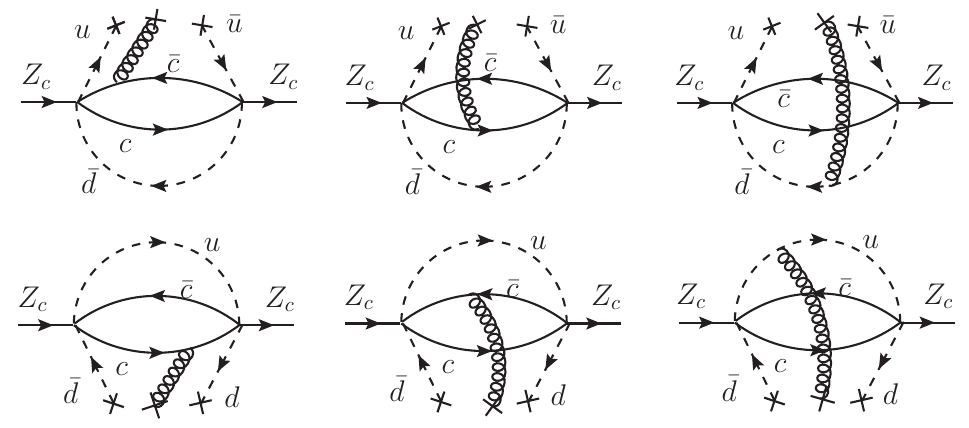}
	}

	\caption{\label{fig:FeynmanDiagrams-LO}
		LO Feynman diagrams of $C_1$ and $C_{i}$. $Z_c$ denotes the interpolating current.}
\end{figure}

\begin{figure}[h]
	\centering
	\vspace{+0.8cm}
	\subfigure[$C_1$-NLO]{
		\includegraphics[scale=1]{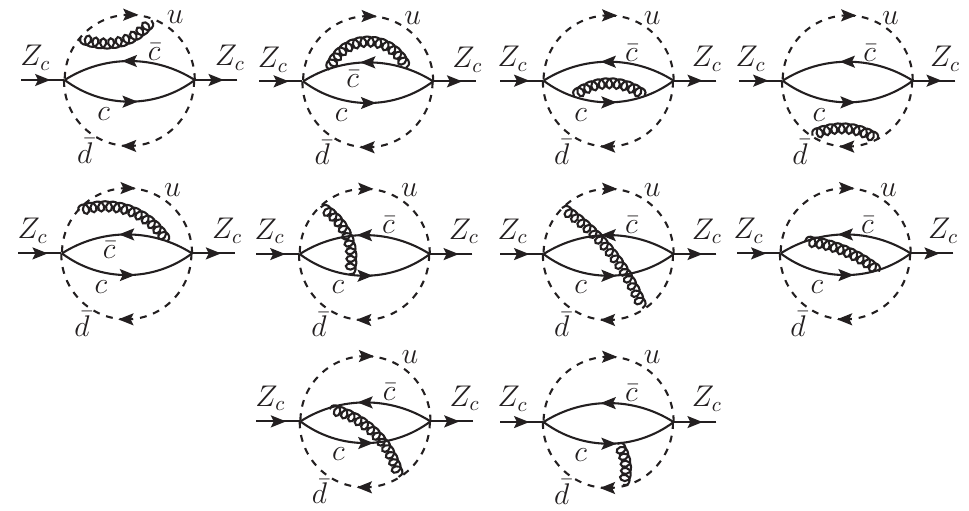}
	}
	\vspace{+0.8 cm}
	\subfigure[$C_1$-NLOct]{
		\includegraphics[scale=1]{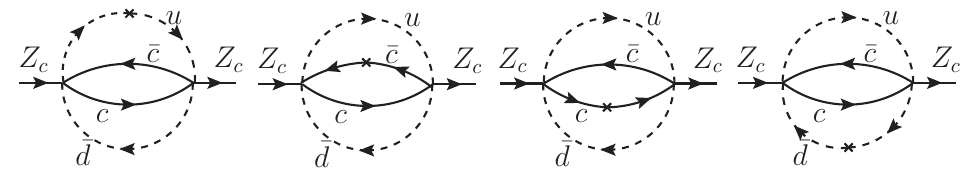}
	}
	\caption{\label{fig:FeynmanDiagrams-NLO}
		NLO Feynman diagrams (including counter terms) of $C_1$. $Z_c$ denotes the interpolating current.}
\end{figure}


\section{Calculation of Wilson coefficients $C_1$ and $C_i$ }\label{sec:Calculation of Ci}

In QCD Sum Rules, there are two kinds of expansions: the OPE and the perturbative expansion in $\alpha_s$. For the OPE, we only consider vacuum condensates up to dimension five (see Eq.~\ref{eq:DisInt}) because other higher dimensional operators are further power suppressed in the OPE, and those condensates values are not well determined. According to Eq.~(\ref{eq:MH}), we need to calculate the imaginary parts of $C_1$ and $C_i$ perturbatively. If the OPE is valid for the correlation function under consideration, then the perturbative contribution $C_{1}$ should dominate the whole expansion. And the next important contribution can be the NLO corrections for $C_1$ or the LO contribution of condensates $C_{i}$. Therefore, the NLO corrections to $C_1$ need to be considered in the calculation in order to reduce  theoretical uncertainties. For convenience, we will call the sum of the LO of $C_1$ and condensates $C_{i}$ as the LO contribution and the NLO corrections to $C_1$ as the NLO contribution in the following.

We use \texttt{FeynArts}~\cite{Kublbeck:1990xc,Hahn:2000kx} to generate Feynman diagrams and Feynman amplitudes of $C_i$. Some representative Feynman diagrams at the LO and the NLO are shown in Fig.~\ref{fig:FeynmanDiagrams-LO} and Fig.~\ref{fig:FeynmanDiagrams-NLO}, respectively. Here we emphasize that Fig.~\ref{fig:FeynmanDiagrams-NLO} includes all NLO diagrams, in particular those one gluon exchange diagrams between $c \overline{d}$ and $\overline{c} u$, without which the gauge invariance can no longer be maintained. This is a crucial point that distinguishes our work from Ref.\cite{Albuquerque:2021tqd,Albuquerque:2022weq}, in which only  the so-called factorized diagrams are considered to intentionally keep the $(c\bar d)_1$-$(\bar cu)_1$ color singlet-singlet configuration for the $c\bar d\bar cu$ system.

Our calculation procedure for $C_1$ and $C_{i}$ are summarized below:
\begin{itemize}
\item
  1. We use \texttt{FeynCalc}~\cite{Mertig:1990an,Shtabovenko:2016sxi} to simplify spinor structures of Feynman amplitudes with the Larin scheme~\cite{Larin:1993tq}.
\item
  2. We use \texttt{Reduze}~\cite{vonManteuffel:2012np} to reduce all loop integrals to linear combinations of a set of simpler integrals,  which are called master integrals (MIs).

 \item
  3. We set up differential equations for MIs~\cite{Kotikov:1990kg,Bern:1992em,Remiddi:1997ny,Gehrmann:1999as} and solve them by $\epsilon$-form\cite{Henn:2013pwa,Lee:2014ioa}. Moreover, we also use auxiliary mass flow method~\cite{Liu:2017jxz,Liu:2021wks,Liu:2022mfb, Liu:2022chg} to check our results.

 \item
  4. Renormalization. There is no infrared divergence in the NLO amplitude of $C_1$. After performing the wave-function and mass renormalization of quarks ($m_Q$ is renormalized in either the $\overline{\text{MS}}$ scheme or the on-shell scheme), the remaining ultraviolet divergences can be removed by the renomalization of the current operators. When there are more than one current operators sharing the same quantum numbers $J^{PC}$, they are usually mixed with each others under the renormalization. We get operator renormalization matrices in the $\overline{\text{MS}}$ scheme, which are shown explicitly in Appendix~\ref{sec:matrix}.
\end{itemize}
Because the expressions of $C_1$ and $C_i$ are too complicated to be shown in the paper, we attach the imaginary part of $C_1$ and $C_i$, which are the only needed information in phenomenological studies, as ancillary files.

\section{Current operators of $Z_c$ system ($\bar{d}c\bar{c}u$) with $(I^G)J^{P} =(1^+) 1^{+}$ }\label{sec:Current operators}
\subsection{Meson-meson type and diquark-antidiquark type operators}
For $Z_c^+(3900)$, which has the quantum numbers of $(I^G)J^{P} =(1^+) 1^{+}$, there are eight independent four-quark interpolating currents. The operator basis, written in terms of the meson-meson type currents, can be chosen as~\footnote{For convenience, lorentz index $\mu$ of the axial vector operator symbol $J$ is omitted hereafter.}
\begin{align}\label{eq:1+-meson}
\begin{split}
J_{1}^{\text{\text{M-M}}}&= (\bar{c}_a  u_a)(\bar{d}_b \gamma_\mu \gamma_5 c_b)-(\bar{c}_a \gamma_\mu \gamma_5 u_a)(\bar{d}_b c_b)\,, \\
J_{2}^{\text{\text{M-M}}}&= (\bar{c}_a  u_b)(\bar{d}_b \gamma_\mu \gamma_5 c_a)-(\bar{c}_a \gamma_\mu \gamma_5 u_b)(\bar{d}_b c_a)\,, \\

J_{3}^{\text{\text{M-M}}}&= (\bar{c}_a \sigma_{\mu\nu}i\gamma^5 u_a)(\bar{d}_b \gamma^\nu c_b)-(\bar{c}_a \gamma^\nu u_a)(\bar{d}_b \sigma_{\mu\nu}i\gamma^5 c_b)\,, \\
J_{4}^{\text{\text{M-M}}}&= (\bar{c}_a \sigma_{\mu\nu}i\gamma^5 u_b)(\bar{d}_b \gamma^\nu c_a)-(\bar{c}_a \gamma^\nu u_b)(\bar{d}_b \sigma_{\mu\nu}i\gamma^5 c_a)\,, \\

J_{5}^{\text{\text{M-M}}}&= (\bar{c}_a i\gamma_5 u_a)(\bar{d}_b \gamma_\mu  c_b)+(\bar{c}_a \gamma_\mu  u_a)(\bar{d}_b i\gamma_5 c_b)\,, \\
J_{6}^{\text{\text{M-M}}}&= (\bar{c}_a i\gamma_5 u_b)(\bar{d}_b \gamma_\mu c_a)+(\bar{c}_a \gamma_\mu u_b)(\bar{d}_b i\gamma_5 c_a)\,, \\

J_{7}^{\text{\text{M-M}}}&= (\bar{c}_a \sigma_{\mu\nu} u_a)(\bar{d}_b \gamma^\nu \gamma^5 c_b)+(\bar{c}_a \gamma^\nu \gamma^5 u_a)(\bar{d}_b \sigma_{\mu\nu} c_b)\,, \\
J_{8}^{\text{\text{M-M}}}&= (\bar{c}_a \sigma_{\mu\nu} u_b)(\bar{d}_b \gamma^\nu \gamma^5 c_a)+(\bar{c}_a \gamma^\nu \gamma^5 u_b)(\bar{d}_b \sigma_{\mu\nu} c_a)\,, \\

\end{split}
\end{align}
 where $a$ and $b$ represent color indices. Alternatively, one can choose the diquark-antidiquark type currents as the basis, which are given by
 \begin{align}\label{eq:1+-Diquark}
	\begin{split}
		J_{1}^{\text{\text{Di-Di}}}&= (u^T_a \hat{C}  c_b)(\bar{d}_a \gamma_\mu\gamma^5 \hat{C} \bar{c}^T_b)-(u^T_a \hat{C} \gamma_\mu\gamma^5 c_b)(\bar{d}_a \hat{C} \bar{c}^T_b)\,, \\
		J_{2}^{\text{\text{Di-Di}}}&= (u^T_a \hat{C}  c_b)(\bar{d}_b \gamma_\mu\gamma^5 \hat{C} \bar{c}^T_a)-(u^T_a \hat{C} \gamma_\mu\gamma^5 c_b)(\bar{d}_b \hat{C} \bar{c}^T_a)\,, \\
		
		J_{3}^{\text{\text{Di-Di}}}&= (u^T_a \hat{C} \sigma_{\mu\nu}i\gamma^5  c_b)(\bar{d}_a \gamma^\nu \hat{C} \bar{c}^T_b)-(u^T_a \hat{C} \gamma^\nu c_b)(\bar{d}_a  \sigma_{\mu\nu}i\gamma^5 \hat{C} \bar{c}^T_b)\,, \\
		J_{4}^{\text{\text{Di-Di}}}&= (u^T_a \hat{C} \sigma_{\mu\nu}i\gamma^5  c_b)(\bar{d}_b \gamma^\nu \hat{C} \bar{c}^T_a)-(u^T_a \hat{C} \gamma^\nu c_b)(\bar{d}_b  \sigma_{\mu\nu}i\gamma^5 \hat{C} \bar{c}^T_a)\,, \\
		
		J_{5}^{\text{\text{Di-Di}}}&= (u^T_a \hat{C} i\gamma^5 c_b)(\bar{d}_a \gamma_\mu \hat{C} \bar{c}^T_b)-(u^T_a \hat{C} \gamma_\mu c_b)(\bar{d}_a i\gamma^5 \hat{C} \bar{c}^T_b)\,, \\
		J_{6}^{\text{\text{Di-Di}}}&= (u^T_a \hat{C} i\gamma^5  c_b)(\bar{d}_b \gamma_\mu \hat{C} \bar{c}^T_a)-(u^T_a \hat{C} \gamma_\mu c_b)(\bar{d}_b i\gamma^5 \hat{C} \bar{c}^T_a)\,, \\
		
	J_{7}^{\text{\text{Di-Di}}}&= (u^T_a \hat{C} \sigma_{\mu\nu}  c_b)(\bar{d}_a \gamma^\nu \gamma^5 \hat{C} \bar{c}^T_b)-(u^T_a \hat{C} \gamma^\nu\gamma^5 c_b)(\bar{d}_a  \sigma_{\mu\nu} \hat{C} \bar{c}^T_b)\,, \\
	J_{8}^{\text{\text{Di-Di}}}&= (u^T_a \hat{C} \sigma_{\mu\nu}  c_b)(\bar{d}_b \gamma^\nu\gamma^5 \hat{C} \bar{c}^T_a)-(u^T_a \hat{C} \gamma^\nu\gamma^5 c_b)(\bar{d}_b  \sigma_{\mu\nu} \hat{C} \bar{c}^T_a)\,, \\
	\end{split}
\end{align}
where $\hat{C}$ is the charge-conjugation matrix. The two types of bases can be associated with each other by the Fierz transformation in four dimensions,
\begin{equation}\label{eq:1+-FierzTrans}
\vec{J}^{\text{\text{Di-Di}}}=\frac{i}{2}\begin{pmatrix}
  0 \, & \, 0 \, & \, 0 \, & \, 0 \, & \, 0 \, & \,  1 \, & \, 0 \, & \, 1 \\
  0 & 0 & 0 & 0 &  1 & 0 & 1 & 0  \\
  0 & 0 & 0 & 0 & 0 &  3 & 0 & -1 \\
0 & 0 & 0 & 0 &  3 & 0 & -1 & 0  \\
0 &  -1 & 0 & 1 & 0 & 0 & 0 & 0 \\
  -1 & 0 & 1 & 0 & 0 & 0 & 0 & 0 \\
  0 &  3 & 0 & 1 & 0 & 0 & 0 & 0 \\
  3 & 0 & 1 & 0 & 0 & 0 & 0 & 0 \\

\end{pmatrix} \cdot \vec{J}^{\text{\text{M-M}}}\,,
\end{equation}
where we use the column vector $\vec{J}$ to represent the basis in Eq.~(\ref{eq:1+-meson}) or (\ref{eq:1+-Diquark}).

\subsection{Mixed operators}\label{subsec:Mixed operators}

A physical state can couple to any current that shares the same quantum numbers, therefore, it can also couple to a mixture of these currents. And the operator mixing has significant effects on the QCD sum rules calculations, say, for the heavy baryon spectrum~\cite{Wang:2017qvg} and fully heavy tetraquark spectrum\cite{Wu:2022qwd}. However, the LO calculation of the Wilson coefficients alone provides no guidance to pin down the mixing scheme of the operators. Thanks to the NLO calculations, the currents are mixed with each other naturally under the renormalization. If one choose the basis which diagonalizes the anomalous dimension matrix, the dependence on the renormalization scale $\mu$ tends to be cancelled out on the righthand side of Eq.~(\ref{eq:MH}), which is desirable since the left-hand side $M_H^2$ is a physical quantity.

In this mixing scheme, no matter with which type (Eq.~(\ref{eq:1+-meson}) or (\ref{eq:1+-Diquark})) operators basis we start from, we always get the same mixed operators. In this paper, we choose the meson-meson type currents in Eq.~(\ref{eq:1+-meson}) as the operator basis, the operator anomalous dimension matrix $\gamma^{\text{\text{M-M}}}$ can be obtained according to the operator renormalization matrix Eq.~(\ref{eq:1+-meson-renormalization}),
\begin{equation}\label{eq:1+-meson-AnoDim}
	\gamma ^{\text{M-M}}= \frac{\mathrm{d}\, \ln(Z_O^{\text{M-M}})}{\mathrm{d}\, \ln(\mu^2)}=\frac{\delta}{3} \begin{pmatrix}
  48 &  0 & -4&12&0 &0&0&0 \\
 18 &  -6 & 6 &14&0 &0&0&0 \\
 -12 &  36 & -16 &0&0 &0&0&0 \\
 18 &  42 & -18& 38 &0 &0&0&0\\
   0 &  0 & 0 &0&48&0 &4 &-12 \\
     0 &  0 & 0 &0&18& -6 &-6 &-14 \\
     0 &  0 & 0 &0&12&-36 &-16 &0 \\
     0 &  0 & 0 &0&-18& -42 &-18 &38 \\
\end{pmatrix}\,,
\end{equation}
where $\delta=-\frac{\alpha_s}{16 \pi }$.
To diagonalize the matrix in Eq.~(\ref{eq:1+-meson-AnoDim}), one needs the following transformation matrix
\begin{equation}\label{eq:1+-meson-Diagonal}
\mathcal{T}^{\text{\text{Dia}}}=\begin{pmatrix}
  \frac{3}{8} + \frac{45}{8\sqrt{241}}& \frac{3}{\sqrt{241}} &-\frac{1}{8} + \frac{1}{8\sqrt{241}}&\frac{1}{4} + \frac{11}{4\sqrt{241}}&0&0&0&0  \\
  -\frac{3}{8} - \frac{63}{8\sqrt{241}}& \frac{15}{2\sqrt{241}} &\frac{1}{8} - \frac{19}{8\sqrt{241}}&\frac{1}{4} + \frac{11}{4\sqrt{241}}&0&0&0&0 \\
  \frac{3}{8} - \frac{45}{8\sqrt{241}}& -\frac{3}{\sqrt{241}} &-\frac{1}{8} - \frac{1}{8\sqrt{241}}&\frac{1}{4} - \frac{11}{4\sqrt{241}}&0&0&0&0  \\
    -\frac{3}{8} + \frac{63}{8\sqrt{241}}&- \frac{15}{2\sqrt{241}} &\frac{1}{8} + \frac{19}{8\sqrt{241}}&\frac{1}{4} - \frac{11}{4\sqrt{241}}&0&0&0&0 \\
    0&0&0&0& - \frac{3}{8} - \frac{45}{8\sqrt{241}}& -\frac{3}{\sqrt{241}} &-\frac{1}{8} + \frac{1}{8\sqrt{241}}&\frac{1}{4} + \frac{11}{4\sqrt{241}} \\
     0&0&0&0&   \frac{3}{8} + \frac{63}{8\sqrt{241}}&- \frac{15}{2\sqrt{241}} &\frac{1}{8} - \frac{19}{8\sqrt{241}}&\frac{1}{4} + \frac{11}{4\sqrt{241}} \\
      0&0&0&0& -\frac{3}{8} + \frac{45}{8\sqrt{241}}& \frac{3}{\sqrt{241}} &-\frac{1}{8} - \frac{1}{8\sqrt{241}}&\frac{1}{4} - \frac{11}{4\sqrt{241}} \\
       0&0&0&0&  \frac{3}{8} - \frac{63}{8\sqrt{241}}& \frac{15}{2\sqrt{241}} &\frac{1}{8} + \frac{19}{8\sqrt{241}}&\frac{1}{4} - \frac{11}{4\sqrt{241}}\\
\end{pmatrix}\,.
\end{equation}
So, we get the new basis
\begin{equation}\label{eq:1+-diagonal}
\vec{J}^{\text{\text{Dia}}}=\mathcal{T}^{\text{\text{Dia}}}\cdot \vec{J}^{\text{\text{M-M}}}\,.
\end{equation}
And the anomalous dimension matrix of $\vec{J}^{\text{\text{Dia}}}$ is diagonal, which is given by
\begin{equation}\label{eq:1+-Diagonal-AnoDim}
\begin{split}
& \gamma^{\text{\text{Dia}}} =\mathcal{T}^{\text{\text{Dia}}}\cdot \gamma^{\text{\text{M-M}}}\cdot (\mathcal{T}^{\text{\text{Dia}}})^{-1}\,\\
& =\frac{2}{3} \delta \begin{pmatrix}
  17+\sqrt{241} &  0 & 0&0&0&  0 & 0&0 \\
   0 &  -1+\sqrt{241} & 0&0&0 &  0 & 0&0\\
      0 &  0 &  17-\sqrt{241}&0&0&  0 & 0&0 \\
         0 &  0 & 0&-1-\sqrt{241}&0&  0 & 0&0 \\
            0 &  0 & 0&0& 17+\sqrt{241}&  0 & 0&0 \\
             0 &  0 & 0&0&0 &   -1+\sqrt{241} & 0&0 \\
              0 &  0 & 0&0& 0&  0 & 17-\sqrt{241}&0 \\
               0 &  0 & 0&0& 0&  0 & 0&-1-\sqrt{241} \\
\end{pmatrix}\,.
\end{split}
\end{equation}

From Eq.~(\ref{eq:1+-Diagonal-AnoDim}), one can see that the eigenvalues of anomalous dimension matrix are degenerate. Since the operators with the same eigenvalue can be further mixed with each other, we defined the following operators as
 \begin{align}\label{eq:1+-mixing}
	\begin{split}
		J_{1,5}^{\text{\text{Mixed}}}&=\cos(\theta_1) J_{1}^{\text{\text{Dia}}}\, + \, \sin(\theta_1) e^{i\phi_1}\,  J_{5}^{\text{\text{Dia}}}   \,, \\
		J_{2,6}^{\text{\text{Mixed}}}&= \cos(\theta_2) J_{2}^{\text{\text{Dia}}}\, + \, \sin(\theta_2) e^{i\phi_2}\,  J_{6}^{\text{\text{Dia}}}  \,, \\
		J_{3,7}^{\text{\text{Mixed}}}&= \cos(\theta_3) J_{3}^{\text{\text{Dia}}}\, + \, \sin(\theta_3) e^{i\phi_3}\,  J_{7}^{\text{\text{Dia}}}  \,, \\
		J_{4,8}^{\text{\text{Mixed}}}&= \cos(\theta_4) J_{4}^{\text{\text{Dia}}}\, + \, \sin(\theta_4) e^{i\phi_4}\,  J_{8}^{\text{\text{Dia}}}   \,, \\
	\end{split}
\end{align}
where the parameters $\theta_i (i=1,2,3,4)$ and $\phi_i (i=1,2,3,4)$ are real.

In the following, we call $\vec{J}^{\text{\text{Mixed}}}$ as mixed operators.

\section{Phenomenology}\label{sec:Phenomenology}
In our numerical analysis, we choose the following parameters~\cite{Bagan:1992za,Dominguez:1994ce,Dominguez:2014pga,Aoki:2016frl,Wang:2017qvg},
\begin{align}\label{eq:parameters}
\begin{split}
m_c^{\overline{\text{MS}}}(m_c)&=1.27 \pm 0.03\, \,  {\text{GeV}}\,,\\
m_c^{\text{OS}}&=1.46 \pm 0.07\, \,  {\text{GeV}}\,\\
m_u(2\text{GeV})&=2.16^{+0.49}_{-0.26}\, \,  {\text{MeV}}\,,\\
m_d(2\text{GeV})&=4.67^{+0.48}_{-0.17}\, \,  {\text{MeV}}\,,\\
\langle \bar{q}q \rangle(2\text{GeV})&=-(0.280\pm 0.017 \text{GeV})^3\, \,  ,\\
\langle g_s^2 GG \rangle &= 4\pi^2(0.037\pm0.015) \, \, {\text{GeV}}^4 \,,\\
\langle g_s \bar{q}q G \rangle(2\text{GeV})&=(0.8\pm 0.2 \text{GeV}^2)\times \langle \bar{q}q \rangle(2\text{GeV})\, \,  ,\\
\alpha_s(m_Z&=91.1876 ~{\text{GeV}})=0.1181\,.
\end{split}
\end{align}
where $q$ denotes light quark $u, d$. Moreover, it is worth emphasizing that consistent with our calculation, $\alpha_{s}(\mu)$ and the heavy quark mass $m_c^{\overline{\text{MS}}}(\mu)$ are obtained through two-loop running, where $\mu$ is the renormalization scale. $\langle \bar{q}q \rangle(\mu)$ and $\langle g_s \bar{q}q G \rangle(\mu)$ are obtained through one-loop running, and their anomalous dimensions are given by~\cite{Albuquerque:2012jbz}
\begin{align}\label{eq:qq-anomalousD}
	\begin{split}
		\gamma_{ \langle \bar{q}q \rangle}&=-\gamma_{m_q}\, \, ,\\
		\gamma_{\langle g_s \bar{q}q G \rangle}&=-\frac{\gamma_{m_q}}{6}\, \, ,\\
		\gamma_{m_q}&=-\frac{3 C_F \alpha_s}{4\pi}\, ,
	\end{split}
\end{align}
Since the anomalous dimension vanishes at one-loop level, we don't need to consider the running of the $GG$ condensate $\langle g_s^2 GG \rangle$. As a typical choice, we set $\mu=M_B$ in our phenomenological analysis~\cite{Shifman:1978bx,Bertlmann:1981he}, but the renormalization scale dependence will also be discussed.

According to Eq.~(\ref{eq:MH}), numerical result $M_H$ also depends on other two parameters: $s_0$ and $M_B^2$. However, the  physical value of $M_H$ should be independent of any artificial parameters. So a credible result should be obtained from an appropriate region where the dependence of $s_0$ and $M_B^2$ is weak. On the other hand, the choice of $M_B^2$ and $s_0$ should ensure the validity of the OPE and ground-state contribution dominance, which constrain the two parameters to be within the so-called ``Borel window''. Within the Borel window, one should find the region, the so-called ``Borel platform'', in which $M_H$ depends on $s_0$ and $M_B^2$  weakly.

To search for the Borel window, we define the relative contributions of the condensate and continuum as
\begin{align}\label{eq:borelwindow}

r_{i} =\frac{\langle O_i \rangle\int_{s_{\text {th}}}^{\infty} \mathrm{d} s\, \, \rho_{i}(s)\, e^{-\frac{s}{M_B^2}}}{\int_{s_{\text{th}}}^{\infty} \mathrm{d} s\, \, \rho_{1}(s)\, e^{-\frac{s}{M_B^2}}}\,,  \quad
r_{\text{cont}} =\frac{\int_{s_{0}}^{\infty} \mathrm{d} s\, \, \rho_{1}(s)\, e^{-\frac{s}{M_B^2}}}{\int_{s_{\text {th}}}^{\infty} \mathrm{d} s\, \, \rho_{1}(s)\, e^{-\frac{s}{M_B^2}}}\,,

\end{align}
where $\langle O_i \rangle \in \{ \langle \bar{q} q \rangle,\,  \langle g_s^2 GG \rangle ,\,  \langle g_s \bar{q} q G \rangle \}$~\footnote{Here, $\bar{q} q$ means the sum of $\bar{u} u$ and $\bar{d} d$.}. And we impose the following constraints:
\begin{align}\label{eq:BWcondition}
|r_{i}|\leq t_r, \, \quad \, | {\textstyle \sum_{i}} r_{i}|\leq t_r,\, \quad \, |r_{\text{cont}}|\leq t_c \,.
\end{align}
To guarantee the validity of OPE and the ground-state contribution dominance, there are constraints for $t_r$ and $t_c$, that $t_r \le 1 $ and $t_c \le 0.5$. In addition to the conditions given in Eq.~(\ref{eq:BWcondition}), we also impose the following constraint on $s_0$:
\begin{equation}\label{eq:S0condition}
s_0<(M_H+1 \ \text{GeV})^2,
\end{equation}
since, roughly speaking, $s_0$ denotes the energy scale where the continuum spectrum begins to contribute and that the binding energy for a hadron is usually smaller than 1~GeV.

To find the Borel platform, we search for the point where the parametric dependence of $M_H$ is weakest within the Borel window.
More explicitly, we choose the variables as $x=s_0$ and $y=M_B^2$ and define the function
\begin{align}\label{eq:findPlatform}
\Delta(x,y)=\left( \frac{\partial M_H}{\partial x} \right)^2+ \left( \frac{\partial M_H}{\partial y}\right)^2\,.
\end{align}
By minimizing the function $\Delta(x,y)$ within the Borel window and with the constraint Eq.~(\ref{eq:S0condition}), we get a point ($x_0,y_0$), which will be used to calculate the central value of $M_H$. To estimate errors of $M_H$, we vary the values of $s_0$ and $M_B^2$ around the point ($x_0,y_0$) up to 10\% in magnitude. It should be emphasized that the central point ($x_0,y_0$) may lies on the margin of the Borel window in some cases. Therefore, the parameter space used to estimate errors of $M_H$ may exceed the Borel window, and also, the upper and the lower errors are usually asymmetric.

\subsection{Numerical results for meson-meson type and diquark-antidiquark type operators of $Z_c$ system}

The comprehensive results for both meson-meson type (see  Eq.~(\ref{eq:1+-meson})) and diquark-antidiquark type (see  Eq.~(\ref{eq:1+-Diquark})) operators are shown in Tab.~\ref{tab:A-meson-NLOresult-MSbar}--\ref{tab:A-Diquark-NLOresult-OS} in Appendix~\ref{sec:details}, where we include both LO and NLO,  both $\overline{\text{MS}}$ scheme and on-shell scheme for the results of $Z_c^+$ mass $M_H$. In these tables, we set $\mu=M_B$ and thus the errors of $M_H$ are due to choices of $s_0$ and $M_B^2$. Further information of $s_0$ and $M_B^2$ dependence is shown in  Fig.~\ref{fig:Zc-[1+]-M-G-1-NLO-MSbar-OS}--\ref{fig:Zc-[1+]-Di-G-8-NLO-MSbar-OS} in Appendix~\ref{sec:details}. In these plots, a black dot denotes the central point ($x_0,y_0$), and shadows denote the Borel window determined by Eq.~(\ref{eq:BWcondition-meson-diquark}).

\begin{figure}[htb]
	\vspace{-0.3cm}
	\begin{center}
		\includegraphics[scale=0.47]{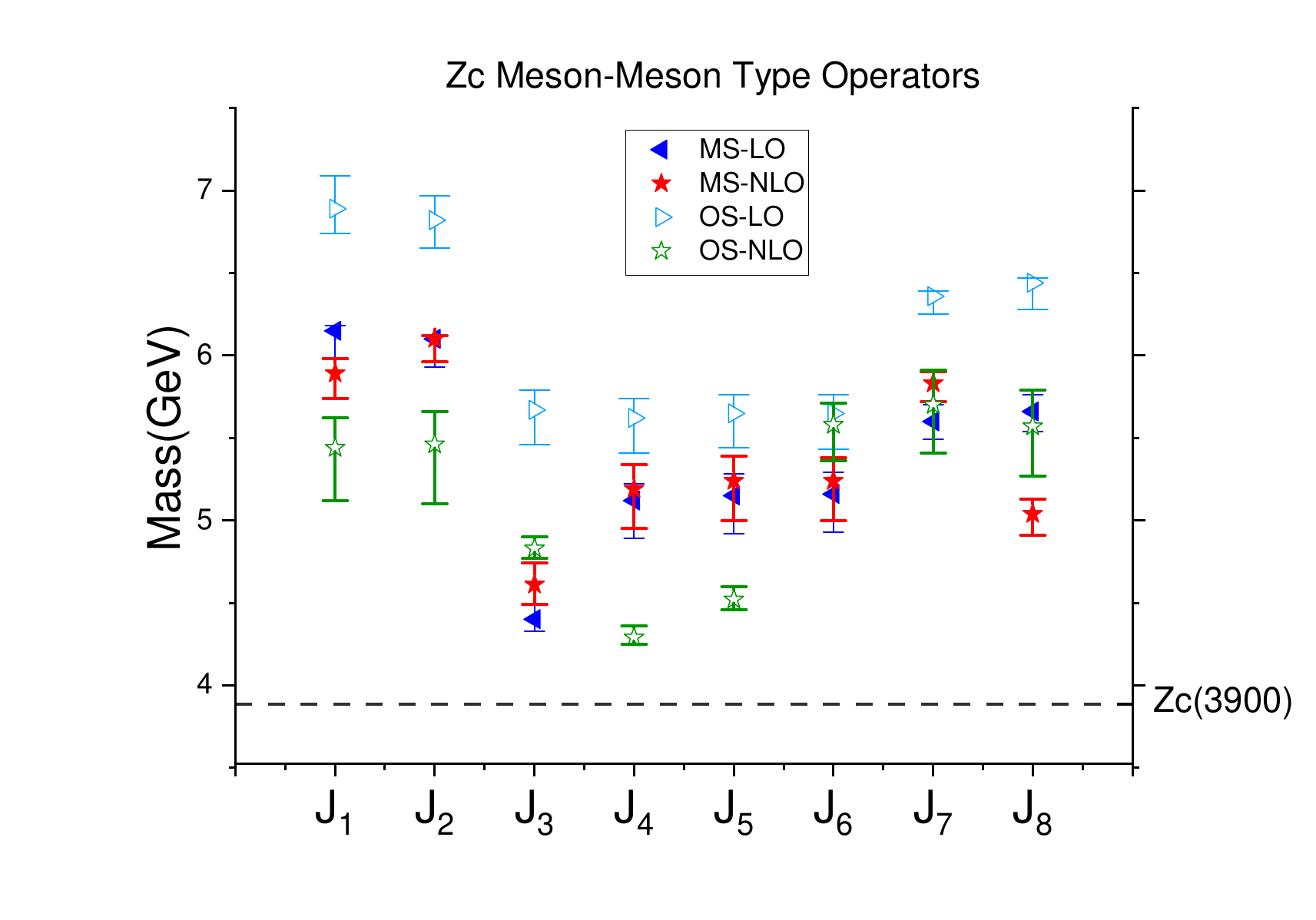}
		\vspace{-0.7cm}
		\caption{\label{fig:Zc-meson-Mass-Spectrum}
			The LO and NLO mass spectra of meson-meson type operators of $Z_c$ system in the $\overline{\text{MS}}$ and $\text{OS}$ schemes. The errors of masses shown in this figure just come from the parametric dependence on $s_0$ and $M_B^2$, listed in Tab.~\ref{tab:A-meson-NLOresult-MSbar}--\ref{tab:A-meson-NLOresult-OS} in Appendix~\ref{sec:details}.}
	\end{center}
\end{figure}

\begin{figure}[htb]
	\vspace{-0.6cm}
	\begin{center}
		\includegraphics[scale=0.47]{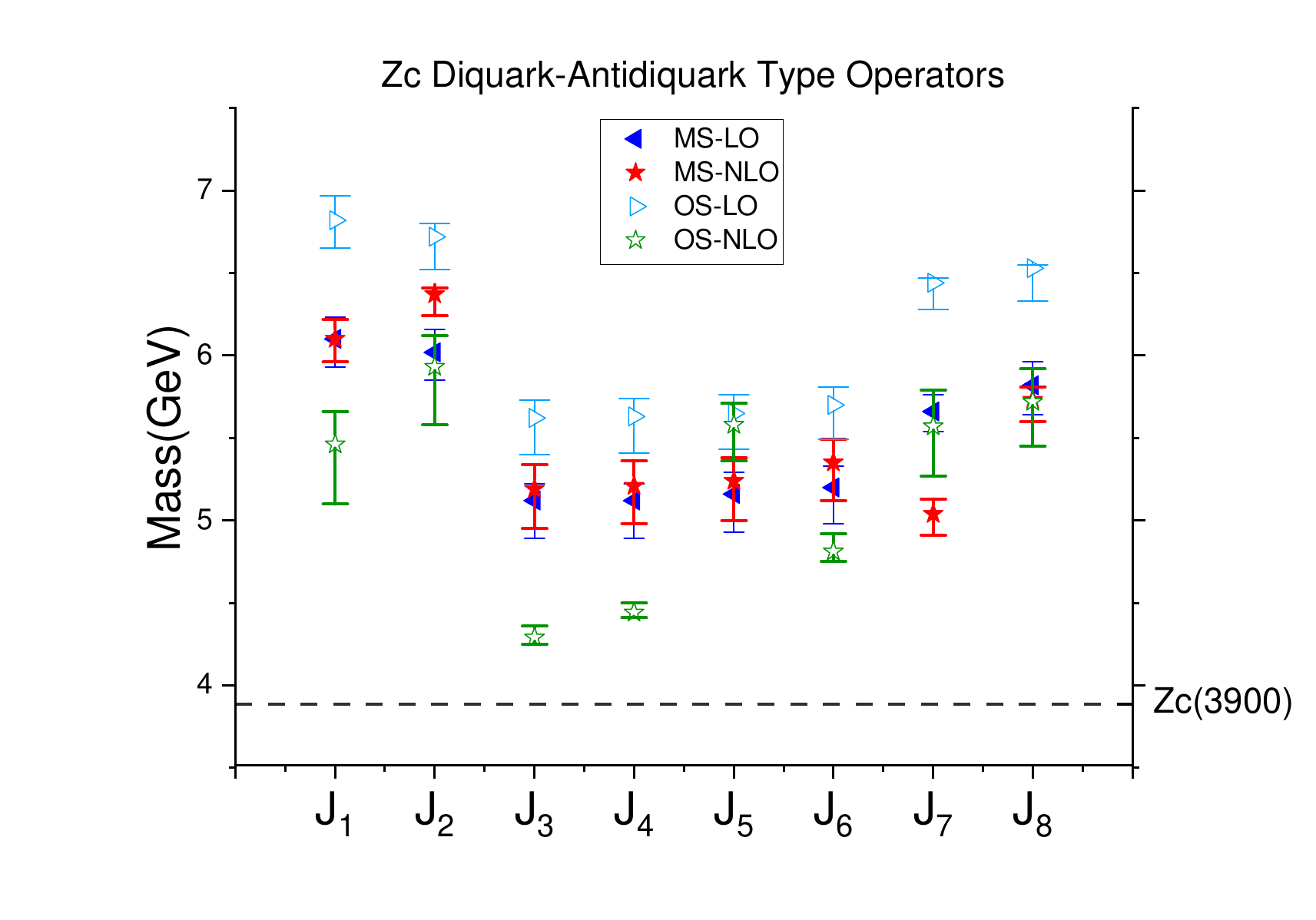}
		\vspace{-0.7cm}
		\caption{\label{fig:Zc-diquark-Mass-Spectrum}
			The LO and NLO mass spectra of diquark-antidiquark type operators of $Z_c$ system in the $\overline{\text{MS}}$ and $\text{OS}$ schemes. The errors of masses shown in this figure just come from the parametric dependence on $s_0$ and $M_B^2$, listed in Tab.~\ref{tab:A-Diquark-NLOresult-MSbar}--\ref{tab:A-Diquark-NLOresult-OS} in Appendix~\ref{sec:details}.}
	\end{center}
\end{figure}

We also summarize the results of Tab.~\ref{tab:A-meson-NLOresult-MSbar}--\ref{tab:A-meson-NLOresult-OS} (for meson-meson type operators) and Tab.~\ref{tab:A-Diquark-NLOresult-MSbar}--\ref{tab:A-Diquark-NLOresult-OS} (for  diquark-antidiquark type operators) in Fig.~\ref{fig:Zc-meson-Mass-Spectrum} and Fig.~\ref{fig:Zc-diquark-Mass-Spectrum}, respectively. Through these figures, one can see that the NLO corrections are important, especially for the on-shell scheme. However, we should emphasize that to obtain the Borel windows for both meson-meson type and diquark-antidiquark type operators, one needs very loose constraints for the Borel conditions given in Eq.~(\ref{eq:borelwindow}), which are corresponding to
\begin{align}\label{eq:BWcondition-meson-diquark}
	|r_{i}|\leq 100\%, \, \quad \, | {\textstyle \sum_{i}} r_{i}|\leq 100\%,\, \quad \, |r_{\text{cont}}|\leq 50\% \,.
\end{align}
Thus, the convergence of the OPE may be bad, and the results shown in Fig.~\ref{fig:Zc-meson-Mass-Spectrum} and Fig.~\ref{fig:Zc-diquark-Mass-Spectrum} may suffer severe theoretical uncertainties. Furthermore, the qualities of these Borel platforms are bad (see Fig.~\ref{fig:Zc-[1+]-M-G-1-NLO-MSbar-OS}--\ref{fig:Zc-[1+]-Di-G-8-NLO-MSbar-OS}), which cause large errors of $M_H$'s from their dependence on $s_0$ and $M_B^2$ (see Tab.~\ref{tab:A-meson-NLOresult-MSbar}--\ref{tab:A-Diquark-NLOresult-OS}). In addition, although the NLO corrections tend to reduce the mass gap between $\overline{\text{MS}}$ scheme and on-shell scheme, the NLO mass difference  $ \left|M_H^{\rm{NLO}\mbox{-}\rm{OS}} - M_H^{\rm{NLO}\mbox{-}\overline{\rm{MS}}} \right|> 0.5$~GeV for most of these operators (see Fig.~\ref{fig:Zc-meson-Mass-Spectrum} and Fig.~\ref{fig:Zc-diquark-Mass-Spectrum}), which indicates that the perturbative convergence of the results for these operators may be bad.

Due to large uncertainties of these results, we will not use the masses shown in Fig.~\ref{fig:Zc-meson-Mass-Spectrum} and Fig.~\ref{fig:Zc-diquark-Mass-Spectrum} to do further phenomenological analysis, although we have shown the mass of $Z_c(3900)$ as a horizontal line in these figures.

\newpage

\subsection{Numerical results for mixed operators of $Z_c$ system}

As have been mentioned in the subsection \ref{subsec:Mixed operators}, the mixed operators given in Eq.~(\ref{eq:1+-mixing}) are more adequate to be used in the QCD sum rules. On the one hand, due to the $\mu$-dependence tends to be cancelled out in the righthand side of Eq.~(\ref{eq:MH}) for the mixed operator, the corresponding results are expected to have better perturbative convergence compared to those given in last subsection. On the other hand, one can reduce the $s_0$- and $M_B^2$-dependence of the results by choosing ideally mixing parameters $\theta_i$ and $\phi_i$ defined in Eq.~(\ref{eq:1+-mixing}) to get Borel platforms with high quality.

We scan in the  parameter spaces of $\theta_i$ and $\phi_i$, and determine the following mixing parameters through minimizing $s_0$- and $M_B^2$-dependence of the results:
\begin{align}\label{eq:mixedparameters}
	\begin{split}
		\theta_1 &= 48^{\circ}, \, \quad  \quad \, \phi_1 = 20^{\circ}	\, , \\
		\theta_2 &= 48^{\circ}, \, \quad  \quad \, \phi_2 = 0^{\circ}	\, , \\
		\theta_3 &= 46^{\circ}, \, \quad  \quad \, \phi_3 = 220^{\circ}	\, , \\
		\theta_4 &= 42^{\circ}, \, \quad  \quad \, \phi_4 = 30^{\circ}	\, , \\
	
	\end{split}
\end{align}
Thus we determine four ideally mixed operators by inserting Eq.~(\ref{eq:mixedparameters}) into Eq.~(\ref{eq:1+-mixing}). Comparing with the un-mixed operators given in  Eq.~(\ref{eq:1+-meson}) and  Eq.~(\ref{eq:1+-Diquark}), for these ideally mixed operators we can find better Borel windows even if we choose relative strict constraints
\begin{align}\label{eq:BWcondition-mixed}
	|r_{i}|\leq 60\%, \, \quad \, | {\textstyle \sum_{i}} r_{i}|\leq 60\%,\, \quad \, |r_{\text{cont}}|\leq 30\% \,.
\end{align}
Thus, both the convergence of the OPE and  the validity of the ground state dominance are guaranteed better after considering mixed effect.

\begin{figure}[htb]
	\vspace{-0.3cm}
	\setlength{\belowcaptionskip}{-0.4cm}
	\begin{center}
		\includegraphics[scale=0.5]{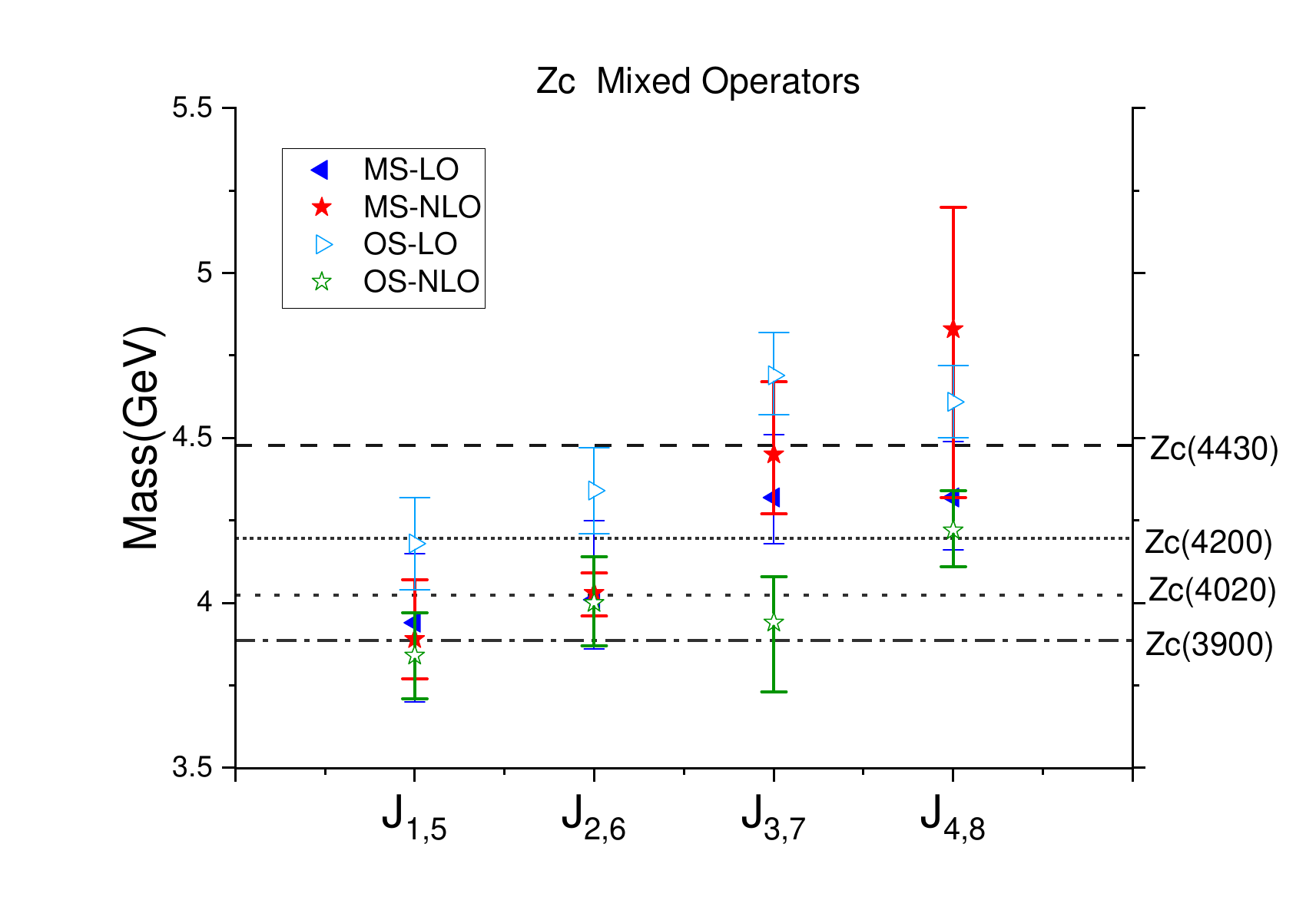}
		\vspace{-0.6cm}
		\caption{\label{fig:Zc-mixed-Mass-Spectrum}
			The LO and NLO mass spectra of mixed operators of $Z_c$ system in the $\overline{\text{MS}}$ and $\text{OS}$ schemes. The errors of masses shown in this figure just come from the parametric dependence on $s_0$, $M_B^2$, heavy quark mass $m_Q$ and renormalization scale $\mu$, which are shown in the Tab.\ref{tab:tabel-Mixed-1-MSbar-OS}-\ref{tab:tabel-Mixed-4-MSbar-OS}.}
	\end{center}
\end{figure}
The results for these ideally mixed operators are listed in Tab.~\ref{tab:Zc-Mixed-MSbar}-\ref{tab:Zc-Mixed-OS} in Appendix~\ref{sec:details}, where we include both LO and NLO,  both $\overline{\text{MS}}$ scheme and on-shell scheme for the results of $Z_c^+$ mass $M_H$. In these tables, again, we set $\mu=M_B$ and thus errors of $M_H$ are only due to choices of $s_0$ and $M_B^2$. Further information of $s_0$ and $M_B^2$ dependence is shown in  Fig.~\ref{fig:Zc-[1+]-M-Mixed-1-NLO-MSbar-OS}--\ref{fig:Zc-[1+]-M-Mixed-4-NLO-MSbar-OS} in Appendix~\ref{sec:details}. In these plots, a black dot denotes the central point ($x_0,y_0$), and shadows denote the Borel window determined by Eq.~(\ref{eq:BWcondition-meson-diquark}). Form Fig.~\ref{fig:Zc-[1+]-M-Mixed-1-NLO-MSbar-OS}--\ref{fig:Zc-[1+]-M-Mixed-4-NLO-MSbar-OS}, one can see the quality of the Borel platforms are evidently better than those for the un-mixed operators, thus, the errors of the results listed in Tab.~\ref{tab:Zc-Mixed-MSbar}-\ref{tab:Zc-Mixed-OS} are vastly smaller than those in Tab.~\ref{tab:A-meson-NLOresult-MSbar}--\ref{tab:A-Diquark-NLOresult-OS}.

\begin{figure}[H]
	\centering
	\subfigure[$J_{1,5}^{\text{Mixed}}$]{
		\includegraphics[scale=0.27]{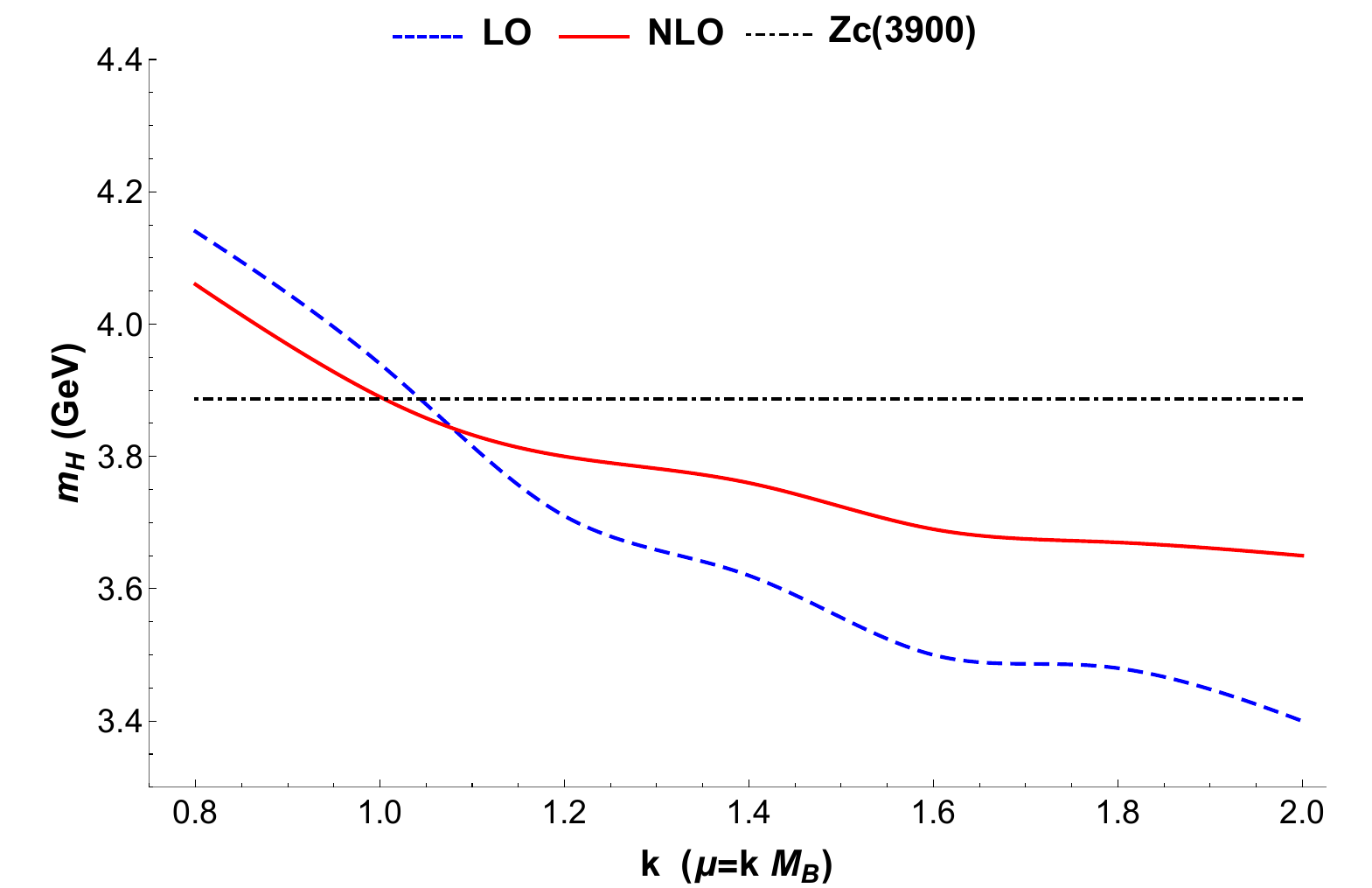}
	}
	\subfigure[$J_{2,6}^{\text{Mixed}}$]{
		\includegraphics[scale=0.27]{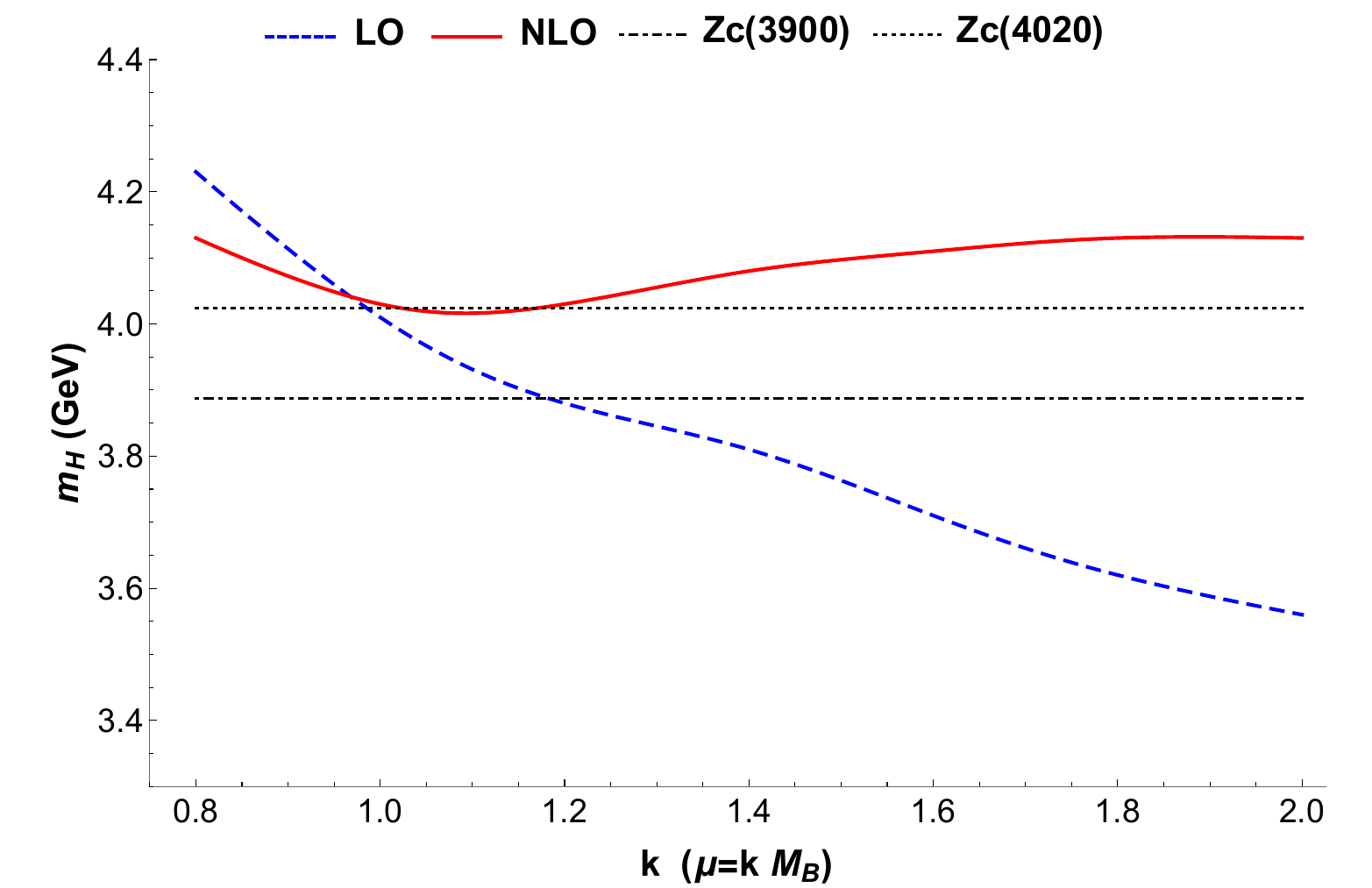}
	}\\
	\subfigure[$J_{3,7}^{\text{Mixed}}$]{
		\includegraphics[scale=0.21]{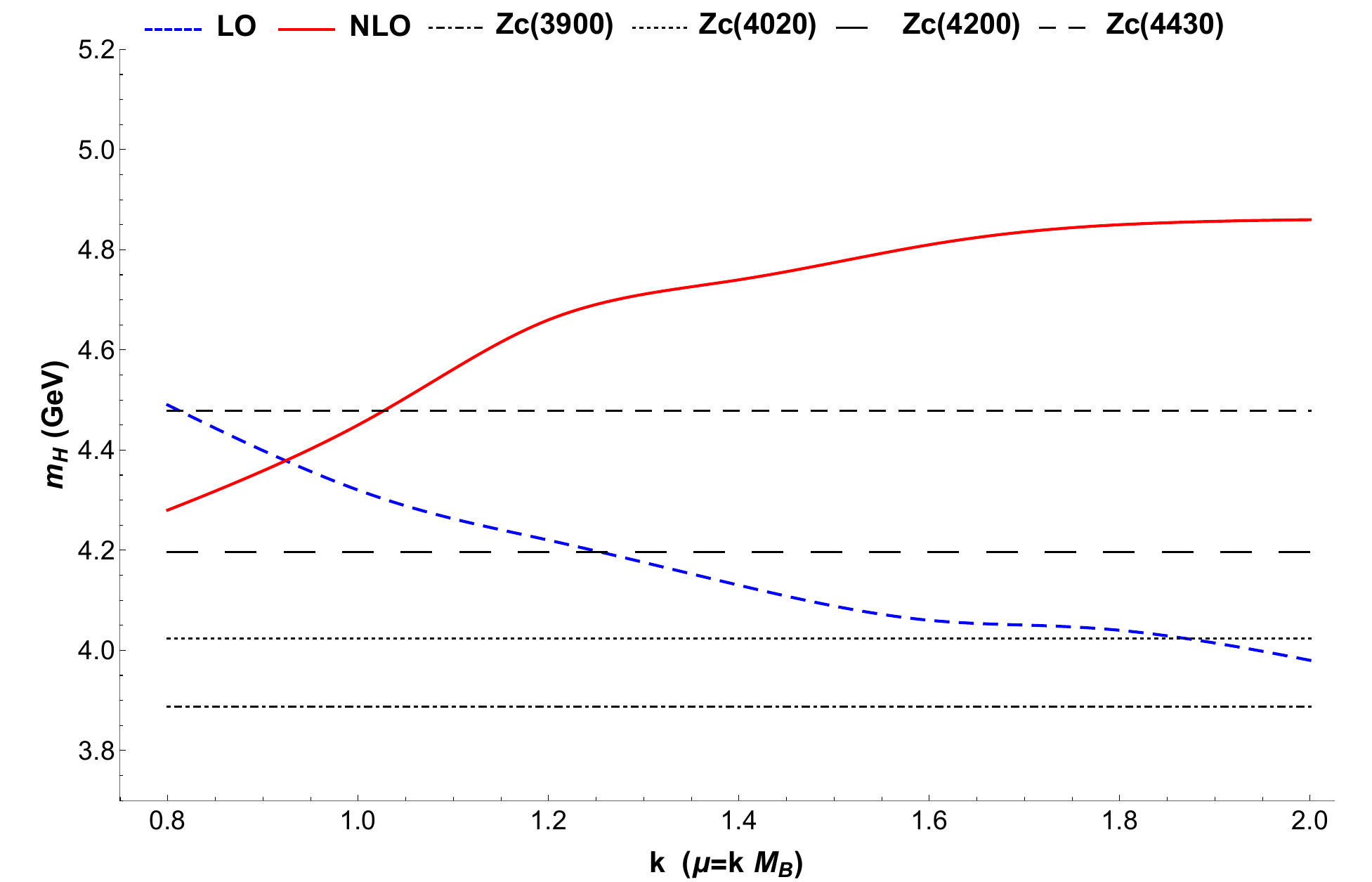}
	}
	\subfigure[$J_{4,8}^{\text{Mixed}}$]{
		\includegraphics[scale=0.21]{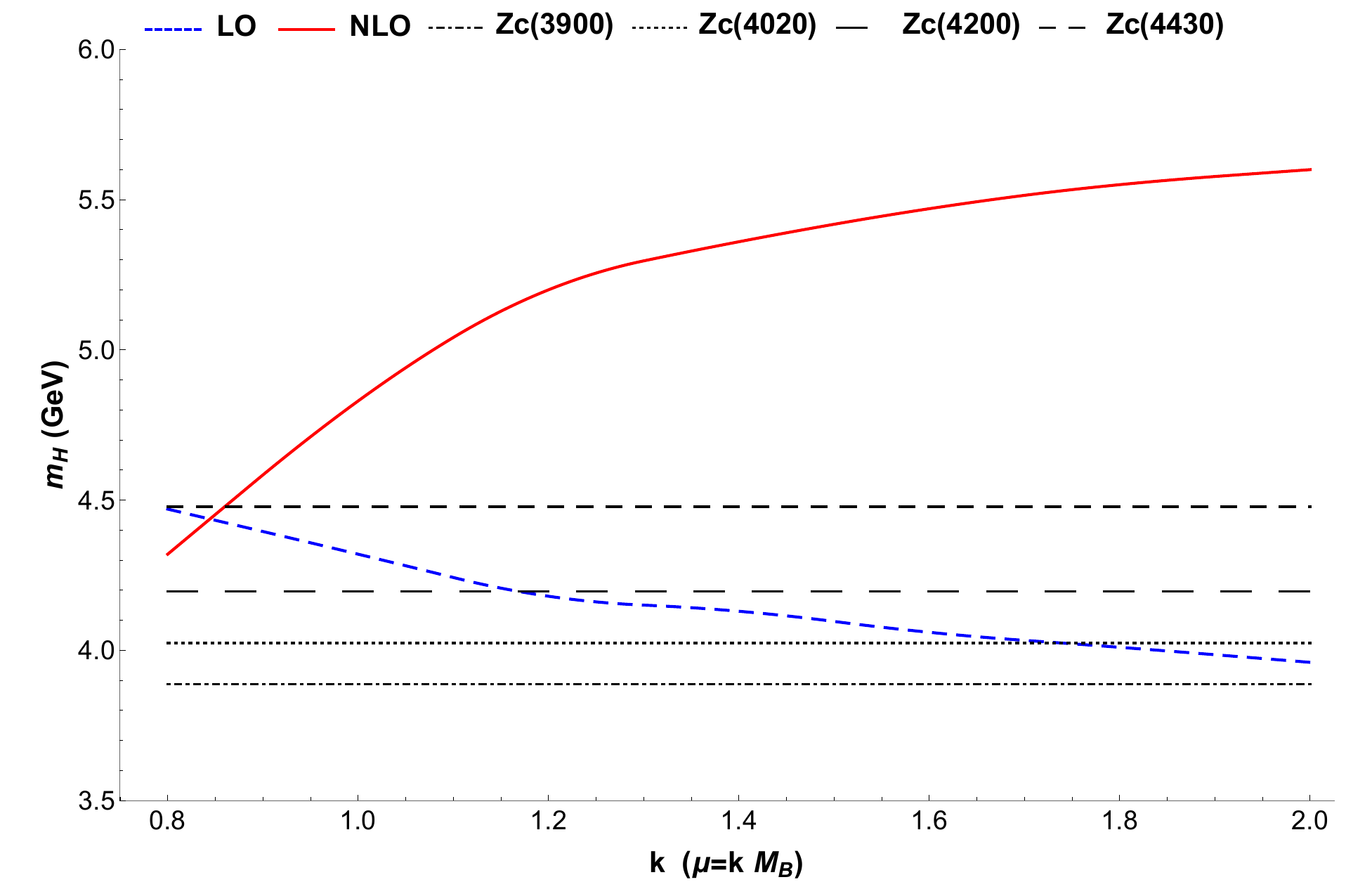}
	}
	\caption{\label{fig:mu-dependence-mixed}The $\mu$-dependence of the LO and NLO results for mixed operators $J_{i}^{\text{Mixed}}$ of $Z_c$ system in the $\overline{\rm{MS}}$ scheme. (a) for $J_{1,5}^{\text{Mixed}}$, (b) for $J_{2,6}^{\text{Mixed}}$, (c) for $J_{3,7}^{\text{Mixed}}$ and (d) for $J_{4,8}^{\text{Mixed}}$.}
\end{figure}

We also explore the $m_c$- and the $\mu$-dependence of the above results, and the corresponding errors coming from these dependence are listed in Tab.~\ref{tab:tabel-Mixed-1-MSbar-OS}-\ref{tab:tabel-Mixed-4-MSbar-OS} for the four ideally mixed operators, respectively. To explore the $\mu$-dependent of the $\overline{\text{MS}}$ mass $M_H^{\overline{\rm{MS}}}$, we set $\mu=k M_B$ and vary the value of $k$ from $0.8$ to $2.0$. The $\mu$-dependent of both the LO and NLO $\overline{\text{MS}}$ masses are shown in Fig.~\ref{fig:mu-dependence-mixed}. Furthermore, we also summarize the $M_H$ results of Tab.~\ref{tab:tabel-Mixed-1-MSbar-OS}-\ref{tab:tabel-Mixed-4-MSbar-OS} with errors in Fig.~\ref{fig:Zc-mixed-Mass-Spectrum}.

\begin{table}[H]
	\vspace{-0.3cm}
	\renewcommand\arraystretch{1.5}
	\begin{center}
		\setlength{\tabcolsep}{1.2 mm}
		\begin{tabular}{cccccccccc}
			\hline\hline
			Current& Order&   $M_H$ (GeV)   &  $s_0$ (${\text{GeV}}^2$)  & $M_B^2$ (${\text{GeV}}^2$)  & \makecell{Error from \\$s_0$ and $M_B^2$ } & \makecell{Error from \\$m_c$ } & \makecell{Error from \\$\mu$} \\ \hline
			
			\multirow{4}{*}{$J_{1,5}^{\text{Mixed}}$} & LO($\overline{\text{MS}}$) &    $3.94^{+0.21}_{-0.24}$   &    $24(\pm 10\%)$   &   $1.30(\pm 10\%)$   &  $^{+0.02}_{-0.02}$ &  $^{+0.07}_{-0.08}$ &  $^{+0.2}_{-0.23}$ \\
			
			&NLO($\overline{\text{MS}}$) &   $3.89^{+0.18}_{-0.12}$   &    $22(\pm 10\%)$   &   $1.10(\pm 10\%)$    &  $^{+0.02}_{-0.02}$ &  $^{+0.07}_{-0.07}$ &  $^{+0.17}_{-0.09}$ \\
			
			&LO(OS)     &         $4.18^{+0.14}_{-0.14}$         &      $23(\pm 10\%)$    &   $1.30(\pm 10\%)$    & $^{+0.04}_{-0.04}$ &  $^{+0.13}_{-0.13}$ & \\
			
			&NLO(OS)     &         $3.84^{+0.13}_{-0.13}$           &   $17(\pm 10\%)$     &   $1.00(\pm 10\%)$    & $^{+0.07}_{-0.09}$ &  $^{+0.11}_{-0.1}$ &\\
			\hline\hline
		\end{tabular}
		\caption{The LO and NLO results for the mass of $J_{1,5}^{\text{Mixed}}$ of $Z_c$ system in $\overline{\text{MS}}$ and On-Shell schemes. Here the errors for $M_H$ are from $s_0, M_B$, the charm quark mass, and the renormalization scale $\mu$ with $\mu=kM_B$ and $k \in (0.8, 1.2)$ (the central values correspond to $\mu=M_B$).}
		\label{tab:tabel-Mixed-1-MSbar-OS}
	\end{center}
\end{table}

\begin{table}[H]
	\vspace{-0.4cm}
	\renewcommand\arraystretch{1.5}
	\begin{center}
		\setlength{\tabcolsep}{1.2 mm}
		\begin{tabular}{cccccccccc}
			\hline\hline
			Current& Order&   $M_H$ (GeV)   &  $s_0$ (${\text{GeV}}^2$)  & $M_B^2$ (${\text{GeV}}^2$)  & \makecell{Error from \\$s_0$ and $M_B^2$ } & \makecell{Error from \\$m_c$ } & \makecell{Error from \\$\mu$} \\ \hline
			
			\multirow{4}{*}{$J_{2,6}^{\text{Mixed}}$} & LO($\overline{\text{MS}}$) &    $4.01^{+0.24}_{-0.15}$   &    $22(\pm 10\%)$   &   $1.50(\pm 10\%)$   &  $^{+0.06}_{-0.05}$ &  $^{+0.06}_{-0.06}$ &  $^{+0.22}_{-0.13}$ \\
			
			&NLO($\overline{\text{MS}}$) &   $4.03^{+0.06}_{-0.07}$   &    $25(\pm 10\%)$   &   $1.20(\pm 10\%)$    &  $^{+0.02}_{-0.02}$ &  $^{+0.06}_{-0.07}$ &  $^{+0.01}_{-0}$ \\
			
			&LO(OS)     &         $4.34^{+0.13}_{-0.13}$         &      $28(\pm 10\%)$    &   $1.40(\pm 10\%)$    & $^{+0.02}_{-0.02}$ &  $^{+0.13}_{-0.13}$ & \\
			
			&NLO(OS)     &         $4.00^{+0.14}_{-0.13}$           &   $22(\pm 10\%)$     &   $1.00(\pm 10\%)$    & $^{+0.02}_{-0.02}$ &  $^{+0.14}_{-0.13}$ &\\
			\hline\hline
		\end{tabular}
		\caption{The LO and NLO results for the mass of $J_{2,6}^{\text{Mixed}}$ of $Z_c$ system in $\overline{\text{MS}}$ and On-Shell schemes. Here the errors for $M_H$ are from $s_0, M_B$, the charm quark mass, and the renormalization scale $\mu$ with $\mu=kM_B$ and $k \in (0.8, 1.2)$ (the central values correspond to $\mu=M_B$).}
		\label{tab:tabel-Mixed-2-MSbar-OS}
	\end{center}
\end{table}

\begin{table}[H]
	\renewcommand\arraystretch{1.5}
	\begin{center}
		\setlength{\tabcolsep}{1.2 mm}
		\begin{tabular}{cccccccccc}
			\hline\hline
			Current& Order&   $M_H$ (GeV)   &  $s_0$ (${\text{GeV}}^2$)  & $M_B^2$ (${\text{GeV}}^2$)  & \makecell{Error from \\$s_0$ and $M_B^2$ } & \makecell{Error from \\$m_c$ } & \makecell{Error from \\$\mu$} \\ \hline
			
			\multirow{4}{*}{$J_{3,7}^{\text{Mixed}}$} & LO($\overline{\text{MS}}$) &    $4.32^{+0.19}_{-0.14}$   &    $26(\pm 10\%)$   &   $1.80(\pm 10\%)$   &  $^{+0.06}_{-0.08}$ &  $^{+0.06}_{-0.05}$ &  $^{+0.17}_{-0.1}$ \\
			
			&NLO($\overline{\text{MS}}$) &   $4.45^{+0.22}_{-0.18}$   &    $28(\pm 10\%)$   &   $1.40(\pm 10\%)$    &  $^{+0.01}_{-0.02}$ &  $^{+0.05}_{-0.04}$ &  $^{+0.21}_{-0.17}$ \\
			
			&LO(OS)     &         $4.69^{+0.13}_{-0.12}$         &      $31(\pm 10\%)$    &   $1.70(\pm 10\%)$    & $^{+0.06}_{-0.03}$ &  $^{+0.11}_{-0.12}$ & \\
			
			&NLO(OS)     &         $3.94^{+0.14}_{-0.21}$           &   $18(\pm 10\%)$     &   $1.30(\pm 10\%)$    & $^{+0.13}_{-0.20}$ &  $^{+0.05}_{-0.06}$ &\\
			\hline\hline
		\end{tabular}
		\caption{The LO and NLO results for the mass of $J_{3,7}^{\text{Mixed}}$ of $Z_c$ system in $\overline{\text{MS}}$ and On-Shell schemes. Here the errors for $M_H$ are from $s_0, M_B$, the charm quark mass, and the renormalization scale $\mu$ with $\mu=kM_B$ and $k \in (0.8, 1.2)$ (the central values correspond to $\mu=M_B$).}
		\label{tab:tabel-Mixed-3-MSbar-OS}
	\end{center}
\end{table}

\begin{table}[H]
	\renewcommand\arraystretch{1.5}
	\begin{center}
		\setlength{\tabcolsep}{1.2 mm}
		\begin{tabular}{cccccccccc}
			\hline\hline
			Current& Order&   $M_H$ (GeV)   &  $s_0$ (${\text{GeV}}^2$)  & $M_B^2$ (${\text{GeV}}^2$)  & \makecell{Error from \\$s_0$ and $M_B^2$ } & \makecell{Error from \\$m_c$ } & \makecell{Error from \\$\mu$} \\ \hline
			
			\multirow{4}{*}{$J_{4,8}^{\text{Mixed}}$} & LO($\overline{\text{MS}}$) &    $4.32^{+0.17}_{-0.16}$   &    $27(\pm 10\%)$   &   $1.80(\pm 10\%)$   &  $^{+0.06}_{-0.06}$ &  $^{+0.05}_{-0.06}$ &  $^{+0.15}_{-0.14}$ \\
			
			&NLO($\overline{\text{MS}}$) &   $4.83^{+0.37}_{-0.51}$   &    $34(\pm 10\%)$   &   $1.90(\pm 10\%)$    &  $^{+0.03}_{-0.02}$ &  $^{+0.04}_{-0.03}$ &  $^{+0.37}_{-0.51}$ \\
			
			&LO(OS)     &         $4.61^{+0.11}_{-0.11}$         &      $28(\pm 10\%)$    &   $1.80(\pm 10\%)$    & $^{+0.07}_{-0.07}$ &  $^{+0.09}_{-0.09}$ & \\
			
			&NLO(OS)     &         $4.22^{+0.12}_{-0.11}$           &   $27(\pm 10\%)$     &   $1.20(\pm 10\%)$    & $^{+0.02}_{-0.02}$ &  $^{+0.12}_{-0.11}$ &\\
			\hline\hline
		\end{tabular}
		\caption{The LO and NLO results for the mass of $J_{4,8}^{\text{Mixed}}$ of $Z_c$ system in $\overline{\text{MS}}$ and On-Shell schemes. Here the errors for $M_H$ are from $s_0, M_B$, the charm quark mass, and the renormalization scale $\mu$ with $\mu=kM_B$ and $k \in (0.8, 1.2)$ (the central values correspond to $\mu=M_B$).}
		\label{tab:tabel-Mixed-4-MSbar-OS}
	\end{center}
\end{table}

For the first two ideally mixed operators $J_{1,5}^{\text{Mixed}}$ and $J_{2,6}^{\text{Mixed}}$, one can find that the NLO corrections are important for both $\overline{\text{MS}}$ and on-shell schemes, and the NLO corrections largely reduce the mass gap between the two schemes. To be specific, for these two operators (see Tab.~\ref{tab:tabel-Mixed-1-MSbar-OS}-\ref{tab:tabel-Mixed-2-MSbar-OS}), the LO mass central value difference of  the two schemes $ \left| M_H^{\rm{LO}\mbox{-}\rm{OS}} - M_H^{\rm{LO}\mbox{-}\overline{\rm{MS}}} \right | \sim 0.3$~GeV, while the NLO mass difference $ \left | M_H^{\rm{NLO}\mbox{-}\rm{OS}} - M_H^{\rm{NLO}\mbox{-}\overline{\rm{MS}}} \right | \leq 0.05$~GeV. This indicates that the perturbative convergence for the results of these two operators are very good, which is also reflected in the weaker $\mu$-dependence of the NLO $\overline{\text{MS}}$ masses than that of the LO ones of these two operators (see Fig.~\ref{fig:mu-dependence-mixed} (a) and (b)).

As for the last two ideally mixed operators $J_{3,7}^{\text{Mixed}}$ and $J_{4,8}^{\text{Mixed}}$, with the NLO corrections the mass gap between the two schemes are not reduced, and is even enlarged for the $J_{4,8}^{\text{Mixed}}$ operator. Correspondingly, the $\mu$-dependence is not improved with the NLO corrections for these two operators (see Fig.~\ref{fig:mu-dependence-mixed} (c) and (d)), and is even worse for the $J_{4,8}^{\text{Mixed}}$ operator. All the above indicate that the perturbative convergence for the results of these two operators are bad, and the higher order corrections may be important for these two operators, which are beyond the scope of this paper.

Phenomenologically, the NLO mass of $J_{1,5}^{\text{Mixed}}$ in the $\overline{\text{MS}}$ scheme is given by $3.89^{+0.18}_{-0.12}$~GeV (see Tab.~\ref{tab:tabel-Mixed-1-MSbar-OS}), which is very close to that of $Z_c(3900)$, and the NLO mass of $J_{2,6}^{\text{Mixed}}$ in the $\overline{\text{MS}}$ scheme is given by $4.03^{+0.06}_{-0.07}$~GeV (see Tab.~\ref{tab:tabel-Mixed-2-MSbar-OS}), which is very close to that of $Z_c(4020)$. Thus our results supports that there are two $Z_c$ states with the same quantum numbers $(I^G)J^{P} =(1^+) 1^{+}$. But it is not sure if there are $Z_c$ states other than $Z_c(3900)$ and $Z_c(4020)$ through the NLO calculations in QCD sum rules, since the results other than those for $J_{1,5}^{\text{Mixed}}$ and $J_{2,6}^{\text{Mixed}}$ suffer large uncertainties and thus are not reliable to make any predictions. Even for the operators $J_{1,5}^{\text{Mixed}}$ and $J_{2,6}^{\text{Mixed}}$, the uncertainties of the NLO masses due to renormalization scheme dependences still remain significant, at approximately 0.05 GeV and 0.03 GeV (see Tab.~\ref{tab:tabel-Mixed-1-MSbar-OS} and Tab.~\ref{tab:tabel-Mixed-2-MSbar-OS}), respectively. This may introduce potential systematic errors into our predictions of mass at the NLO level. Nevertheless, it is conceivable that the discrepancy between the $\overline{\text{MS}}$ scheme and the on-shell scheme could be further mitigated by higher-order QCD corrections.

One may think that the threshold parameter $s_0$ is a little bit large for the NLO result for $J_{1,5}^{\text{Mixed}}$ (or $J_{2,6}^{\text{Mixed}}$) in the $\overline{\text{MS}}$ scheme. That is, for the central values of $M_H^{\rm{NLO}\mbox{-}\overline{\rm{MS}}}$ and the corresponding $s_0$ (see Tab.~\ref{tab:tabel-Mixed-1-MSbar-OS}), $\sqrt{s_0}-m_H=0.80$~GeV, which is larger than the typical energy gap $\Delta(\sim 0.5~{\rm GeV})$ between the ground state and the first radial exited state for a hadronic system. However, one can see from Tab.~\ref{tab:tabel-Mixed-1-MSbar-OS} that our prediction on the NLO-$\overline{\text{MS}}$ mass is not very sensitive to the threshold parameter, which can also be seen from Fig.~\ref{fig:Zc-[1+]-M-Mixed-1-NLO-MSbar-OS}(a). That means, if we choose $s_0=19.8~{\rm GeV}^2$, then the central value of the mass is about $M_H=3.87~{\rm GeV}$, and $\sqrt{s_0}-m_H=0.58$~GeV, which approaches to the value of $\Delta$.

\subsection{Numerical results for mixed operators of $Z_{cs}$ system}

In this paper, we regard $Z_{cs}^+$(3985)/$Z_{cs}^+(4000)$ and $Z_{cs}^+$(4220) as the partners of $Z_c^+$(3900) and $Z_c^+$(4020), respectively. Thus, all the calculation details about $Z_{cs}$ system are almost the same as $Z_c$ system. One needs just replace down quark $d$ with the strange quark $s$ in both the operators and the quark condensate. For numerical analysis associating to $s$ quark, we choose the following parameters~\cite{Aoki:2016frl}
\begin{align}\label{eq:squark-parameters}
	\begin{split}
		m_{ud}& =\frac{1}{2}\left( m_u(2\text{GeV})+ m_d(2 \text{GeV}) \right)\, , \\
		m_s(2\text{GeV})&= (27.43 \pm 0.31)\times m_{ud}\, ,\\
		\langle \bar{s}s \rangle(2\text{GeV})&= (0.6 \pm 0.1)\times\langle \bar{q}q \rangle(2\text{GeV})\, \, ,\\
		
		\langle g_s \bar{s}s G \rangle(2\text{GeV})&=(0.8\pm 0.2 \text{GeV}^2)\times \langle \bar{s}s \rangle(2\text{GeV})\, \, .
	\end{split}
\end{align}
where $\langle \bar{q}q \rangle(2\text{GeV})$ has been listed in Eq.~(\ref{eq:parameters}).

Similar to the case of $Z_c$ system, the QCD sum rules for the un-mixed operators of the $Z_{cs}$ system are not reliable even with NLO corrections. Firstly, to obtain the Borel windows, one needs very loose constraints as those given in Eq.~(\ref{eq:BWcondition-meson-diquark}), and therefore the convergence of OPE is bad and the theoretical uncertainties may be very large for these operators. Secondly, the quality of the Borel platforms are bad and the results are sensitive to the choice of $s_0$ and $M_B^2$. Thirdly, the perturbative convergence for the results of these operator are bad. Thus, we will not list the results for these unmixed operators here.

\begin{figure}[htb]
	\vspace{-0.5cm}
	\begin{center}
		\includegraphics[scale=0.5]{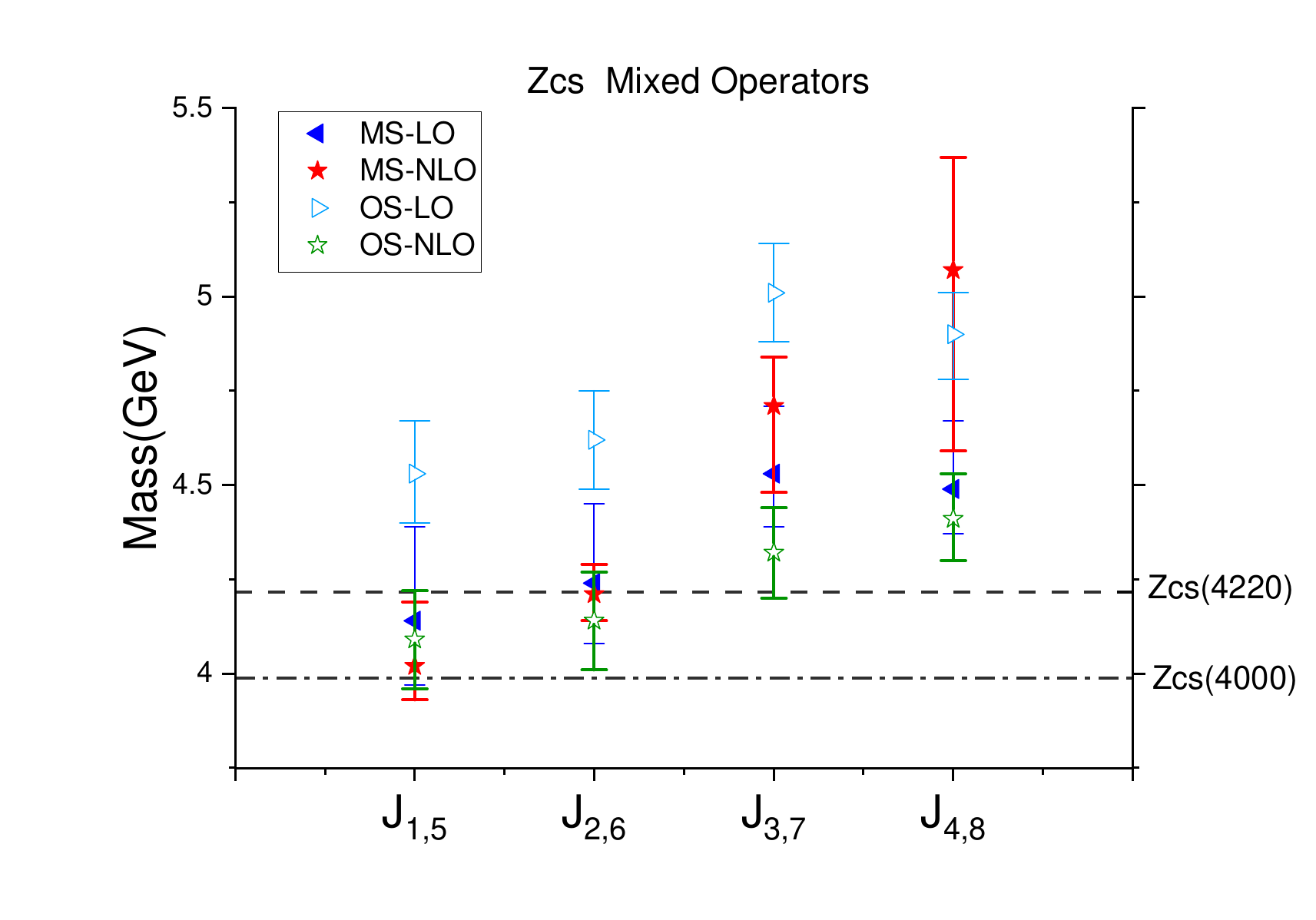}
		\vspace{-0.8cm}
		\caption{\label{fig:Zcs-mixed-Mass-Spectrum}
			The LO and NLO mass spectra of mixed operators of $Z_{cs}$ system in the $\overline{\text{MS}}$ and $\text{OS}$ schemes. The errors of masses shown in this figure just come from the parametric dependence on $s_0$, $M_B^2$, heavy quark mass $m_Q$ and renormalization scale $\mu$, which are show in the Tab.\ref{tab:Zcs-tabel-Mixed-1-MSbar-OS}-\ref{tab:Zcs-tabel-Mixed-4-MSbar-OS}.}
	\end{center}
\end{figure}

\begin{figure}[H]
	\centering
	\subfigure[$J_{1,5}^{\text{Mixed}}$]{
		\includegraphics[scale=0.27]{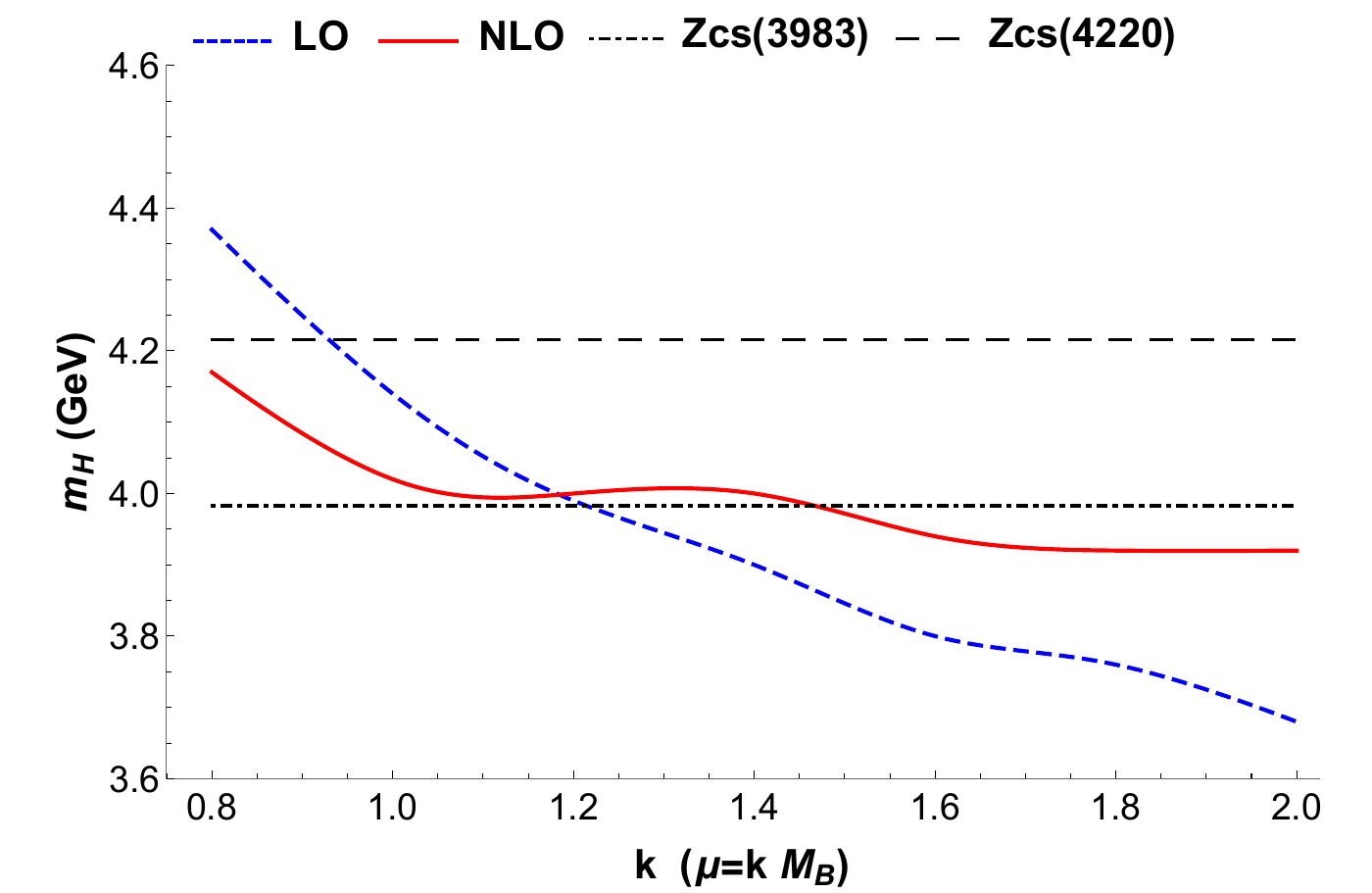}
	}
	\subfigure[$J_{2,6}^{\text{Mixed}}$]{
		\includegraphics[scale=0.28]{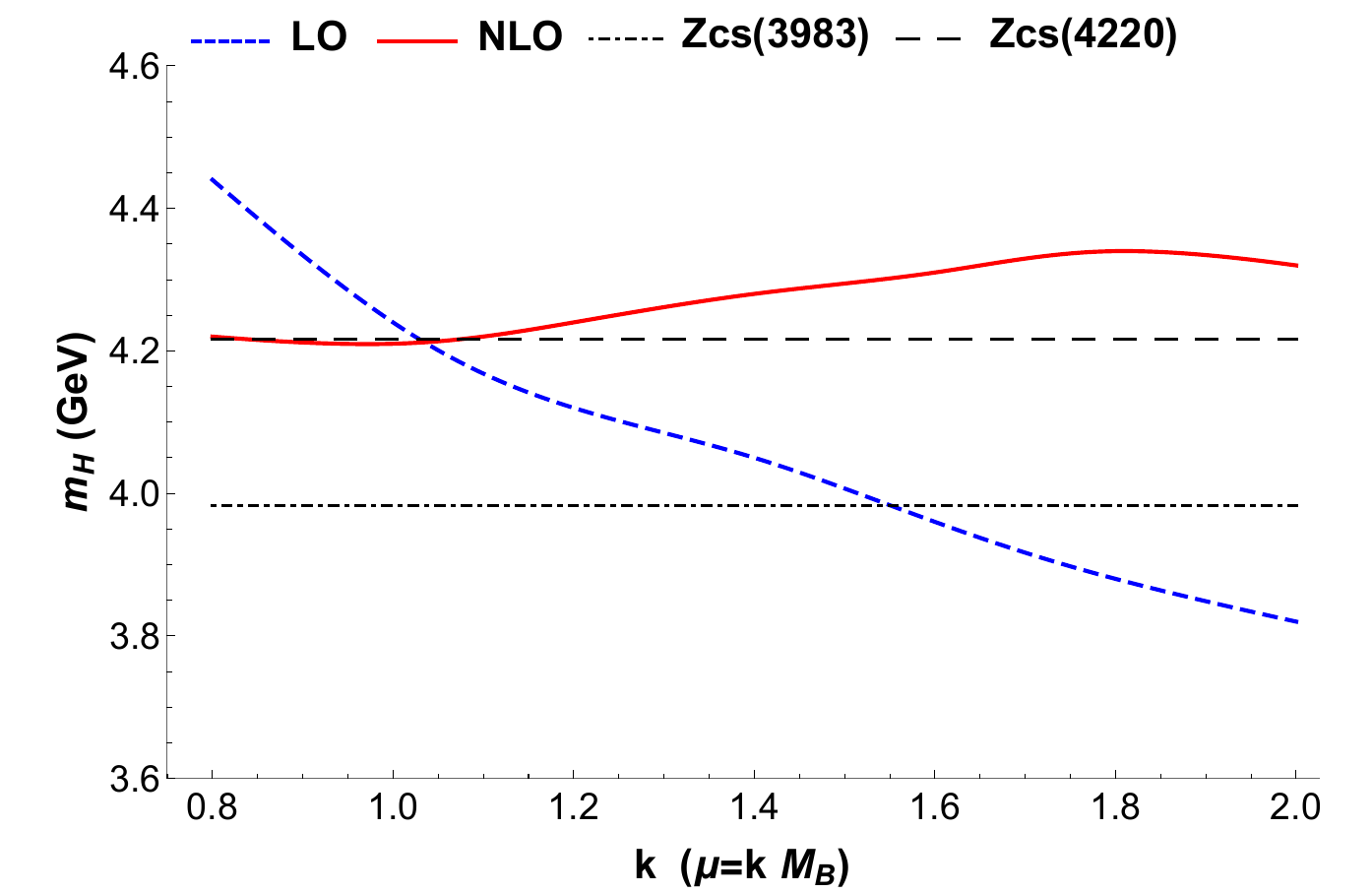}
	}\\
	\subfigure[$J_{3,7}^{\text{Mixed}}$]{
		\includegraphics[scale=0.28]{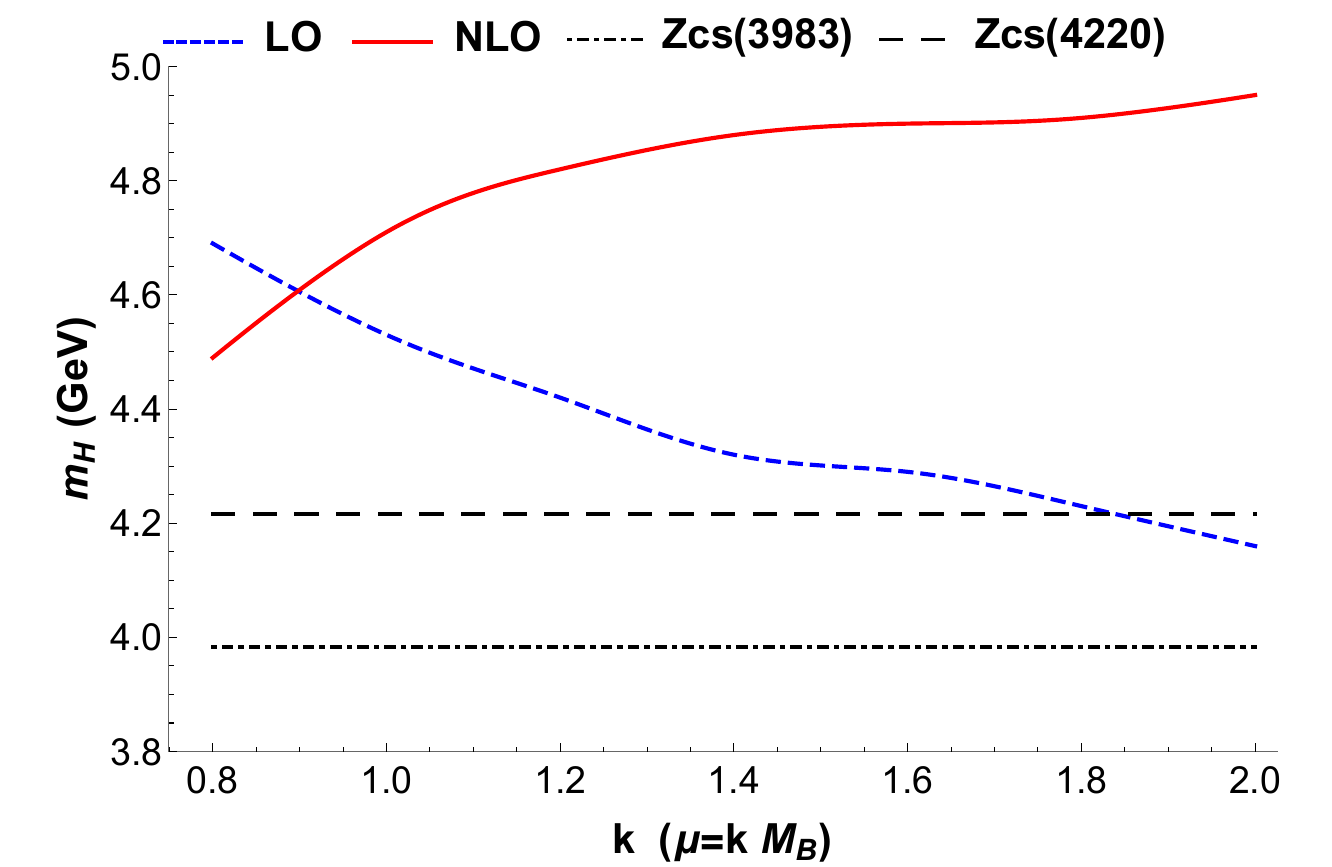}
	}
	\subfigure[$J_{4,8}^{\text{Mixed}}$]{
		\includegraphics[scale=0.28]{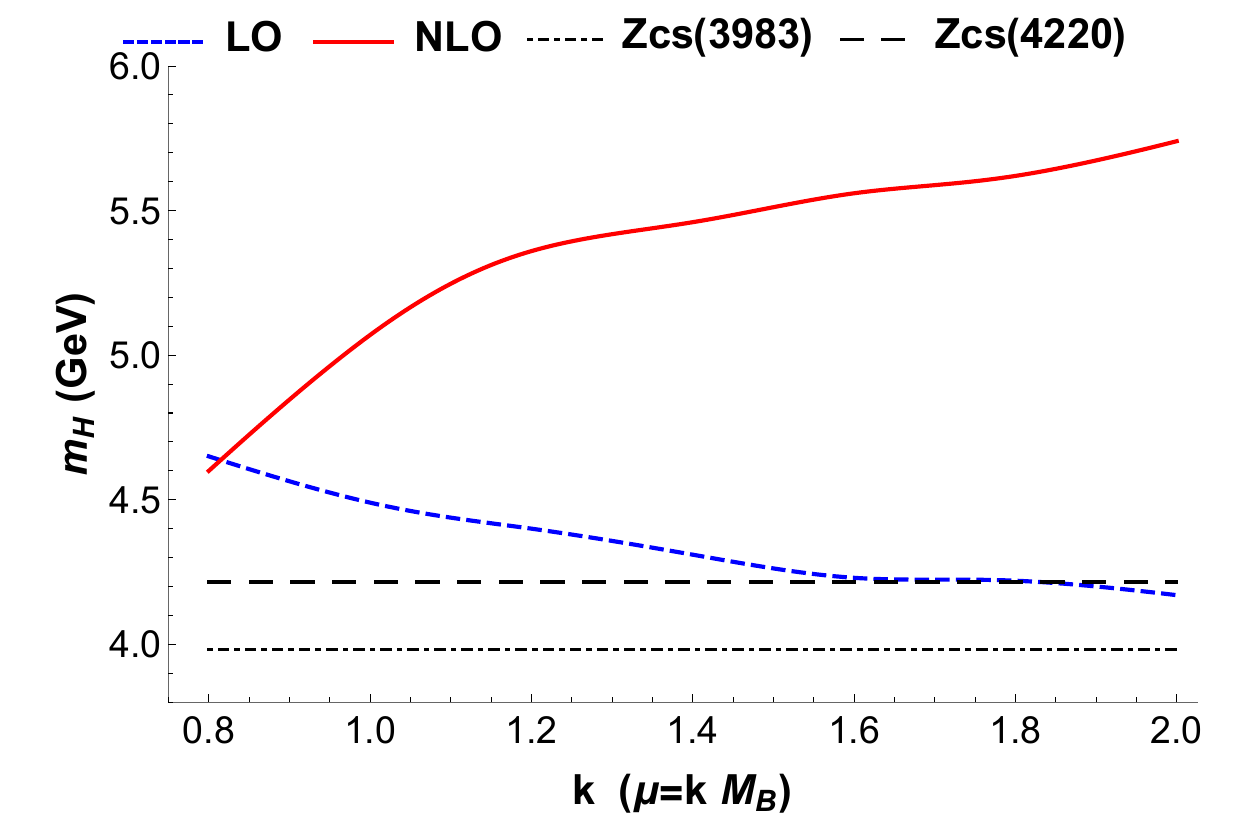}
	}
	\caption{\label{fig:Zcs-mu-dependence-mixed}The $\mu$-dependence of the LO and NLO results for mixed operators $J_{i}^{\text{Mixed}}$ of $Z_{cs}$ system in the $\overline{\rm{MS}}$ scheme. (a) for $J_{1,5}^{\text{Mixed}}$, (b) for $J_{2,6}^{\text{Mixed}}$, (c) for $J_{3,7}^{\text{Mixed}}$ and (d) for $J_{4,8}^{\text{Mixed}}$.}
\end{figure}

As for the mixed operators, we choose the same mixing parameters as those given in Eq.~(\ref{eq:mixedparameters}) and the same Borel conditions as those given in Eq.~(\ref{eq:BWcondition-mixed}). The results are listed in Tab.~\ref{tab:Zcs-tabel-Mixed-1-MSbar-OS}-\ref{tab:Zcs-tabel-Mixed-4-MSbar-OS} for the four ideally mixed operators, respectively. For $\mu=M_B$, the curves of the Borel platforms are shown in Fig.~\ref{fig:Zcs-[1+]-M-Mixed-1-NLO-MSbar-OS}--\ref{fig:Zcs-[1+]-M-Mixed-4-NLO-MSbar-OS} in Appendix~\ref{sec:details}. In these plots, again, a black dot denotes the central point ($x_0,y_0$), and shadows denote the Borel window determined by Eq.~(\ref{eq:BWcondition-mixed}). The quality of the platforms are quite good, thus, the errors of $M_H$ form $s_0$ and $M_B^2$ are very small (see Tab.~\ref{tab:Zcs-tabel-Mixed-1-MSbar-OS}-\ref{tab:Zcs-tabel-Mixed-4-MSbar-OS}). In Tab.~\ref{tab:Zcs-tabel-Mixed-1-MSbar-OS}-\ref{tab:Zcs-tabel-Mixed-4-MSbar-OS}, we also list the errors coming from $m_c$ and $\mu$. Again, to explore the $\mu$-dependence of $M_H$, we set $\mu=k M_B$ and vary the value of $k$ from $0.8$ to $2.0$. The $\mu$-dependent of both the LO and NLO $\overline{\text{MS}}$ masses are shown in Fig.~\ref{fig:Zcs-mu-dependence-mixed}. Furthermore, we also summarize the $M_H$ results with errors in Fig.~\ref{fig:Zcs-mixed-Mass-Spectrum}.

\begin{table}[H]
	\vspace{-0.3cm}
	\renewcommand\arraystretch{1.5}
	\begin{center}
		\setlength{\tabcolsep}{1.2 mm}
		\begin{tabular}{cccccccccc}
			\hline\hline
			Current& Order&   $M_H$ (GeV)   &  $s_0$ (${\text{GeV}}^2$)  & $M_B^2$ (${\text{GeV}}^2$)  & \makecell{Error from \\$s_0$ and $M_B^2$ } & \makecell{Error from \\$m_c$ } & \makecell{Error from \\$\mu$} \\ \hline
			
			\multirow{4}{*}{$J_{1,5}^{\text{Mixed}}$} & LO($\overline{\text{MS}}$) &    $4.14^{+0.25}_{-0.17}$   &    $24(\pm 10\%)$   &   $1.7(\pm 10\%)$   &  $^{+0.06}_{-0.05}$ &  $^{+0.06}_{-0.06}$ &  $^{+0.23}_{-0.15}$ \\
			
			&NLO($\overline{\text{MS}}$) &   $4.02^{+0.17}_{-0.09}$   &    $21(\pm 10\%)$   &   $1.4(\pm 10\%)$    &  $^{+0.05}_{-0.06}$ &  $^{+0.06}_{-0.06}$ &  $^{+0.15}_{-0.02}$ \\
			
			&LO(OS)     &         $4.53^{+0.14}_{-0.13}$         &      $28(\pm 10\%)$    &   $1.7(\pm 10\%)$    & $^{+0.06}_{-0.05}$ &  $^{+0.13}_{-0.12}$ & \\
			
			&NLO(OS)     &         $4.09^{+0.13}_{-0.13}$           &   $23(\pm 10\%)$     &   $1.1(\pm 10\%)$    & $^{+0.03}_{-0.02}$ &  $^{+0.13}_{-0.13}$ &\\
			\hline\hline
		\end{tabular}
		\caption{The LO and NLO results for the mass of $J_{1,5}^{\text{Mixed}}$ of $Z_{cs}$ system in $\overline{\text{MS}}$ and On-Shell schemes. Here the errors for $M_H$ are from $s_0, M_B$, the charm quark mass, and the renormalization scale $\mu$ with $\mu=kM_B$ and $k \in (0.8, 1.2)$ (the central values correspond to $\mu=M_B$).}
		\label{tab:Zcs-tabel-Mixed-1-MSbar-OS}
	\end{center}
\end{table}

\begin{table}[H]
	\renewcommand\arraystretch{1.5}
	\begin{center}
		\setlength{\tabcolsep}{1.2 mm}
		\begin{tabular}{cccccccccc}
			\hline\hline
			Current& Order&   $M_H$ (GeV)   &  $s_0$ (${\text{GeV}}^2$)  & $M_B^2$ (${\text{GeV}}^2$)  & \makecell{Error from \\$s_0$ and $M_B^2$ } & \makecell{Error from \\$m_c$ } & \makecell{Error from \\$\mu$} \\ \hline
			
			\multirow{4}{*}{$J_{2,6}^{\text{Mixed}}$} & LO($\overline{\text{MS}}$) &    $4.24^{+0.21}_{-0.16}$   &    $25(\pm 10\%)$   &   $1.8(\pm 10\%)$   &  $^{+0.05}_{-0.08}$ &  $^{+0.05}_{-0.06}$ &  $^{+0.2}_{-0.12}$ \\
			
			&NLO($\overline{\text{MS}}$) &   $4.21^{+0.08}_{-0.07}$   &    $24(\pm 10\%)$   &   $1.5(\pm 10\%)$    &  $^{+0.05}_{-0.04}$ &  $^{+0.06}_{-0.06}$ &  $^{+0.03}_{-0}$ \\
			
			&LO(OS)     &         $4.62^{+0.13}_{-0.13}$         &      $29(\pm 10\%)$    &   $1.8(\pm 10\%)$    & $^{+0.06}_{-0.06}$ &  $^{+0.12}_{-0.11}$ & \\
			
			&NLO(OS)     &         $4.14^{+0.13}_{-0.13}$           &   $22(\pm 10\%)$     &   $1.2(\pm 10\%)$    & $^{+0.04}_{-0.04}$ &  $^{+0.12}_{-0.12}$ &\\
			\hline\hline
		\end{tabular}
		\caption{The LO and NLO results for the mass of $J_{2,6}^{\text{Mixed}}$ of $Z_{cs}$ system in $\overline{\text{MS}}$ and On-Shell schemes. Here the errors for $M_H$ are from $s_0, M_B$, the charm quark mass, and the renormalization scale $\mu$ with $\mu=kM_B$ and $k \in (0.8, 1.2)$ (the central values correspond to $\mu=M_B$).}
		\label{tab:Zcs-tabel-Mixed-2-MSbar-OS}
	\end{center}
\end{table}

\begin{table}[H]
	\renewcommand\arraystretch{1.5}
	\begin{center}
		\setlength{\tabcolsep}{1.2 mm}
		\begin{tabular}{cccccccccc}
			\hline\hline
			Current& Order&   $M_H$ (GeV)   &  $s_0$ (${\text{GeV}}^2$)  & $M_B^2$ (${\text{GeV}}^2$)  & \makecell{Error from \\$s_0$ and $M_B^2$ } & \makecell{Error from \\$m_c$ } & \makecell{Error from \\$\mu$} \\ \hline
			
			\multirow{4}{*}{$J_{3,7}^{\text{Mixed}}$} & LO($\overline{\text{MS}}$) &    $4.53^{+0.18}_{-0.14}$   &    $29(\pm 10\%)$   &   $2.2(\pm 10\%)$   &  $^{+0.06}_{-0.07}$ &  $^{+0.05}_{-0.05}$ &  $^{+0.16}_{-0.11}$ \\
			
			&NLO($\overline{\text{MS}}$) &   $4.71^{+0.13}_{-0.23}$   &    $31(\pm 10\%)$   &   $2.1(\pm 10\%)$    &  $^{+0.05}_{-0.05}$ &  $^{+0.06}_{-0.06}$ &  $^{+0.11}_{-0.22}$ \\
			
			&LO(OS)     &         $5.01^{+0.13}_{-0.13}$         &      $36(\pm 10\%)$    &   $2.2(\pm 10\%)$    & $^{+0.05}_{-0.05}$ &  $^{+0.12}_{-0.12}$ & \\
			
			&NLO(OS)     &         $4.32^{+0.12}_{-0.12}$           &   $23(\pm 10\%)$     &   $1.4(\pm 10\%)$    & $^{+0.05}_{-0.06}$ &  $^{+0.11}_{-0.1}$ &\\
			\hline\hline
		\end{tabular}
		\caption{The LO and NLO results for the mass of $J_{3,7}^{\text{Mixed}}$ of $Z_{cs}$ system in $\overline{\text{MS}}$ and On-Shell schemes. Here the errors for $M_H$ are from $s_0, M_B$, the charm quark mass, and the renormalization scale $\mu$ with $\mu=kM_B$ and $k \in (0.8, 1.2)$ (the central values correspond to $\mu=M_B$).}
		\label{tab:Zcs-tabel-Mixed-3-MSbar-OS}
	\end{center}
\end{table}

\begin{table}[H]
	\renewcommand\arraystretch{1.5}
	\begin{center}
		\setlength{\tabcolsep}{1.2 mm}
		\begin{tabular}{cccccccccc}
			\hline\hline
			Current& Order&   $M_H$ (GeV)   &  $s_0$ (${\text{GeV}}^2$)  & $M_B^2$ (${\text{GeV}}^2$)  & \makecell{Error from \\$s_0$ and $M_B^2$ } & \makecell{Error from \\$m_c$ } & \makecell{Error from \\$\mu$} \\ \hline
			
			\multirow{4}{*}{$J_{4,8}^{\text{Mixed}}$} & LO($\overline{\text{MS}}$) &    $4.49^{+0.18}_{-0.12}$   &    $29(\pm 10\%)$   &   $2.1(\pm 10\%)$   &  $^{+0.06}_{-0.06}$ &  $^{+0.05}_{-0.05}$ &  $^{+0.16}_{-0.09}$ \\
			
			&NLO($\overline{\text{MS}}$) &   $5.07^{+0.3}_{-0.48}$   &    $35(\pm 10\%)$   &   $2.5(\pm 10\%)$    &  $^{+0.04}_{-0.06}$ &  $^{+0.05}_{-0.05}$ &  $^{+0.29}_{-0.47}$ \\
			
			&LO(OS)     &         $4.9^{+0.11}_{-0.12}$         &      $34(\pm 10\%)$    &   $2.1(\pm 10\%)$    & $^{+0.05}_{-0.04}$ &  $^{+0.1}_{-0.11}$ & \\
			
			&NLO(OS)     &         $4.41^{+0.12}_{-0.11}$           &   $29(\pm 10\%)$     &   $1.4(\pm 10\%)$    & $^{+0.02}_{-0.02}$ &  $^{+0.12}_{-0.11}$ &\\
			\hline\hline
		\end{tabular}
		\caption{The LO and NLO results for the mass of $J_{4,8}^{\text{Mixed}}$ of $Z_{cs}$ system in $\overline{\text{MS}}$ and On-Shell schemes. Here the errors for $M_H$ are from $s_0, M_B$, the charm quark mass, and the renormalization scale $\mu$ with $\mu=kM_B$ and $k \in (0.8, 1.2)$ (the central values correspond to $\mu=M_B$).}
		\label{tab:Zcs-tabel-Mixed-4-MSbar-OS}
	\end{center}
\end{table}

Similar to the case of $Z_c$ system, for the first two ideally mixed operators $J_{1,5}^{\text{Mixed}}$ and $J_{2,6}^{\text{Mixed}}$ in $Z_{cs}$ system, the results show very good perturbative convergence. On the one hand, the mass gaps between the $\overline{\rm{MS}}$ and the on-shell schemes are largely reduced, the LO mass difference of the central values in the two scheme $ \left| M_H^{\rm{LO}\mbox{-}\rm{OS}} - M_H^{\rm{LO}\mbox{-}\overline{\rm{MS}}} \right | \sim 0.4$~GeV while the NLO mass difference $ \left | M_H^{\rm{NLO}\mbox{-}\rm{OS}} - M_H^{\rm{NLO}\mbox{-}\overline{\rm{MS}}} \right | \sim 0.07$~GeV. On the other hand, the $\mu$-dependence of the $\overline{\text{MS}}$ masses are improved evidently with the NLO corrections for these two operators (see Fig.~\ref{fig:Zcs-mu-dependence-mixed} (a) and (b)). As for the last two mixed operators $J_{3,7}^{\text{Mixed}}$ and $J_{4,8}^{\text{Mixed}}$, perturbative convergence of the results are bad, and the $\mu$-dependence is not improved evidently with the NLO corrections for these two operators (see Fig.~\ref{fig:mu-dependence-mixed} (c) and (d)), and is even worse for the $J_{4,8}^{\text{Mixed}}$ operator. Thus, based on our NLO calculations, only the results of operators $J_{1,5}^{\text{Mixed}}$ and $J_{2,6}^{\text{Mixed}}$ are reliable for the $Z_{cs}$ system.

Phenomenologically, the NLO mass of $J_{1,5}^{\text{Mixed}}$ in the $\overline{\text{MS}}$ scheme is given by $4.02^{+0.17}_{-0.09}$~GeV (see Tab.~\ref{tab:Zcs-tabel-Mixed-1-MSbar-OS}), which is very close to that of $Z_{cs}$(3985)/$Z_{cs}(4000)$, and the NLO mass of $J_{2,6}^{\text{Mixed}}$ in the $\overline{\text{MS}}$ scheme is given by $4.21^{+0.08}_{-0.07}$~GeV (see Tab.~\ref{tab:Zcs-tabel-Mixed-2-MSbar-OS}), which is very close to that of $Z_c(4220)$~\footnote{The LHCb data about $Z_{cs}$(4000) and $Z_{cs}$(4220)~\cite{LHCb:2021uow} can also be accounted for by two near-threshold states with masses of about 4.0 and 4.1 GeV, respectively, in a coupled-channels analysis~\cite{Ortega:2021enc}. The latter two states are also predicted by the hadronic molecule model~\cite{Yang:2020nrt}}. Thus, our results support to assign $Z_{cs}$(3985)/$Z_{cs}(4000)$ and $Z_{cs}(4220)$ as the partners of $Z_c(3900)$ and $Z_c(4020)$, respectively.

It should be mentioned here that, in our calculation, we set $m_s=0$ in evaluation of the perturbation part $C_1$, the non-zero $m_s$ given in Eq.~(\ref{eq:squark-parameters}) is only used to evaluate the condensate parts $C_{\bar{s}s}\langle \bar{s}s \rangle$ and $C_{\bar{s}s G}\langle g_s \bar{s}s G \rangle$. And we choose $r_{sq}=\langle \bar{s}s  \rangle /\langle \bar{q}q \rangle =0.6\pm0.1$ to approximately fit the mass of  $Z_{cs}$(3985)/$Z_{cs}(4000)$ or that of $Z_{cs}(4220)$. In Ref.~\cite{Ioffe:1981kw}, the ratio $r_{sq}$ is fitted to be $0.8\pm0.1$ by the mass difference between $\Sigma^{*}$ and $\Delta$ baryons with a fixed strange quark mass $m_s=0.15$~GeV, which is different from our choice in Eq.~(\ref{eq:squark-parameters}). If we choose $r_{sq}=0.8\pm0.1$, the central values of the $M_H$ for the mixed operators will be increased by around $0.15$~GeV.

In the above calculation, we have taken $Z_{cs}$ as the partner of $Z_c$. If we deal with $Z_{cs}$ independently, the number of operators in the basis will be doubled since there are no constraint from $G$-parity as that for the $Z_c$ system (see  Eq.~(\ref{eq:1+-meson}) or Eq.~(\ref{eq:1+-Diquark})). Thus, for $J^P=1^+$ $\bar{c}u\bar{s}c$ system, there are 16 operators in the basis, and the diagonalized anomalous dimension matrix is fourfold degenerated. As a result, the mixing of the operators will be complex for this system.  For example, there will be two extra operators which can be mixed with $J_{1,5}^{\text{Mixed}}$ further. However, the operator $J_{1,5}^{\text{Mixed}}$ is still corresponding to a special choice of the mixing parameters in the four operator mixing space, which can give result with good perturbative convergence as one had seen. In this aspect, our calculation still supports to assign $Z_{cs}$ as the partner of $Z_c$.

\section{Summary}

In this paper, we study the NLO corrections to the spectrum of $\bar{d}c\bar{c}u$ tetraquark system with $(I^G)J^P=(1^+)1^+$, i.e., the $Z_c$ system, in QCD sum rules. As operators with the same quantum numbers can mix with each other under renormalization, we recombine the original operators, either in meson-meson type or diquark-antidiquark type, by diagonalizing the anomalous dimension matrix. However, different from the case of fully heavy tetraquark system~\cite{Wu:2022qwd}, the $\alpha_s$-order anomalous dimension matrix in $Z_c$ system is degenerated and the mixing of operators can not be determined uniquely by diagonalization of the matrix. We then determine the mixing parameters by minimizing the $s_0$- and $M_B^2$-dependence to give the four so-called ideally mixed operators. And we call the operator in the original basis (either in meson-meson type or diquark-antidiquark type) as the unmixed one.

Numerically, we find the results for almost all the unmixed operators may suffer from large theoretical uncertainties caused by both higher order power corrections in OPE and higher order QCD corrections. Thus, these results are not reliable to use in the phenomenological analysis. As for the four ideally mixed operators, we find that the convergence of OPE and the quality of Borel platform are good.  But for the QCD perturbative convergence, it is good only for the first two ideally mixed operators $J_{1,5}^{\text{Mixed}}$ and $J_{2,6}^{\text{Mixed}}$ and not good for the last two $J_{3,7}^{\text{Mixed}}$  and $J_{4,8}^{\text{Mixed}}$. In particular, the first two ideally mixed operators lead to results with  much weaker $\mu$-dependence at NLO than LO.  Phenomenologically, the NLO $\overline{\rm{MS}}$ masses of the first two ideally mixed operators are $3.89^{+0.18}_{-0.12}$~GeV and $4.03^{+0.06}_{-0.07}$~GeV, which are very close to those of $Z_c(3900)$ and $Z_c(4020)$, respectively.

As for the $Z_{cs}$ system with quark content $\bar{d}c\bar{c}u$ and $J^P=1^+$, we study it as a partner of the $Z_c$ system in QCD sum rules. Thus, one needs to just replace the down quark $d$ with the strange quark $s$ in both the operators and the quark condensate. The calculations and conclusions are almost the same as those for the $Z_c$ system. We find that both the convergence of OPE and the convergence of QCD perturbation corrections are good for the ideally mixed operators $J_{1,5}^{\text{Mixed}}$ and $J_{2,6}^{\text{Mixed}}$ in the $Z_{cs}$ system, and they give NLO $\overline{\rm{MS}}$ masses $4.02^{+0.17}_{-0.09}$~GeV and $4.21^{+0.08}_{-0.07}$~GeV in turn, which are very close to $Z_{cs}(3985)$/$Z_{cs}(4000)$ and $Z_{cs}(4220)$, respectively. Thus, our results support to assign $Z_{cs}$(3985)/$Z_{cs}(4000)$ and $Z_{cs}(4220)$ as the partners of $Z_c(3900)$ and $Z_c(4020)$, respectively.

Moreover, we also study the $Z_b$ system, however, both the perturbative convergence and the quality of the Borel platforms are bad even for the mixed operators. This case is very similar to those of fully bottom baryon $\Omega_{bbb}$~\cite{Wu:2021tzo} and fully bottom tetraquark $\bar{b}b\bar{b}b$~\cite{Wu:2022qwd}. Thus, we can not give credible results for $Z_b$ system at present.

Finally, as having been emphasized in fully heavy tetraquark system~\cite{Wu:2022qwd}, we would like to emphasize again the importance of NLO contributions, especially in the operator mixing or color configuration mixing for multiquark systems. (i) As a key NLO contribution, the one-gluon exchange is crucial even for charmonium and bottomonium states, because it provides the color Coulomb interaction between $Q$ and $\bar{Q}$, which is the most important short-range attractive force to form a heavy quarkonium.
(ii) In the $\bar{d}c\bar{c}u$ tetraquark system discussed in this paper, if one starts from a color-singlet current-current operator, the one-gluon exchange will change it to the color-octet current-current operator, therefore leads to the operator mixing. As already shown by our result, the operator mixing induced by renormalization at NLO is inevitable and has very important consequences in the QCD sum rule calculations. (iii) In the literature some works use the color-singlet current-current local operators to describe physical hadronic molecules. However, due to the operator mixing, the color structure of the local operators must be mixed with both color-singlet and color-octet current-current configurations. It is impossible to keep the color-singlet structure unchanged if a complete NLO QCD contribution is seriously considered. In fact, a physical molecule state means that it contains two well separated color-singlet mesons at long-distances mediated by one-meson or two-meson exchanges. And a physical molecule may not be necessarily ascribed to the color-singlet current-current local operators, which only describe the very short-distance behavior of the tetraquark and are subjected to the color configuration mixing. The description for hadronic molecules needs to understand the long-distance dynamics beyond color confinement.

\acknowledgments

The figures in this paper are drawn by using the Origin and Mathematica software. The work is supported in part by the National Natural Science Foundation of China (Grants No. 11875071, No. 11975029, No. 11745006, No. 12325503), the National Key Research and Development Program of China under Contracts No. 2020YFA0406400, and the High-performance Computing Platform of Peking University.

\newpage
\appendix

\section{ Operator Renormalization Matrices}
\label{sec:matrix}
\subsection{Calculation Of Operator Renormalization Matrices }
We present the calculation of operator renormalization matrices of meson-meson type operators. The operator renormalization matrices of diquark-antidiquark type operators then follow from a Fierz transformation.

A general meson-meson type operator where four quarks have distinct flavors can be defined as
\begin{align}
\mathcal{O}_{\Gamma_1,\Gamma_2}=\left(\bar{q}_1^i \Gamma_1 q_2^j \right) \left(\bar{q}_3^k \Gamma_2 q_4^l \right)\ ,
\end{align}
which has two independent color configurations,
\begin{align}\label{eq:Operatorcolor}
\mathcal{O}_{\Gamma_1,\Gamma_2,[1]}&=\left(\bar{q}_1^i \Gamma_1 q_2^j \right) \left(\bar{q}_3^k \Gamma_2 q_4^l \right) \delta_{ij}\delta_{kl}\ ,\\
\mathcal{O}_{\Gamma_1,\Gamma_2,[2]}&=\left(\bar{q}_1^i \Gamma_1 q_2^j \right) \left(\bar{q}_3^k \Gamma_2 q_4^l \right) \delta_{il}\delta_{kj}\ ,
\end{align}
where $\mathcal{O}_{\Gamma_1,\Gamma_2,[1]}$ is called a color-singlet operator. We can also using following relation,
\begin{align}
T^a_{ij}T^a_{kl}=\frac{1}{2}\left( \delta_{il}\delta_{kj} - \frac{1}{N_c}\delta_{ij}\delta_{kl} \right)\ ,
\end{align}
to obtain the color-octet operator,
\begin{align}
\mathcal{O}_{\Gamma_1,\Gamma_2,[8]}&=\left(\bar{q}_1^i \Gamma_1 T^a_{ij} q_2^j \right) \left(\bar{q}_3^k \Gamma_2 T^a_{kl} q_4^l \right)\\
&=\frac{1}{2}\left( \mathcal{O}_{\Gamma_1,\Gamma_2,[2]} - \frac{1}{N_c}\mathcal{O}_{\Gamma_1,\Gamma_2,[1]}                    \right)\,.
\end{align}
For convenience, we choose $\mathcal{O}_{\Gamma_1,\Gamma_2,[1]}$ and $\mathcal{O}_{\Gamma_1,\Gamma_2,[2]}$ as basis in our calculation.

Let us first suppress the dependence of color configuration for operators. According to the definition of our operators and their renormalization matrix,

	\begin{align}\label{eq:opReno}
		\mathcal{O}^B=\left(\sqrt{Z_2}\right)^4 \  \overline{\mathcal{O}}^{B}\, = {Z_{\mathcal{O}}}\  \mathcal{O}^{R} \ ,
	\end{align}
	where $\mathcal{O}^B=\left(\bar{q}_1 \Gamma_1 q_2 \right) \left(\bar{q}_3 \Gamma_2 q_4\right)$ denotes the bare operator, $\overline{\mathcal{O}}^{B}=\left(\bar{q}_1^R \Gamma_1 q_2^R \right) \left(\bar{q}_3^R \Gamma_2 q_4^R\right)$ denotes the bare operator with bare fields replaced by renormalized fields, and $\mathcal{O}^{R}$ denotes the renormalized operator. In our NLO calculation, we directly calculate $\overline{\mathcal{O}}^{B}$, and thus quark self-energy diagrams are cancelled by counter term diagrams. The remaining diagrams can be divided into three parts,
	\begin{align}
		A=\int \frac{\mathrm{d}^{D} p}{(2\pi)^D}(A_1+A_2+A_3)\ ,
	\end{align}
	where $A_1$ denotes the contribution of gluon exchange between $q_1$ and $q_2$, $A_2$ denotes the contribution of gluon exchange between $q_3$ and $q_4$, and $A_3$ denotes other contributions e.g.\ contributions of gluon exchange between $q_1$ and $q_3$, $q_1$ and $q_4$ and so on.
	Because all infrared divergences will be cancelled, we just need to consider ultraviolet (UV) divergences therein. Therefore, the mass terms in quark propagators can be discarded. Explicitly, we have
	\begin{align}
		A_{1} & = \left[i g_s \gamma_{\mu}  \frac{i\slashed{p}}{p^{2}}\Gamma_{1}  \frac{i\slashed{p}}{p^{2}} i g \gamma^{\mu} \left(T^{a}\right)_{i^{\prime} i}\left(T^{a}\right)_{j j^{\prime}}\right]\left[\Gamma_{2} \delta_{k^{\prime} k} \delta_{l l^{\prime}}\right]  \frac{-i}{p^{2}}\ ,
	\end{align}
	\begin{align}
		A_{2} & = \left[\Gamma_{1} \delta_{i^{\prime} i} \delta_{j j^{\prime}}\right] \left[i g \gamma_{\mu}  \frac{i \slashed{p}}{p^{2}}\Gamma_{2}  \frac{i \slashed{p}}{p^{2}} i g_s \gamma^{\mu} \left(T^{a}\right)_{k^{\prime} k}\left(T^{a}\right)_{l l^{\prime}} \right] \frac{-i}{p^{2}}\ ,
	\end{align}
	\begin{align}
		\begin{split}
		A_{3}  = &\left[i g_s \gamma_{\mu} \frac{i \slashed{p}}{p^{2}}\Gamma_{1} \left(T^{a}\right)_{i^{\prime} i} \delta_{j j^{\prime}} + \Gamma_{1} \frac{-i \slashed{p}}{p^{2}} i g_s \gamma_{\mu} \delta_{i^{\prime} i}\left(T^{a}\right)_{j j^{\prime}}  \right]\times \\ &\left[i g_s \gamma^{\mu} \frac{-i\slashed{p}}{p^{2}}\Gamma_{2} \left(T^{a}\right)_{k^{\prime} k} \delta_{l l^{\prime}}+\Gamma_{2} \frac{i \slashed{p}}{p^{2}} i g_s \gamma^{\mu} \delta_{k^{\prime} k}\left(T^{a}\right)_{l l^{\prime}}\right] \frac{-i}{p^{2}}\,.
		\end{split}
	\end{align}
	After a simple manipulation, we get
	\begin{align}\label{eq:UV}
		A=-ig_s^2 \frac{1}{D} \int \frac{\mathrm{d}^D p}{(2\pi)^D}\ \frac{1}{(p^2)^2}\ B\ ,
	\end{align}
	where
	\begin{align}\label{eq:UVB}
		\begin{split}
			B &=\left(\gamma_{\mu} \gamma_{\nu} \Gamma_{1} \gamma^{\nu} \gamma^{\mu}\right)\left(\Gamma_{2}\right)\left(T^{a}\right)_{i^{\prime} i}\left(T^{a}\right)_{j j^{\prime}} \delta_{k^{\prime} k} \delta_{l l^{\prime}}+\left(\Gamma_{1}\right)\left(\gamma_{\mu} \gamma_{v} \Gamma_{2} \gamma^{\nu} \gamma^{\mu}\right) \delta_{i^{\prime} i} \delta_{j j^{\prime}}\left(T^{a}\right)_{k^{\prime} k}\left(T^{a}\right)_{l l^{\prime}} \\
			&+(-D)\left(\Gamma_{1}\right)\left(\Gamma_{2}\right)\left[\left(T^{a}\right)_{i^{\prime} i} \delta_{j j^{\prime}}-\delta_{i^{\prime} i}\left(T^{a}\right)_{j j^{\prime}}\right]\left[\left(T^{a}\right)_{k^{\prime} k} \delta_{l l^{\prime}}-\delta_{k^{\prime} k}\left(T^{a}\right)_{l l^{\prime}}\right] \\
			&+\frac{1}{4}\left(\left\{\sigma_{\mu \nu}, \Gamma_{1}\right\}\right)\left(\left\{\sigma_{\mu \nu}, \Gamma_{2}\right\}\right)\left[\left(T^{a}\right)_{i^{\prime} i} \delta_{j j^{\prime}}+\delta_{i^{\prime} i}\left(T^{a}\right)_{j j^{\prime}}\right]\left[\left(T^{a}\right)_{k^{\prime} k} \delta_{l l^{\prime}}+\delta_{k^{\prime} k}\left(T^{a}\right)_{l l^{\prime}}\right] \\
			&+\frac{1}{4}\left(\left[\sigma_{\mu \nu}, \Gamma_{1}\right]\right)\left(\left[\sigma_{\mu \nu}, \Gamma_{2}\right]\right)\left[\left(T^{a}\right)_{i^{\prime} i} \delta_{j j^{\prime}}-\delta_{i^{\prime} i}\left(T^{a}\right)_{j j^{\prime}}\right]\left[\left(T^{a}\right)_{k^{\prime} k} \delta_{l l^{\prime}}-\delta_{k^{\prime} k}\left(T^{a}\right)_{l l^{\prime}}\right]\,.
		\end{split}
	\end{align}
	
	According to Eq.~(\ref{eq:UV}), we get the UV divergences term
	\begin{align}
		A_{\text{UV}} & = -\left.i g_s^{2} \frac{1}{4} \frac{i}{(4 \pi)^{2}} \frac{1}{\varepsilon} B\right|_{D  = 4}  = \frac{\alpha_{s}}{\varepsilon} \frac{\left.B\right|_{D  = 4}}{16\pi} \,.
	\end{align}
	For operators with definite color configuration $\mathcal{O}_{\Gamma_1,\Gamma_2,[c]}$, we need to multiply the corresponding color configuration ($\delta_{ij}\delta_{kl}$ or $\delta_{il}\delta_{kj}$) in Eq.~(\ref{eq:UVB}).
	
	According to Eq.~(\ref{eq:opReno}), to use the renormalized operator we should multiply our result $A_{\text{LO}}+A_{\text{UV}}+\cdots$ by $Z_2^{2}Z_{\mathcal{O}}^{-1}  \approx1-\delta Z_{\mathcal{O}}+2\delta Z_2  $, where $A_{\text{LO}}=1$ is the LO amplitude and $Z_2=1+\delta Z_2$ and $Z_{\mathcal{O}}=1+\delta Z_{\mathcal{O}}$. Demanding that final results are free of UV divergences, we get
	\begin{align}
		\delta Z_{\mathcal{O}}=A_{\text{UV}}+ 2\  \delta {Z}_2\ ,
	\end{align}
	with
	\begin{align}
		\delta {Z}_2 & = -\frac{\alpha_s}{3\pi \varepsilon} \,,
	\end{align}
	in the $\overline{\text{MS}}$ scheme.

\subsection{The operator renormalization matrix}

For meson-meson type operator bases Eq.(\ref{eq:1+-meson}), the corresponding operator renormalization matrix is given by,

\begin{equation}\label{eq:1+-meson-renormalization}
  \delta Z_{O}^{\text{\text{M-M}}}=\frac{\alpha_s}{16 \pi} \delta_{\overline{\text{MS}}}\begin{pmatrix}
  \frac{6(N_c^2-1)}{N_c}&0 &  -\frac{4}{N_c}&4&0&0&0&0 \\
6& -\frac{6}{N_c}&2 &\frac{2(N_c^2-2)}{N_c}&0&0&0&0 \\
 - \frac{12}{N_c}&12 &  -\frac{2(N_c^2-1)}{N_c}&0&0&0&0&0 \\
 6& \frac{6(N_c^2-2)}{N_c}&-6 &\frac{2(2 N_c^2+1)}{N_c}&0&0&0&0 \\
  0&0&0&0& \frac{6(N_c^2-1)}{N_c}&0 &  \frac{4}{N_c}&-4 \\
0&0&0&0& 6& -\frac{6}{N_c}&-2 &-\frac{2(N_c^2-2)}{N_c} \\
0&0&0&0& \frac{12}{N_c}&-12 &  -\frac{2(N_c^2-1)}{N_c}&0\\
0&0&0&0& -6& -\frac{6(N_c^2-2)}{N_c}&-6 &\frac{2(2 N_c^2+1)}{N_c} \\
\end{pmatrix}
\,,
\end{equation}
where $\delta_{\overline{\text{MS}}}=\frac{1}{\epsilon}+\ln(4\pi)-\gamma_E$.

For diquark-antidiquark type operator bases Eq.(\ref{eq:1+-Diquark}), the corresponding operator renormalization matrix is given by,
\begin{equation}\label{eq:1+-Diquark-renormalization}
	\delta Z_{O}^{\text{\text{Di-Di}}}=\frac{\alpha_s}{16 \pi} \delta_{\overline{\text{MS}}}\begin{pmatrix}
		\frac{6}{N_c}&-6 & -2 N_c &2&0&0&0&0 \\
		-6& \frac{6}{N_c}& -2 &2 N_c&0&0&0&0 \\
		- 6 N_c&6 &  -\frac{2(5-2N_c^2)}{N_c}&6&0&0&0&0 \\
		-6& 6 N_c & 6 &-\frac{2(5-2N_c^2)}{N_c}&0&0&0&0 \\
		0&0&0&0& 	\frac{6}{N_c}&-6 & 2 N_c &-2\\
		0&0&0&0& 	-6& \frac{6}{N_c}& 2 &-2 N_c\\
		0&0&0&0&  6 N_c&-6 &  -\frac{2(5-2N_c^2)}{N_c}&6\\
		0&0&0&0& 6& -6 N_c & 6 & -\frac{2(5-2N_c^2)}{N_c} \\
	\end{pmatrix}
	\,,
\end{equation}

\newpage
\section{Details for Various Operators}
\label{sec:details}

\subsection{Numerical results for meson-meson type operators of $Z_c$ system}

\begin{table}[H]
	\renewcommand\arraystretch{1.3}
	\begin{center}
		\setlength{\tabcolsep}{4 mm}
		\begin{tabular}{|c|c|c|c|c|c|c|c|c|c|}
			\hline
			\multirow{3}{*}{Current} &
			\multicolumn{3}{c|}{LO}& \multicolumn{3}{c|}{NLO($\overline{\text{MS}}$)} \\ \cline{2-7} &
			\makecell{$M_H$ \\ (GeV)} & \makecell{$s_0$ \\ ($\text{GeV}^2$)} & \makecell{$M_B^2$ \\ ($\text{GeV}^2$)} &
			\makecell{$M_H$ \\ (GeV)} & \makecell{$s_0$ \\ ($\text{GeV}^2$)} & \makecell{$M_B^2$ \\ ($\text{GeV}^2$)} \\ \hline
			
		  $J_{1}^{\text{M-M}}$ &$6.15^{+0.03}_{-0.17}$ &$50.(\pm 10\%)$ &$6.00(\pm 10\%)$    &$5.89^{+0.09}_{-0.15}$ &$47.(\pm 10\%)$ &$4.80(\pm 10\%)$\\ \hline
		$J_{2}^{\text{M-M}}$ &$6.10^{+0.02}_{-0.17}$ &$50.(\pm 10\%)$ &$5.80(\pm 10\%)$    &$6.10^{+0.02}_{-0.14}$ &$50.(\pm 10\%)$ &$5.10(\pm 10\%)$\\ \hline
		$J_{3}^{\text{M-M}}$ &$4.40^{+0.09}_{-0.07}$ &$24.(\pm 10\%)$ &$3.20(\pm 10\%)$    &$4.61^{+0.13}_{-0.12}$ &$27.(\pm 10\%)$ &$3.50(\pm 10\%)$\\ \hline
		$J_{4}^{\text{M-M}}$ &$5.12^{+0.10}_{-0.23}$ &$35.(\pm 10\%)$ &$6.00(\pm 10\%)$    &$5.19^{+0.15}_{-0.24}$ &$36.(\pm 10\%)$ &$6.00(\pm 10\%)$\\ \hline
		$J_{5}^{\text{M-M}}$ &$5.15^{+0.13}_{-0.23}$ &$36.(\pm 10\%)$ &$6.00(\pm 10\%)$    &$5.24^{+0.15}_{-0.24}$ &$37.(\pm 10\%)$ &$6.00(\pm 10\%)$\\ \hline
		$J_{6}^{\text{M-M}}$ &$5.16^{+0.13}_{-0.23}$ &$36.(\pm 10\%)$ &$6.00(\pm 10\%)$    &$5.24^{+0.14}_{-0.24}$ &$37.(\pm 10\%)$ &$6.00(\pm 10\%)$\\ \hline
		$J_{7}^{\text{M-M}}$ &$5.60^{+0.10}_{-0.11}$ &$43.(\pm 10\%)$ &$4.00(\pm 10\%)$    &$5.83^{+0.07}_{-0.11}$ &$46.(\pm 10\%)$ &$4.00(\pm 10\%)$\\ \hline
		$J_{8}^{\text{M-M}}$ &$5.66^{+0.10}_{-0.12}$ &$44.(\pm 10\%)$ &$4.40(\pm 10\%)$    &$5.04^{+0.09}_{-0.13}$ &$31.(\pm 10\%)$ &$2.90(\pm 10\%)$\\ \hline
		\end{tabular}
		\caption{LO and NLO results of meson-meson type operators of $Z_c$ system with $\overline{\text{MS}}$ renormalization scheme. The errors of masses shown in this table just come from the parametric dependence on $s_0$ and $M_B^2$.}
		\label{tab:A-meson-NLOresult-MSbar}
	\end{center}
\end{table}

\begin{table}[H]
	\renewcommand\arraystretch{1.3}
	\begin{center}
		\setlength{\tabcolsep}{4 mm}
		\begin{tabular}{|c|c|c|c|c|c|c|c|c|c|}
			\hline
			\multirow{3}{*}{Current} &
			\multicolumn{3}{c|}{LO}& \multicolumn{3}{c|}{NLO($\text{OS}$)} \\ \cline{2-7} &
			\makecell{$M_H$ \\ (GeV)} & \makecell{$s_0$ \\ ($\text{GeV}^2$)} & \makecell{$M_B^2$ \\ ($\text{GeV}^2$)} &
			\makecell{$M_H$ \\ (GeV)} & \makecell{$s_0$ \\ ($\text{GeV}^2$)} & \makecell{$M_B^2$ \\ ($\text{GeV}^2$)} \\ \hline
			
		 $J_{1}^{\text{M-M}}$ &$6.89^{+0.20}_{-0.15}$ &$55.(\pm 10\%)$ &$6.00(\pm 10\%)$    &$5.44^{+0.18}_{-0.32}$ &$36.(\pm 10\%)$ &$3.60(\pm 10\%)$\\ \hline
		$J_{2}^{\text{M-M}}$ &$6.82^{+0.15}_{-0.17}$ &$55.(\pm 10\%)$ &$6.00(\pm 10\%)$    &$5.46^{+0.20}_{-0.36}$ &$37.(\pm 10\%)$ &$4.00(\pm 10\%)$\\ \hline
		$J_{3}^{\text{M-M}}$ &$5.67^{+0.12}_{-0.21}$ &$41.(\pm 10\%)$ &$6.00(\pm 10\%)$    &$4.83^{+0.07}_{-0.06}$ &$31.(\pm 10\%)$ &$2.50(\pm 10\%)$\\ \hline
		$J_{4}^{\text{M-M}}$ &$5.62^{+0.11}_{-0.22}$ &$41.(\pm 10\%)$ &$6.00(\pm 10\%)$    &$4.29^{+0.07}_{-0.04}$ &$21.(\pm 10\%)$ &$2.00(\pm 10\%)$\\ \hline
		$J_{5}^{\text{M-M}}$ &$5.65^{+0.11}_{-0.21}$ &$42.(\pm 10\%)$ &$6.00(\pm 10\%)$    &$4.52^{+0.08}_{-0.06}$ &$24.(\pm 10\%)$ &$2.50(\pm 10\%)$\\ \hline
		$J_{6}^{\text{M-M}}$ &$5.65^{+0.11}_{-0.22}$ &$42.(\pm 10\%)$ &$6.00(\pm 10\%)$    &$5.58^{+0.13}_{-0.22}$ &$40.(\pm 10\%)$ &$6.00(\pm 10\%)$\\ \hline
		$J_{7}^{\text{M-M}}$ &$6.36^{+0.03}_{-0.11}$ &$54.(\pm 10\%)$ &$4.40(\pm 10\%)$    &$5.70^{+0.21}_{-0.29}$ &$42.(\pm 10\%)$ &$6.00(\pm 10\%)$\\ \hline
		$J_{8}^{\text{M-M}}$ &$6.44^{+0.03}_{-0.16}$ &$55.(\pm 10\%)$ &$4.90(\pm 10\%)$    &$5.57^{+0.22}_{-0.30}$ &$40.(\pm 10\%)$ &$5.90(\pm 10\%)$\\ \hline
		
		\end{tabular}
	\caption{LO and NLO results of meson-meson type operators of $Z_c$ system with $\text{OS}$ renormalization scheme. The errors of masses shown in this table just come from the parametric dependence on $s_0$ and $M_B^2$.}
\label{tab:A-meson-NLOresult-OS}
	\end{center}
\end{table}

\begin{figure}[H]
	\centering
	\subfigure[$\overline{\text{MS}}$]{
		\includegraphics[scale=0.4]{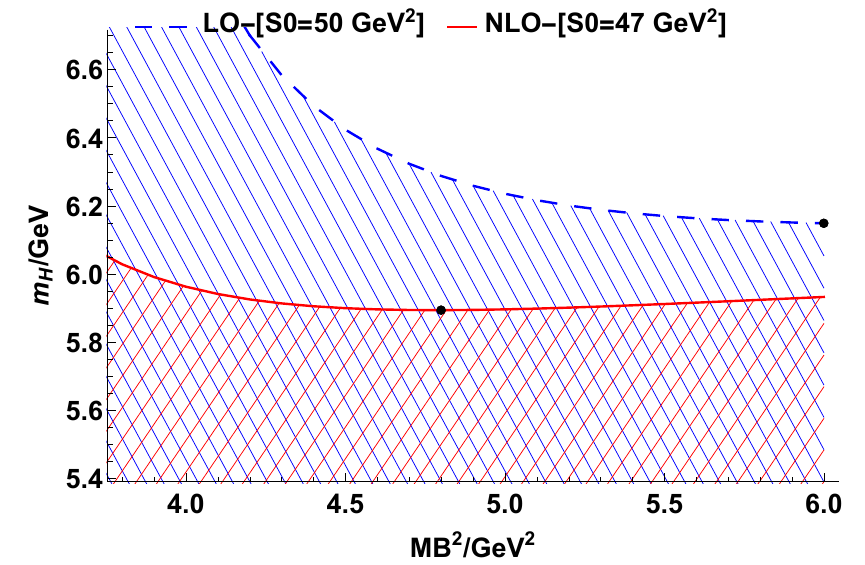}
		\includegraphics[scale=0.4]{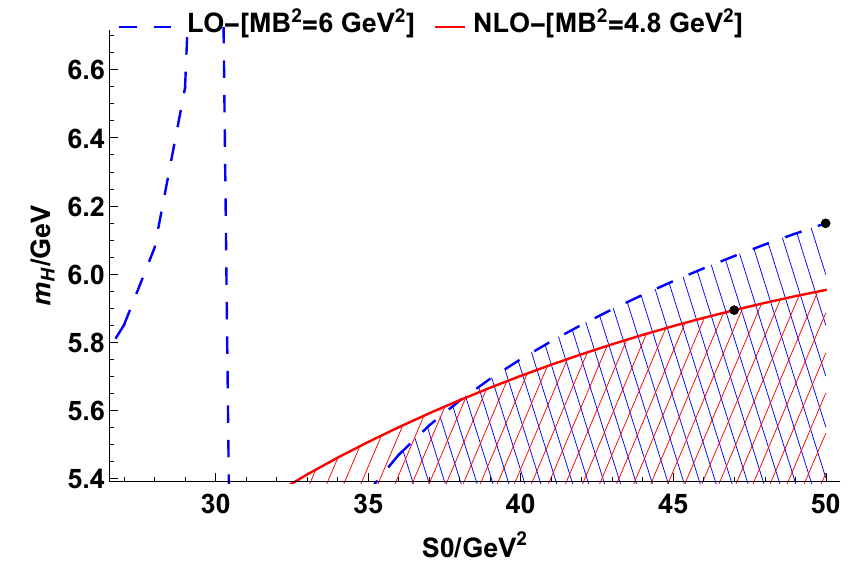}
	}\\
	\subfigure[OS]{
		\includegraphics[scale=0.4]{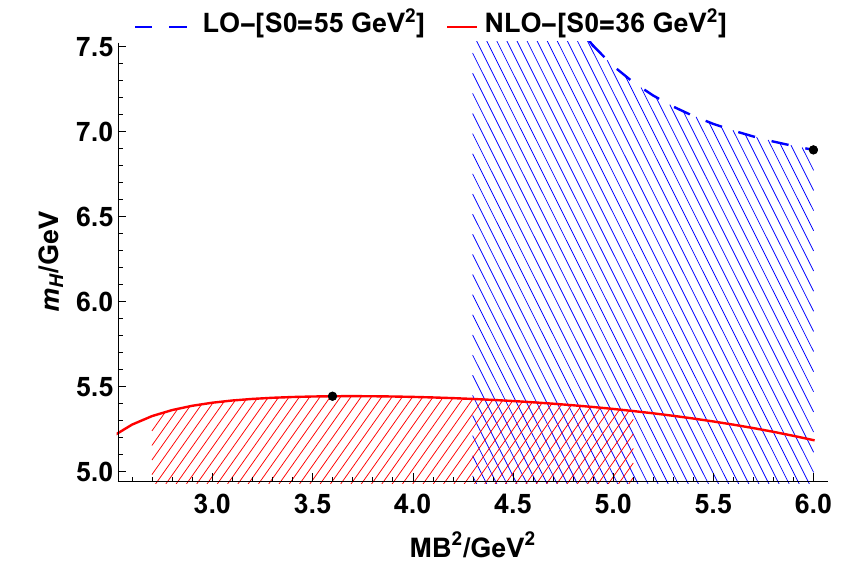}
		\includegraphics[scale=0.4]{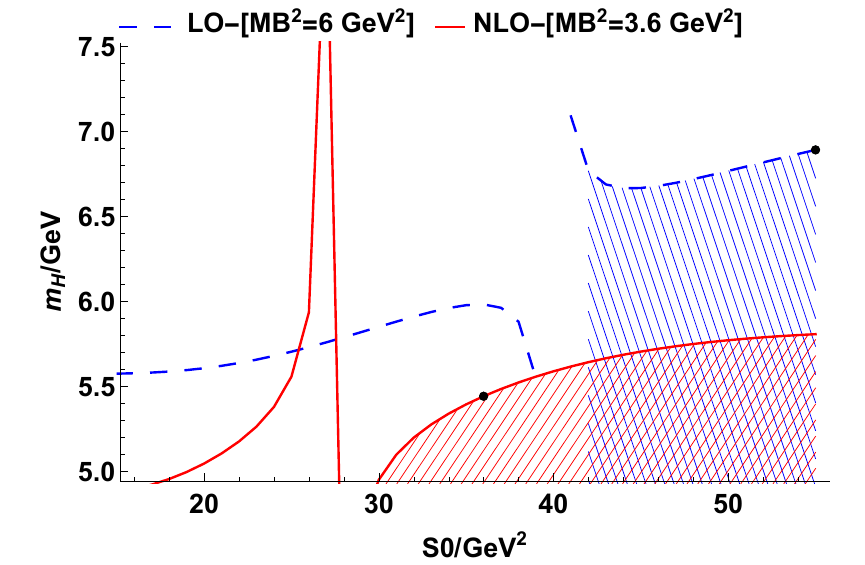}
	}
	\caption{\label{fig:Zc-[1+]-M-G-1-NLO-MSbar-OS}LO and NLO Result of $J_{1}^{\text{M-M}}$ of $Z_c$ system with $\overline{\text{MS}}$ and OS renormalization schemes.}
\end{figure}

\begin{figure}[H]
	\centering
	\subfigure[$\overline{\text{MS}}$]{
		\includegraphics[scale=0.4]{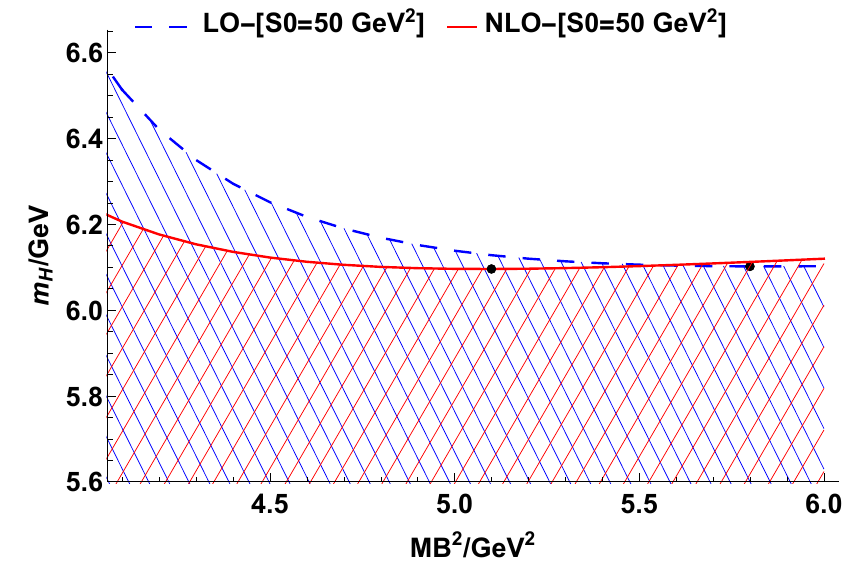}
		\includegraphics[scale=0.4]{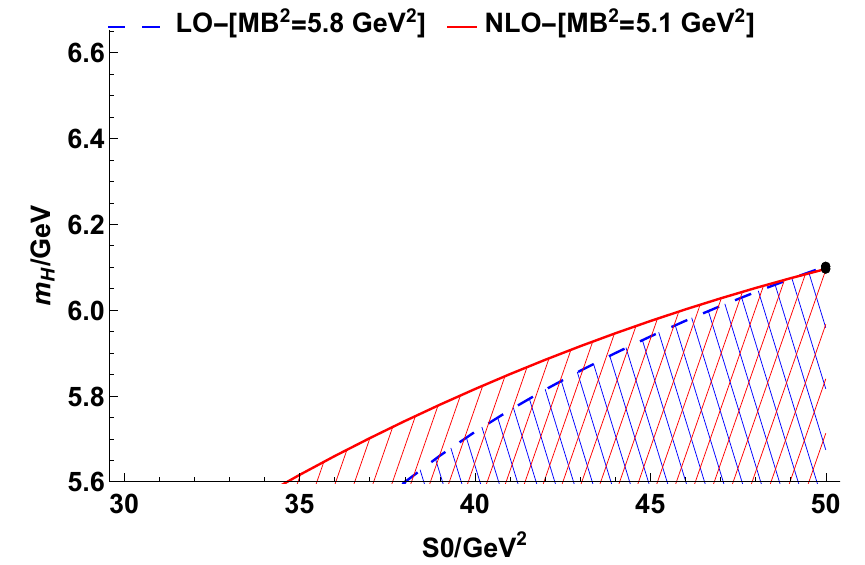}
	}\\
	\subfigure[OS]{
		\includegraphics[scale=0.4]{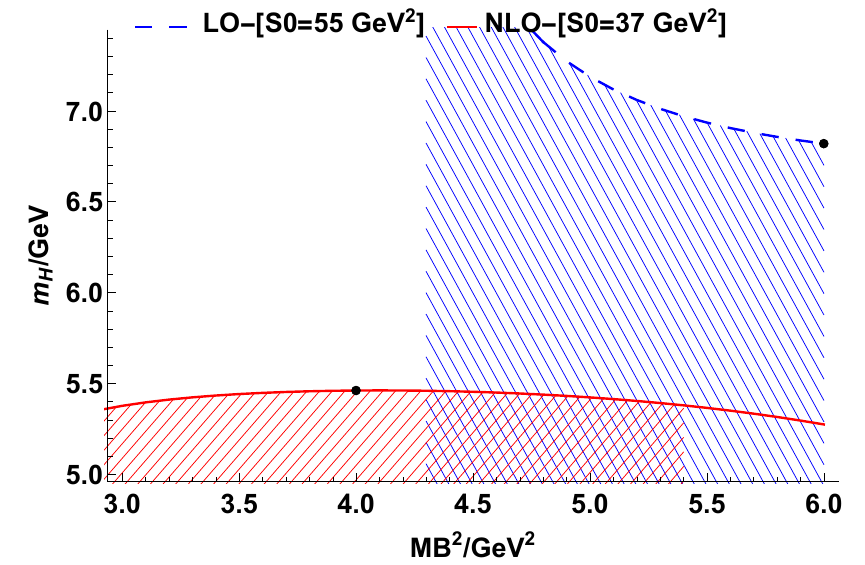}
		\includegraphics[scale=0.4]{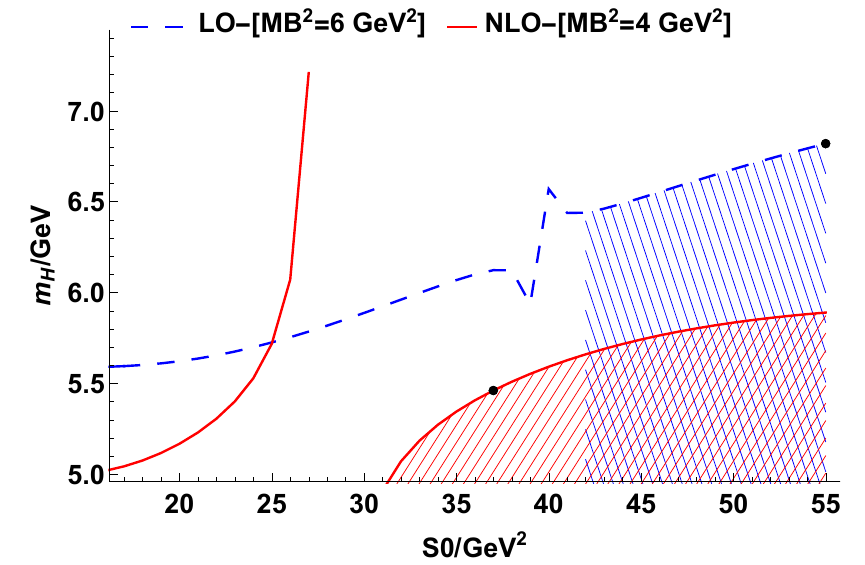}
	}
	\caption{\label{fig:Zc-[1+]-M-G-2-NLO-MSbar-OS}LO and NLO Result of $J_{2}^{\text{M-M}}$ of $Z_c$ system with $\overline{\text{MS}}$ and OS renormalization schemes.}
\end{figure}

\begin{figure}[H]
	\centering
	\subfigure[$\overline{\text{MS}}$]{
		\includegraphics[scale=0.4]{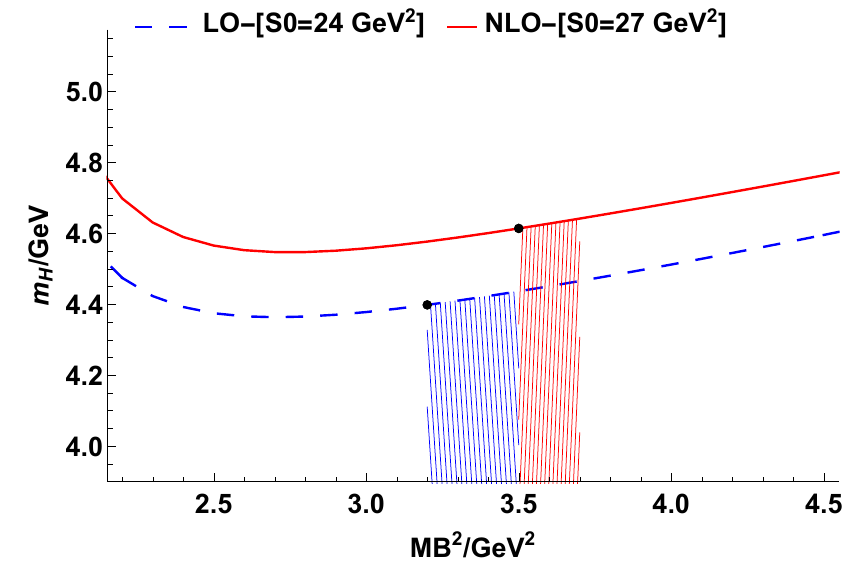}
		\includegraphics[scale=0.4]{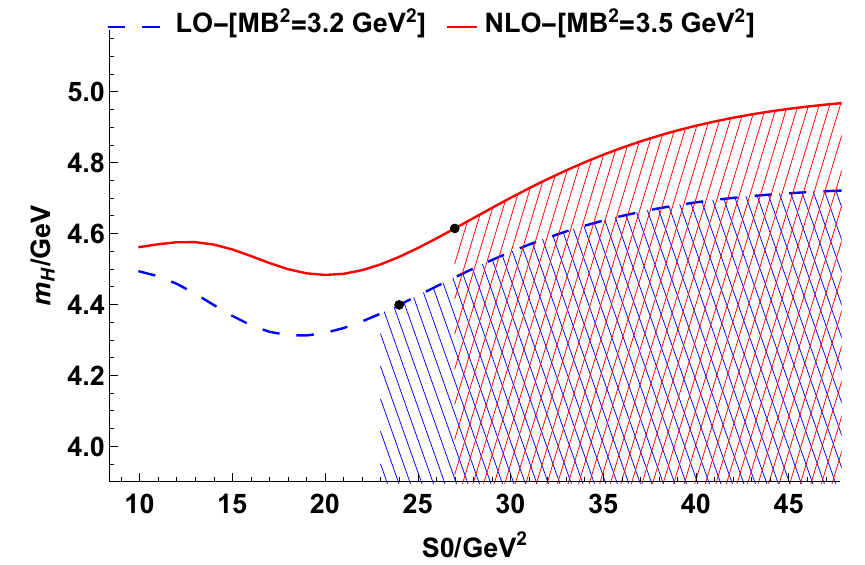}
	}\\
	\subfigure[OS]{
		\includegraphics[scale=0.4]{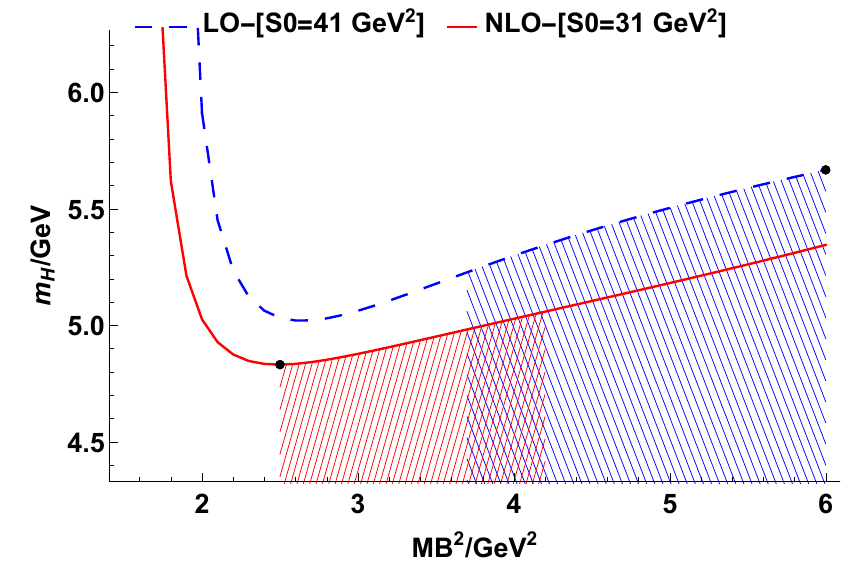}
		\includegraphics[scale=0.4]{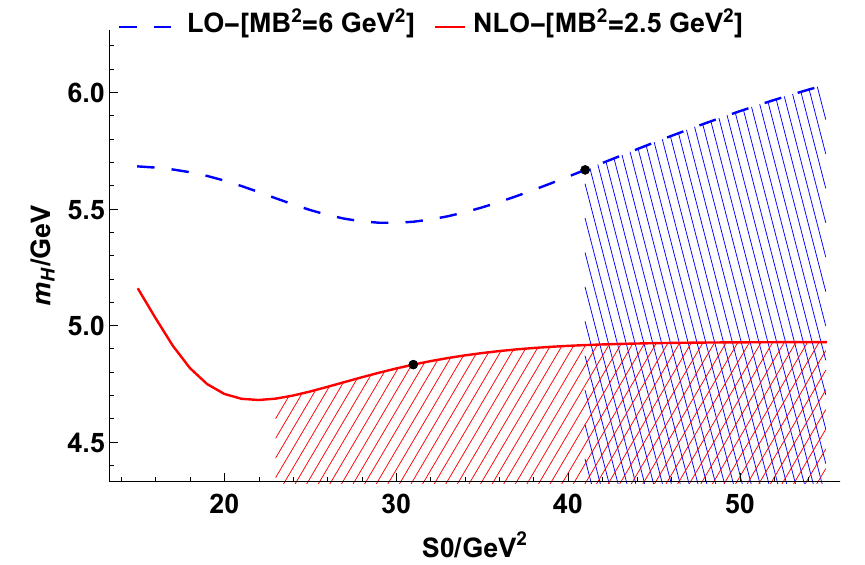}
	}
	\caption{\label{fig:Zc-[1+]-M-G-3-NLO-MSbar-OS}LO and NLO Result of $J_{3}^{\text{M-M}}$ of $Z_c$ system with $\overline{\text{MS}}$ and OS renormalization schemes.}
\end{figure}

\begin{figure}[H]
	\centering
	\subfigure[$\overline{\text{MS}}$]{
		\includegraphics[scale=0.4]{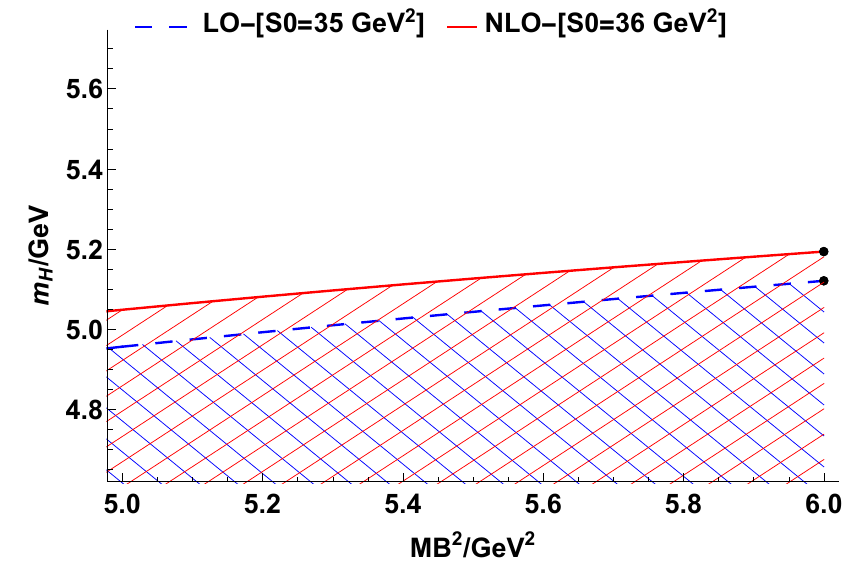}
		\includegraphics[scale=0.4]{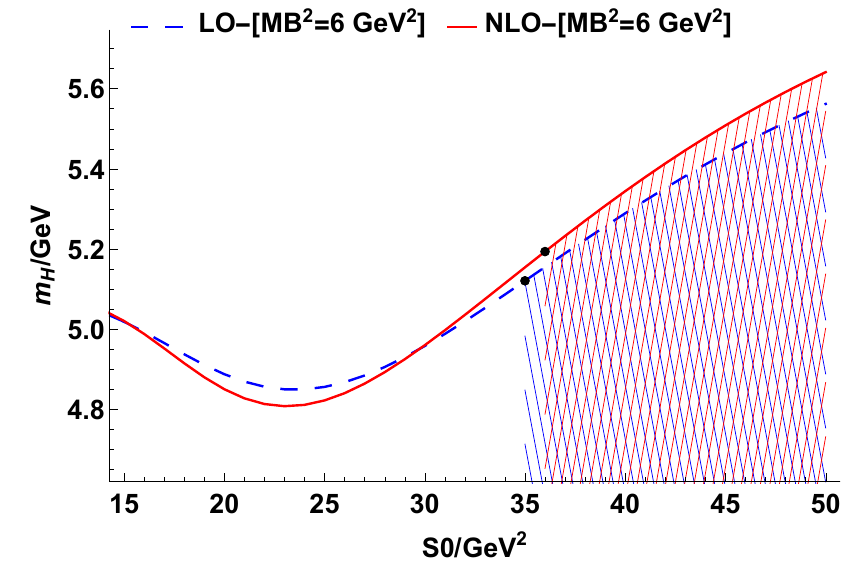}
	}\\
	\subfigure[OS]{
		\includegraphics[scale=0.4]{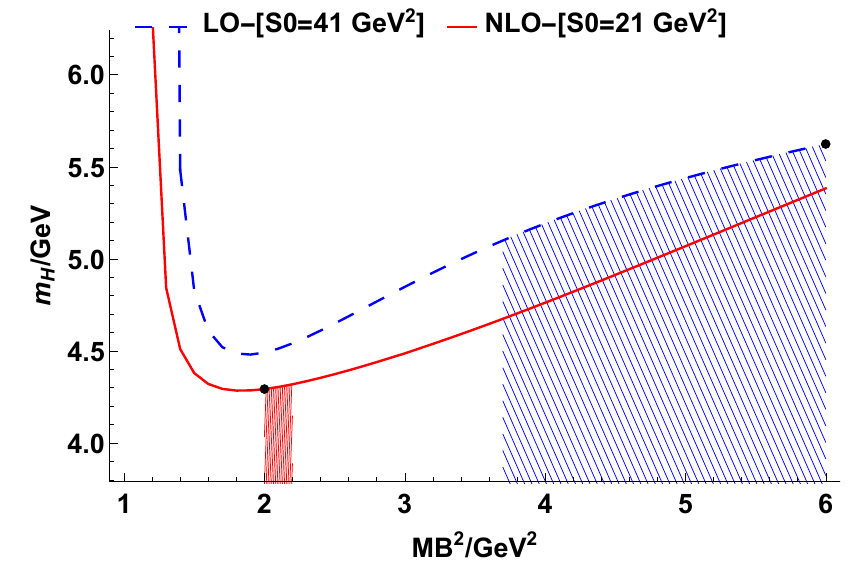}
		\includegraphics[scale=0.4]{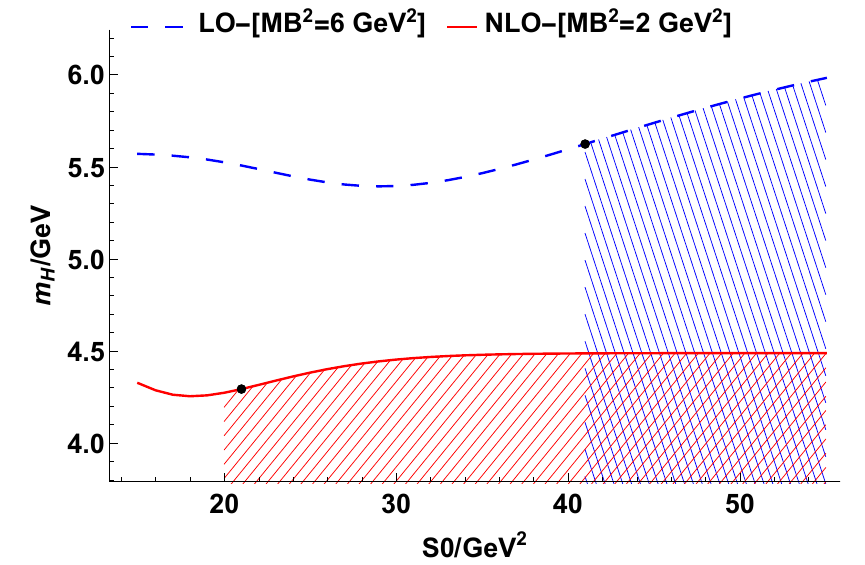}
	}
	\caption{\label{fig:Zc-[1+]-M-G-4-NLO-MSbar-OS}LO and NLO Result of $J_{4}^{\text{M-M}}$ of $Z_c$ system with $\overline{\text{MS}}$ and OS renormalization schemes.}
\end{figure}

\begin{figure}[H]
	\centering
	\subfigure[$\overline{\text{MS}}$]{
		\includegraphics[scale=0.4]{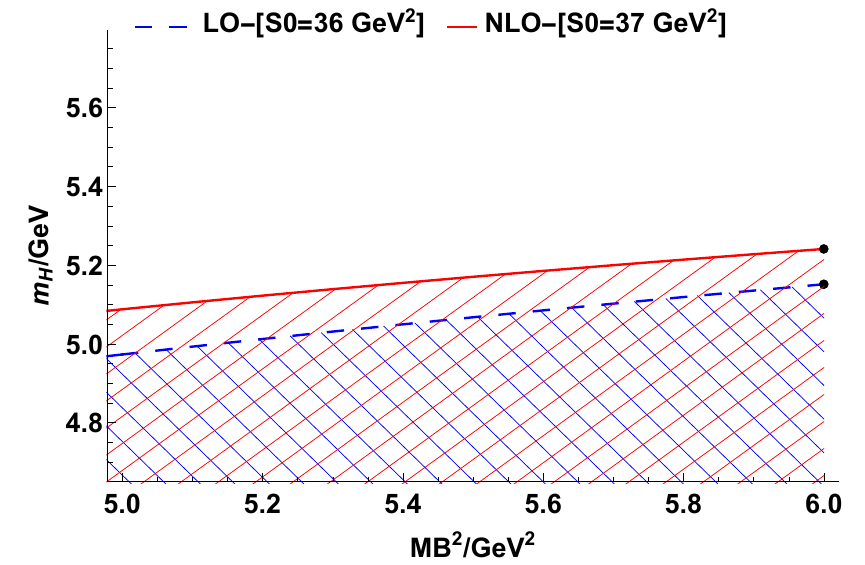}
		\includegraphics[scale=0.4]{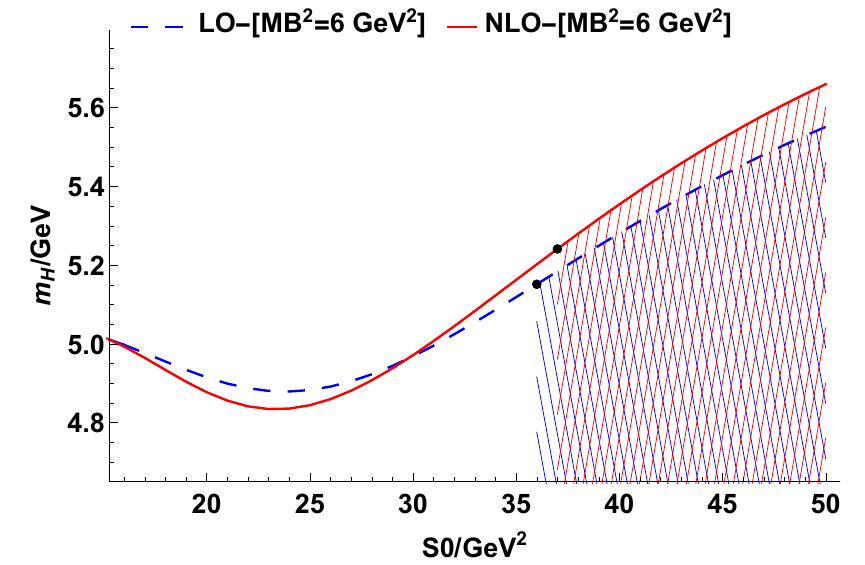}
	}\\
	\subfigure[OS]{
		\includegraphics[scale=0.4]{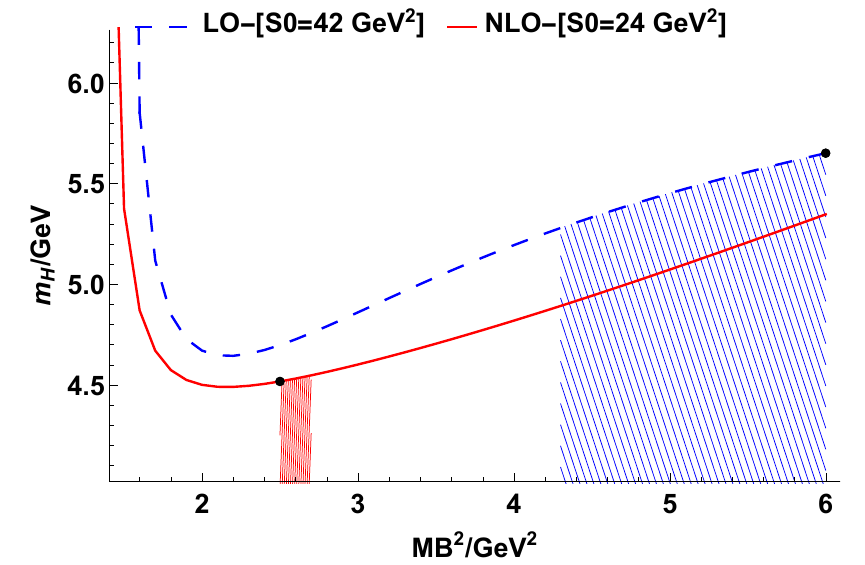}
		\includegraphics[scale=0.4]{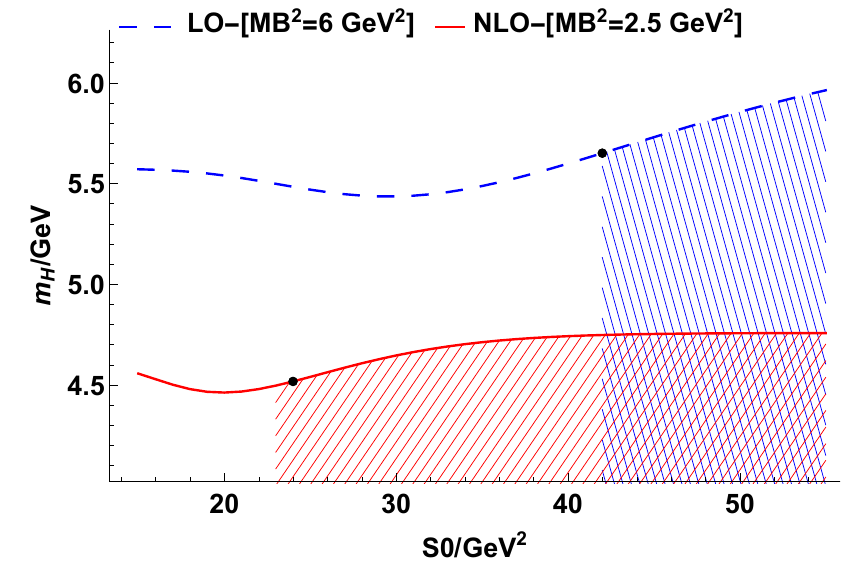}
	}
	\caption{\label{fig:Zc-[1+]-M-G-5-NLO-MSbar-OS}LO and NLO Result of $J_{5}^{\text{M-M}}$ of $Z_c$ system with $\overline{\text{MS}}$ and OS renormalization schemes.}
\end{figure}

\begin{figure}[H]
	\centering
	\subfigure[$\overline{\text{MS}}$]{
		\includegraphics[scale=0.4]{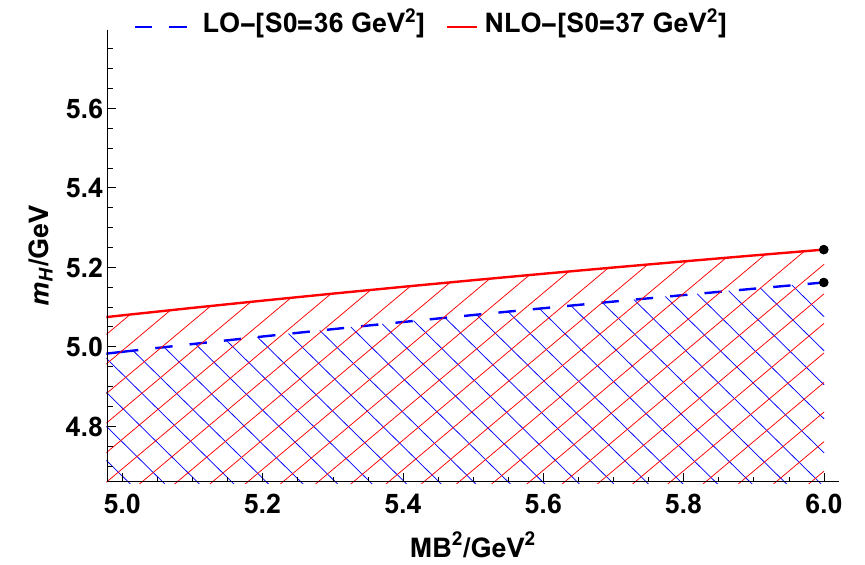}
		\includegraphics[scale=0.4]{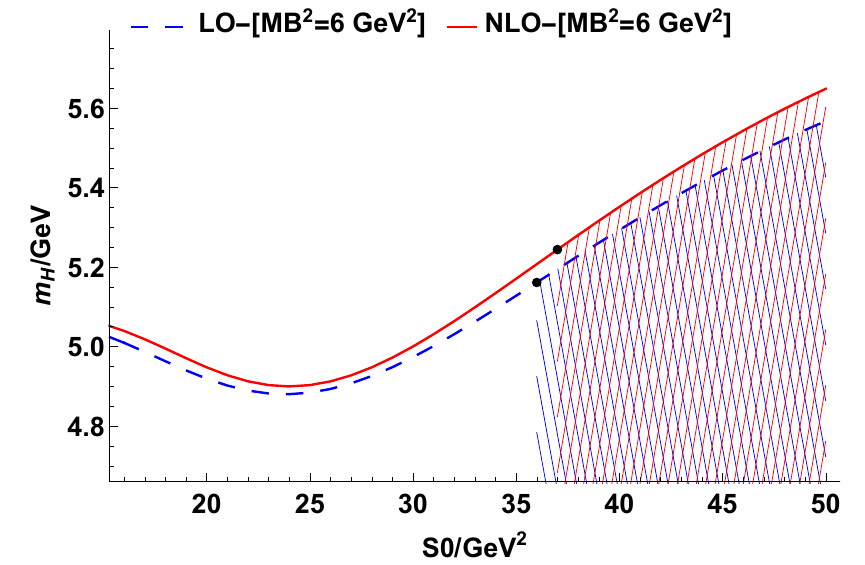}
	}\\
	\subfigure[OS]{
		\includegraphics[scale=0.4]{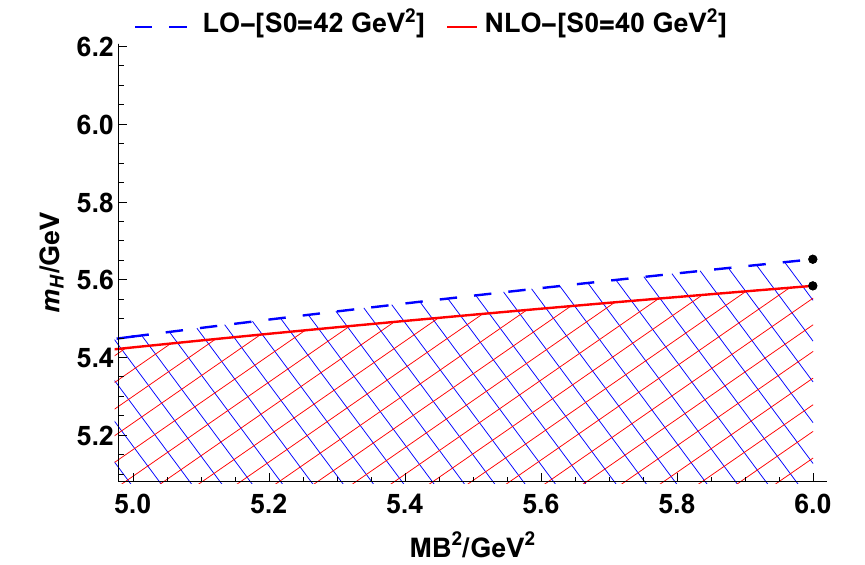}
		\includegraphics[scale=0.4]{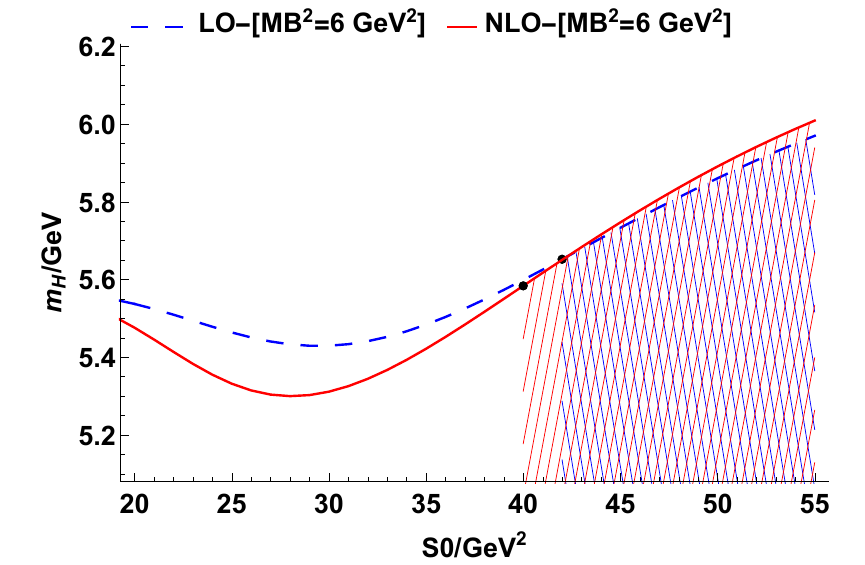}
	}
	\caption{\label{fig:Zc-[1+]-M-G-6-NLO-MSbar-OS}LO and NLO Result of $J_{6}^{\text{M-M}}$ of $Z_c$ system with $\overline{\text{MS}}$ and OS renormalization schemes.}
\end{figure}

\begin{figure}[H]
	\centering
	\subfigure[$\overline{\text{MS}}$]{
		\includegraphics[scale=0.4]{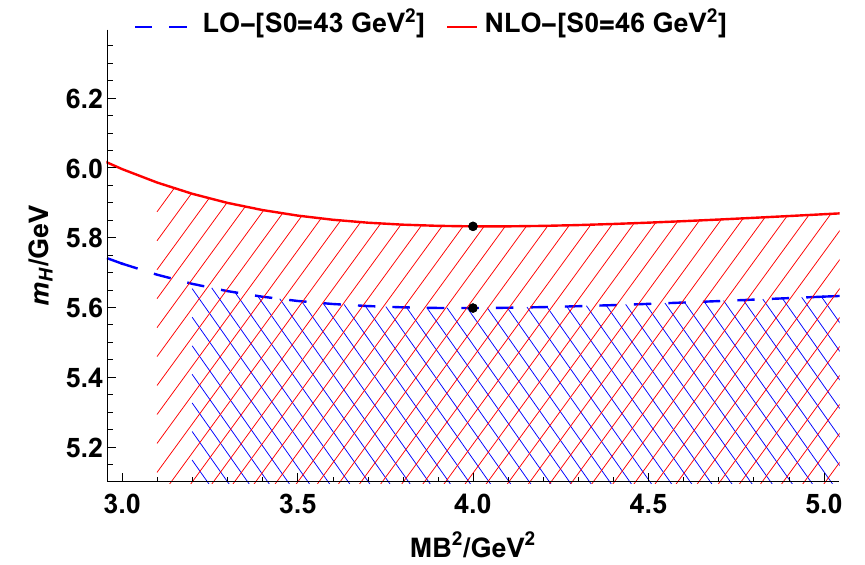}
		\includegraphics[scale=0.4]{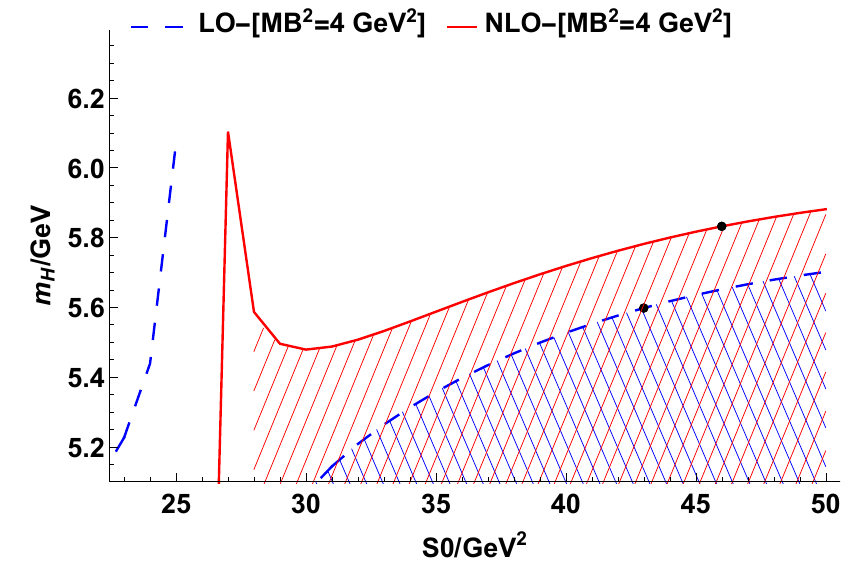}
	}\\
	\subfigure[OS]{
		\includegraphics[scale=0.4]{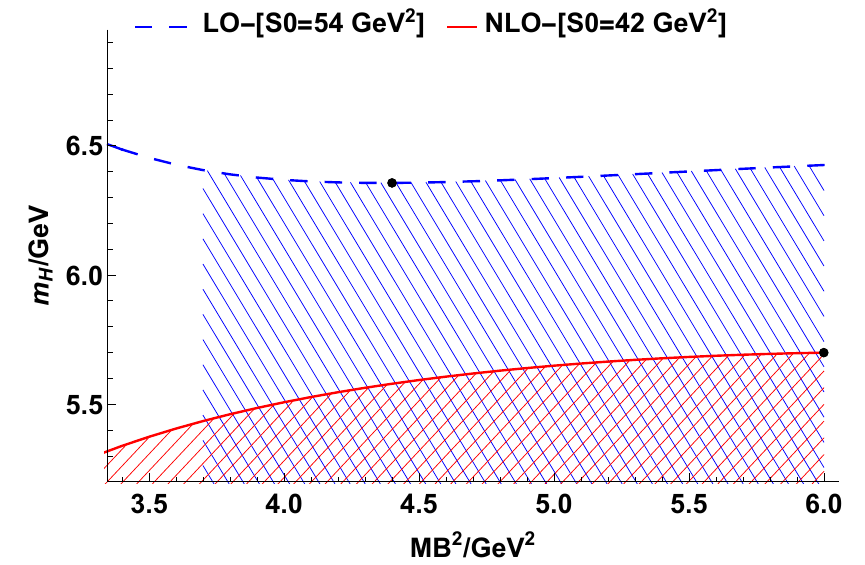}
		\includegraphics[scale=0.4]{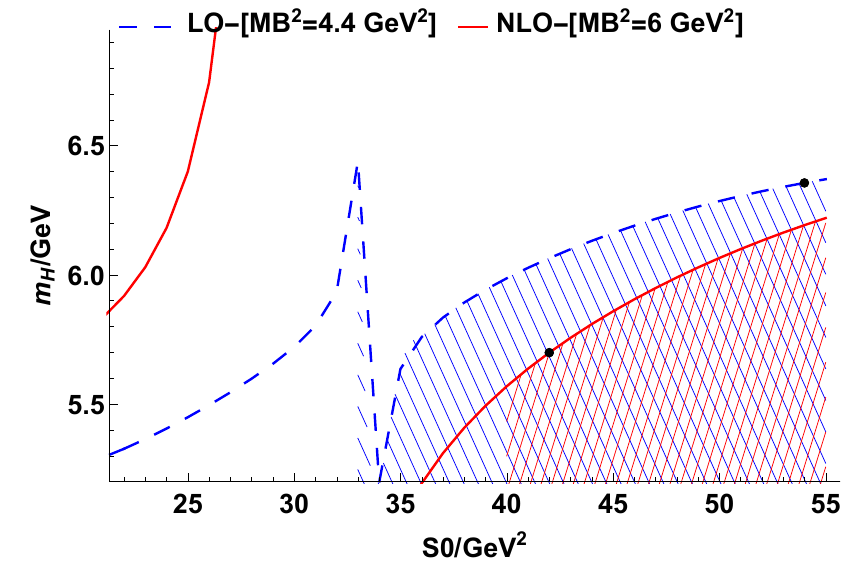}
	}
	\caption{\label{fig:Zc-[1+]-M-G-7-NLO-MSbar-OS}LO and NLO Result of $J_{7}^{\text{M-M}}$ of $Z_c$ system with $\overline{\text{MS}}$ and OS renormalization schemes.}
\end{figure}

\begin{figure}[H]
	\centering
	\subfigure[$\overline{\text{MS}}$]{
		\includegraphics[scale=0.4]{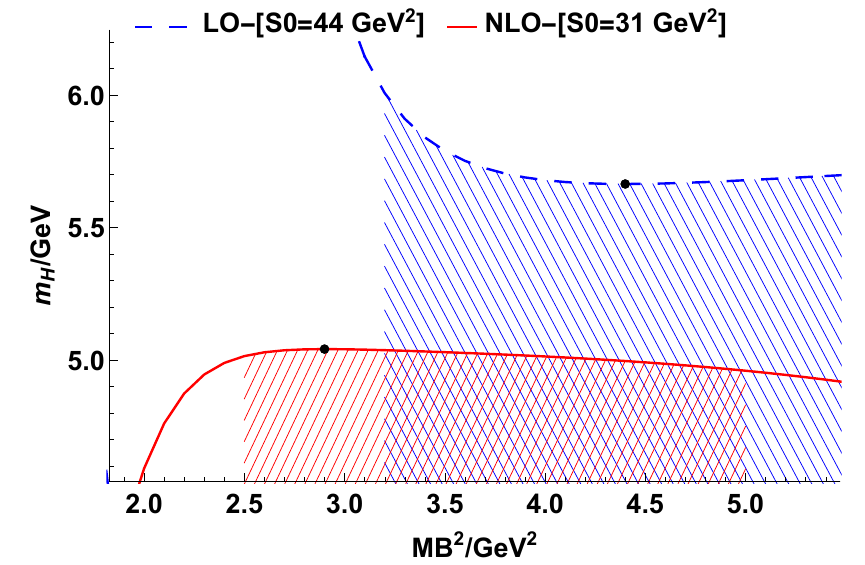}
		\includegraphics[scale=0.4]{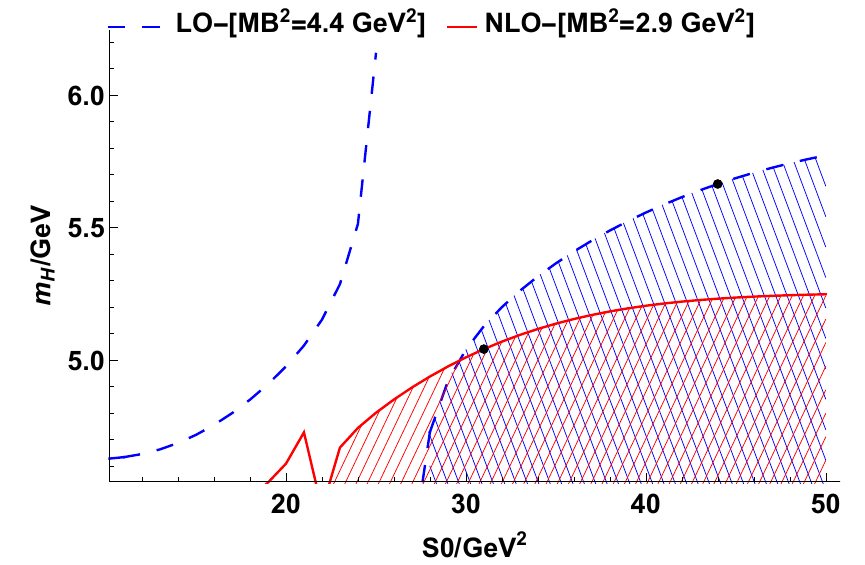}
	}\\
	\subfigure[OS]{
		\includegraphics[scale=0.4]{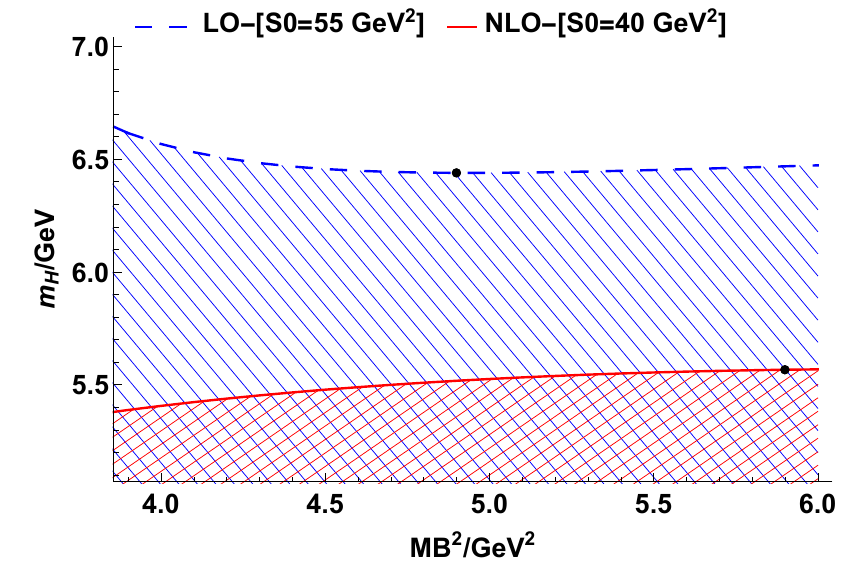}
		\includegraphics[scale=0.4]{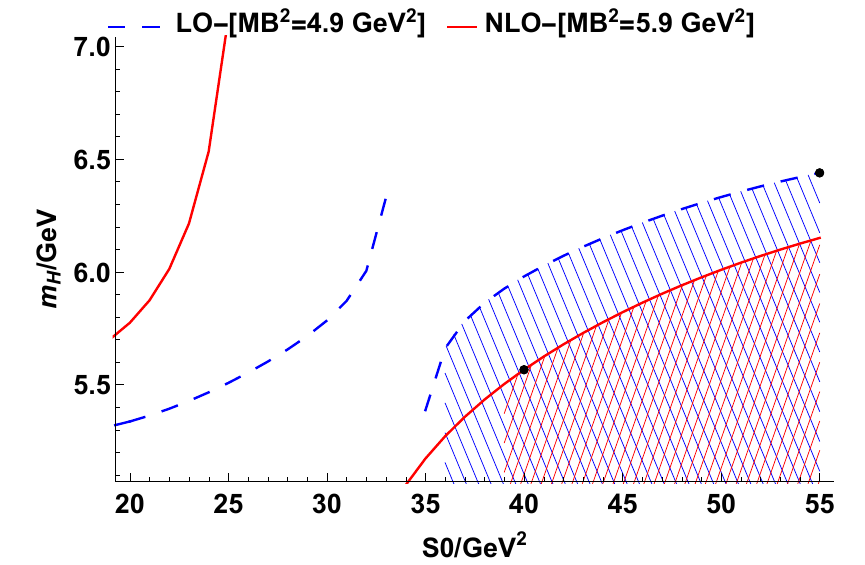}
	}
	\caption{\label{fig:Zc-[1+]-M-G-8-NLO-MSbar-OS}LO and NLO Result of $J_{8}^{\text{M-M}}$ of $Z_c$ system with $\overline{\text{MS}}$ and OS renormalization schemes.}
\end{figure}

\newpage
\subsection{Numerical results for diquark-antidiquark type operators of $Z_c$ system}

\begin{table}[H]
	\renewcommand\arraystretch{1.4}
	\begin{center}
		\setlength{\tabcolsep}{4 mm}
		\begin{tabular}{|c|c|c|c|c|c|c|c|c|c|}
			\hline
			\multirow{3}{*}{Current} &
			\multicolumn{3}{c|}{LO}& \multicolumn{3}{c|}{NLO($\overline{\text{MS}}$)} \\ \cline{2-7} &
			\makecell{$M_H$ \\ (GeV)} & \makecell{$s_0$ \\ ($\text{GeV}^2$)} & \makecell{$M_B^2$ \\ ($\text{GeV}^2$)} &
			\makecell{$M_H$ \\ (GeV)} & \makecell{$s_0$ \\ ($\text{GeV}^2$)} & \makecell{$M_B^2$ \\ ($\text{GeV}^2$)} \\ \hline
			
			 $J_{1}^{\text{Di-Di}}$ &$6.10^{+0.13}_{-0.17}$ &$50.(\pm 10\%)$ &$5.80(\pm 10\%)$    &$6.10^{+0.12}_{-0.14}$ &$50.(\pm 10\%)$ &$5.10(\pm 10\%)$\\ \hline
			$J_{2}^{\text{Di-Di}}$ &$6.02^{+0.14}_{-0.17}$ &$49.(\pm 10\%)$ &$5.30(\pm 10\%)$    &$6.37^{+0.04}_{-0.13}$ &$54.(\pm 10\%)$ &$5.70(\pm 10\%)$\\ \hline
			$J_{3}^{\text{Di-Di}}$ &$5.12^{+0.10}_{-0.23}$ &$35.(\pm 10\%)$ &$6.00(\pm 10\%)$    &$5.19^{+0.15}_{-0.24}$ &$36.(\pm 10\%)$ &$6.00(\pm 10\%)$\\ \hline
			$J_{4}^{\text{Di-Di}}$ &$5.12^{+0.10}_{-0.23}$ &$35.(\pm 10\%)$ &$6.00(\pm 10\%)$    &$5.21^{+0.15}_{-0.23}$ &$36.(\pm 10\%)$ &$6.00(\pm 10\%)$\\ \hline
			$J_{5}^{\text{Di-Di}}$ &$5.16^{+0.13}_{-0.23}$ &$36.(\pm 10\%)$ &$6.00(\pm 10\%)$    &$5.24^{+0.14}_{-0.24}$ &$37.(\pm 10\%)$ &$6.00(\pm 10\%)$\\ \hline
			$J_{6}^{\text{Di-Di}}$ &$5.20^{+0.13}_{-0.22}$ &$36.(\pm 10\%)$ &$6.00(\pm 10\%)$    &$5.35^{+0.14}_{-0.23}$ &$38.(\pm 10\%)$ &$6.00(\pm 10\%)$\\ \hline
			$J_{7}^{\text{Di-Di}}$ &$5.66^{+0.10}_{-0.12}$ &$44.(\pm 10\%)$ &$4.40(\pm 10\%)$    &$5.04^{+0.09}_{-0.13}$ &$31.(\pm 10\%)$ &$2.90(\pm 10\%)$\\ \hline
			$J_{8}^{\text{Di-Di}}$ &$5.82^{+0.14}_{-0.18}$ &$46.(\pm 10\%)$ &$5.00(\pm 10\%)$    &$5.73^{+0.08}_{-0.13}$ &$45.(\pm 10\%)$ &$4.20(\pm 10\%)$\\ \hline
		\end{tabular}
		\caption{LO and NLO results of diquark-antidiquark type operators of $Z_c$ system with $\overline{\text{MS}}$ renormalization scheme. The errors of masses shown in this table just come from the parametric dependence on $s_0$ and $M_B^2$.}
		\label{tab:A-Diquark-NLOresult-MSbar}
	\end{center}
\end{table}

\begin{table}[H]
	\renewcommand\arraystretch{1.4}
	\begin{center}
		\setlength{\tabcolsep}{4 mm}
		\begin{tabular}{|c|c|c|c|c|c|c|c|c|c|}
			\hline
			\multirow{3}{*}{Current} &
			\multicolumn{3}{c|}{LO}& \multicolumn{3}{c|}{NLO($\text{OS}$)} \\ \cline{2-7} &
			\makecell{$M_H$ \\ (GeV)} & \makecell{$s_0$ \\ ($\text{GeV}^2$)} & \makecell{$M_B^2$ \\ ($\text{GeV}^2$)} &
			\makecell{$M_H$ \\ (GeV)} & \makecell{$s_0$ \\ ($\text{GeV}^2$)} & \makecell{$M_B^2$ \\ ($\text{GeV}^2$)} \\ \hline
			
		$J_{1}^{\text{Di-Di}}$ &$6.82^{+0.15}_{-0.17}$ &$55.(\pm 10\%)$ &$6.00(\pm 10\%)$    &$5.46^{+0.20}_{-0.36}$ &$37.(\pm 10\%)$ &$4.00(\pm 10\%)$\\ \hline
		$J_{2}^{\text{Di-Di}}$ &$6.72^{+0.08}_{-0.20}$ &$55.(\pm 10\%)$ &$6.00(\pm 10\%)$    &$5.93^{+0.19}_{-0.35}$ &$45.(\pm 10\%)$ &$6.00(\pm 10\%)$\\ \hline
		$J_{3}^{\text{Di-Di}}$ &$5.62^{+0.11}_{-0.22}$ &$41.(\pm 10\%)$ &$6.00(\pm 10\%)$    &$4.29^{+0.07}_{-0.04}$ &$21.(\pm 10\%)$ &$2.00(\pm 10\%)$\\ \hline
		$J_{4}^{\text{Di-Di}}$ &$5.63^{+0.11}_{-0.21}$ &$41.(\pm 10\%)$ &$6.00(\pm 10\%)$    &$4.44^{+0.06}_{-0.03}$ &$22.(\pm 10\%)$ &$2.20(\pm 10\%)$\\ \hline
		$J_{5}^{\text{Di-Di}}$ &$5.65^{+0.11}_{-0.22}$ &$42.(\pm 10\%)$ &$6.00(\pm 10\%)$    &$5.58^{+0.13}_{-0.22}$ &$40.(\pm 10\%)$ &$6.00(\pm 10\%)$\\ \hline
		$J_{6}^{\text{Di-Di}}$ &$5.70^{+0.11}_{-0.21}$ &$42.(\pm 10\%)$ &$6.00(\pm 10\%)$    &$4.81^{+0.11}_{-0.06}$ &$27.(\pm 10\%)$ &$3.00(\pm 10\%)$\\ \hline
		$J_{7}^{\text{Di-Di}}$ &$6.44^{+0.03}_{-0.16}$ &$55.(\pm 10\%)$ &$4.90(\pm 10\%)$    &$5.57^{+0.22}_{-0.30}$ &$40.(\pm 10\%)$ &$5.90(\pm 10\%)$\\ \hline
		$J_{8}^{\text{Di-Di}}$ &$6.53^{+0.02}_{-0.20}$ &$55.(\pm 10\%)$ &$5.70(\pm 10\%)$    &$5.72^{+0.20}_{-0.27}$ &$42.(\pm 10\%)$ &$6.00(\pm 10\%)$\\ \hline
		\end{tabular}

	\caption{LO and NLO results of diquark-antidiquark type operators of $Z_c$ system with $\text{OS}$ renormalization scheme. The errors of masses shown in this table just come from the parametric dependence on $s_0$ and $M_B^2$.}
	\label{tab:A-Diquark-NLOresult-OS}
	\end{center}
\end{table}

\begin{figure}[H]
	\centering
	\subfigure[$\overline{\text{MS}}$]{
		\includegraphics[scale=0.4]{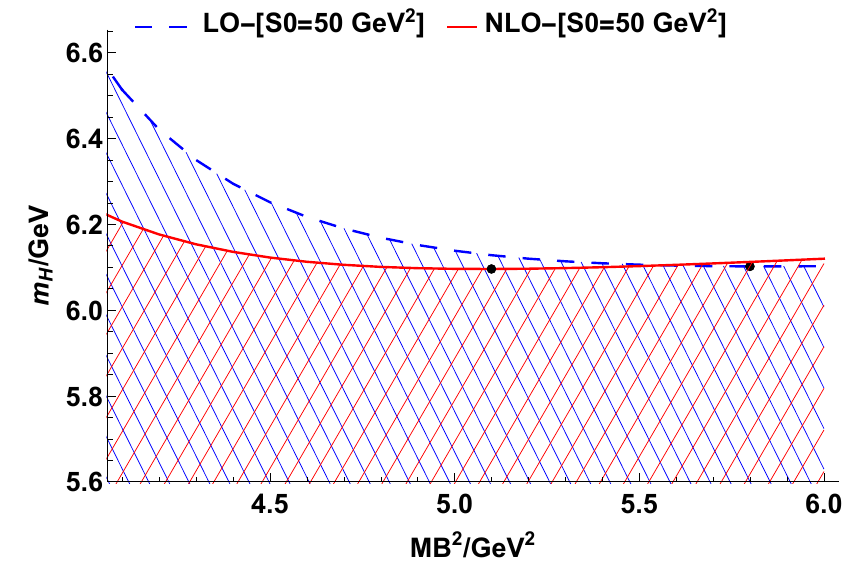}
		\includegraphics[scale=0.4]{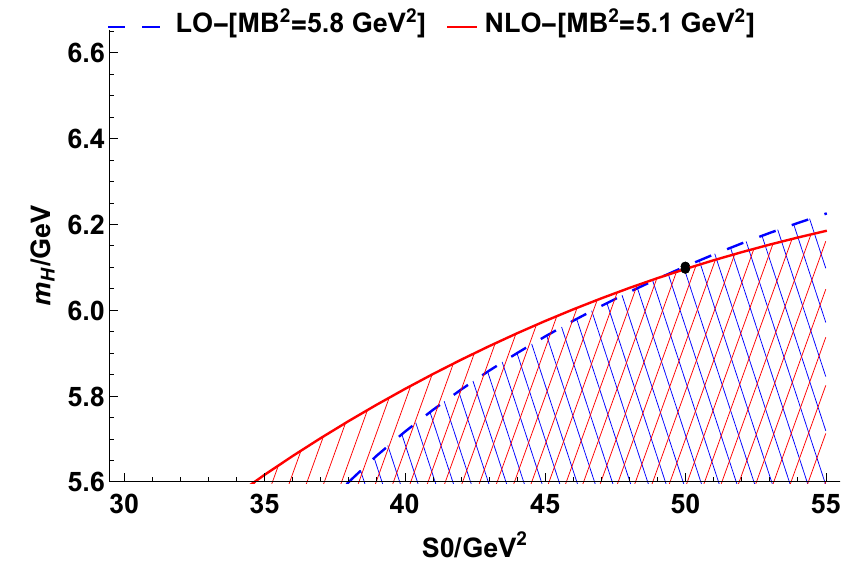}
	}\\
	\subfigure[OS]{
		\includegraphics[scale=0.4]{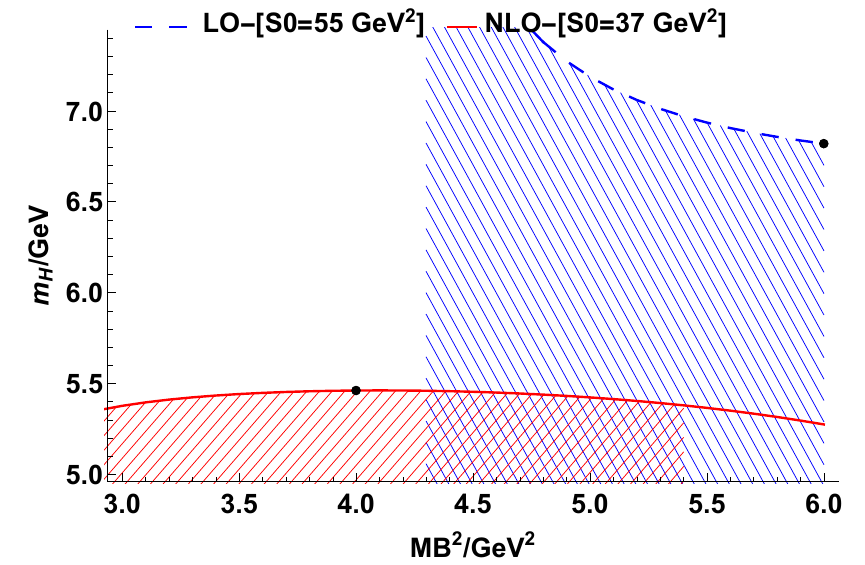}
		\includegraphics[scale=0.4]{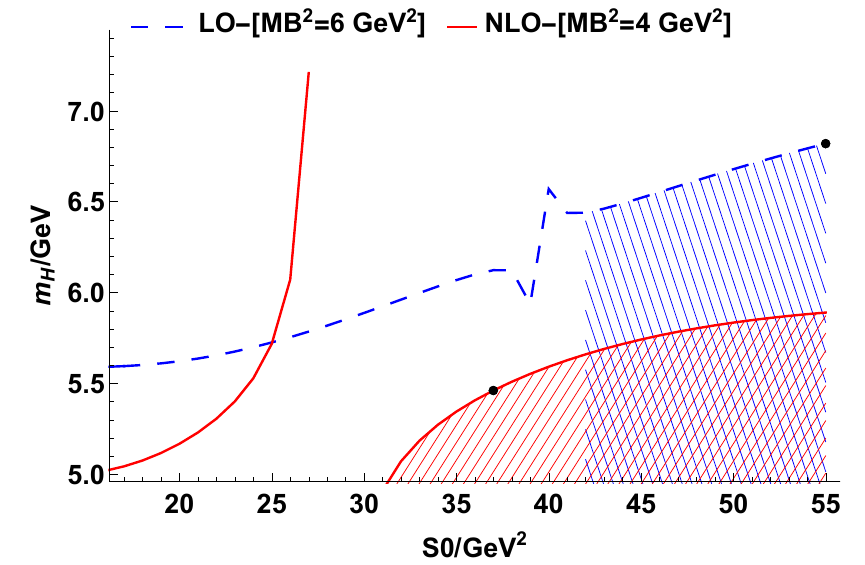}
	}
	\caption{\label{fig:Zc-[1+]-Di-G-1-NLO-MSbar-OS}LO and NLO Result of $J_{1}^{\text{Di-Di}}$ of $Z_c$ system with $\overline{\text{MS}}$ and OS renormalization schemes.}
\end{figure}

\begin{figure}[H]
	\centering
	\subfigure[$\overline{\text{MS}}$]{
		\includegraphics[scale=0.4]{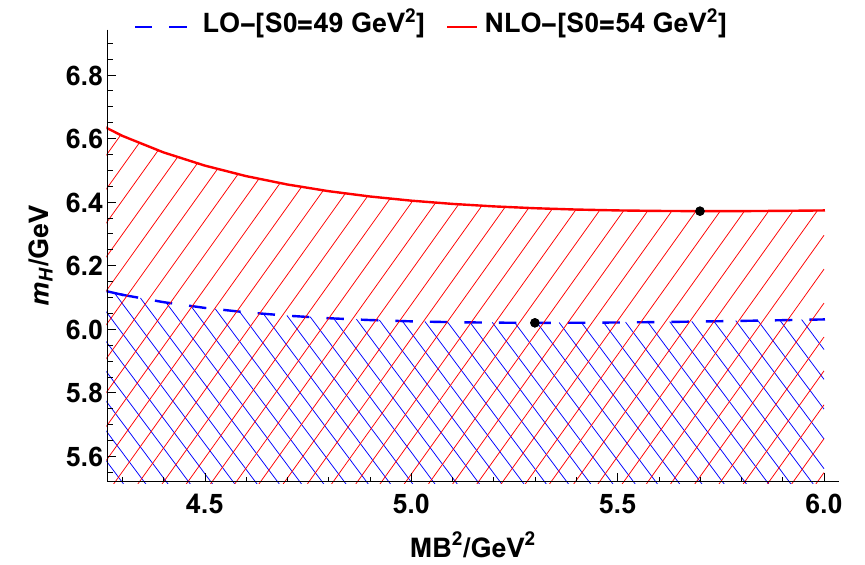}
		\includegraphics[scale=0.4]{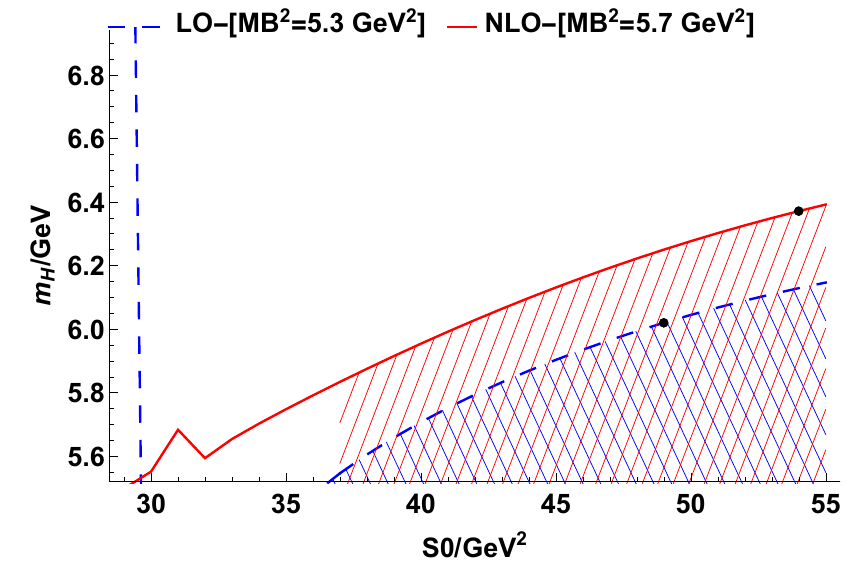}
	}\\
	\subfigure[OS]{
		\includegraphics[scale=0.4]{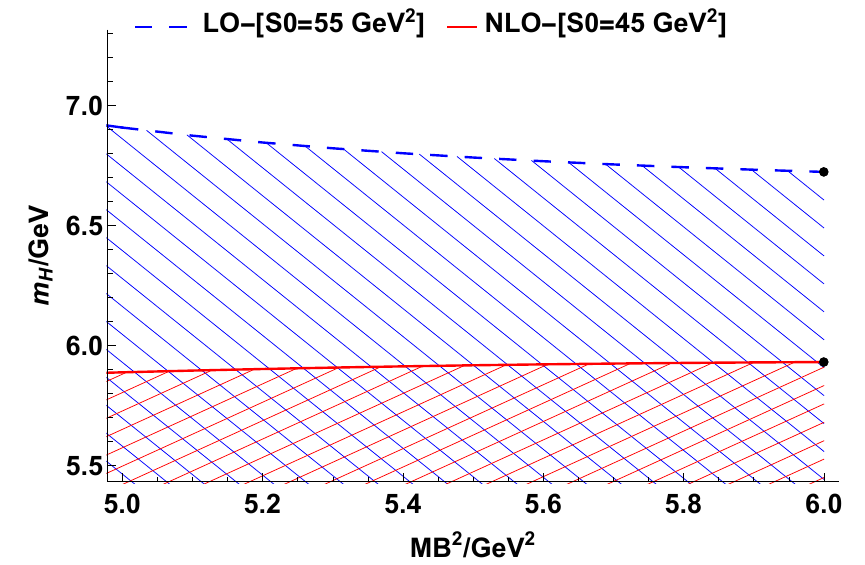}
		\includegraphics[scale=0.4]{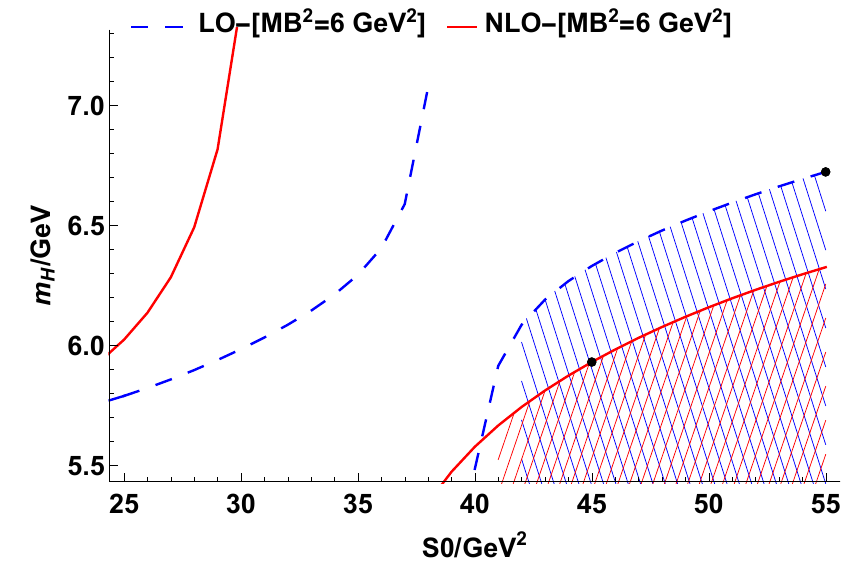}
	}
	\caption{\label{fig:Zc-[1+]-Di-G-2-NLO-MSbar-OS}LO and NLO Result of $J_{2}^{\text{Di-Di}}$ of $Z_c$ system with $\overline{\text{MS}}$ and OS renormalization schemes.}
\end{figure}

\begin{figure}[H]
	\centering
	\subfigure[$\overline{\text{MS}}$]{
		\includegraphics[scale=0.4]{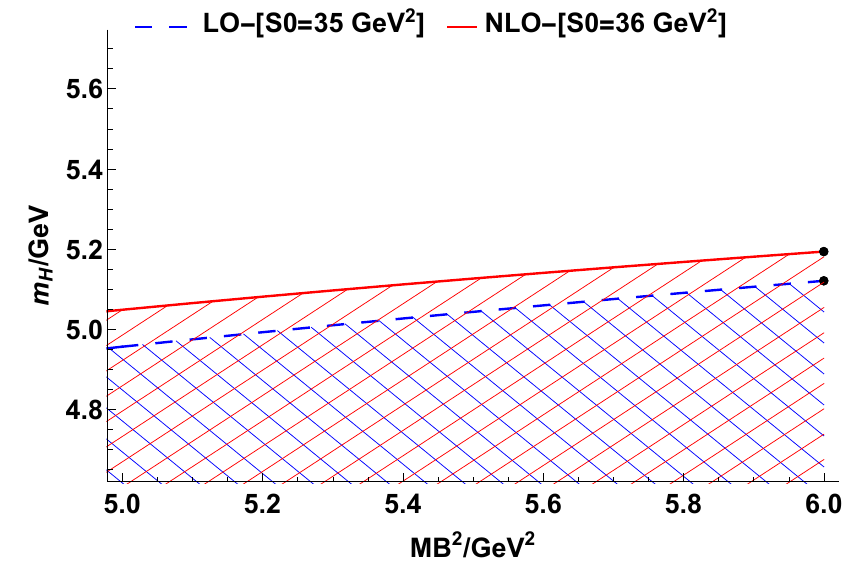}
		\includegraphics[scale=0.4]{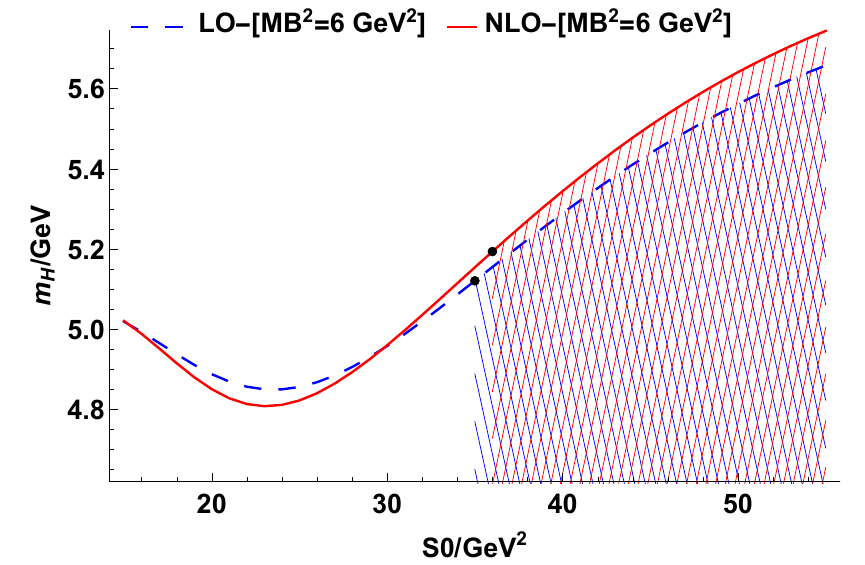}
	}\\
	\subfigure[OS]{
		\includegraphics[scale=0.4]{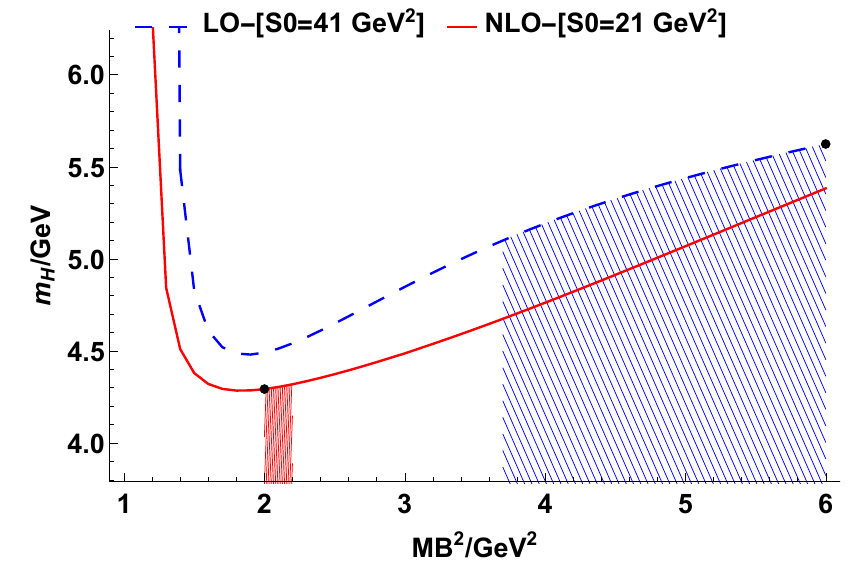}
		\includegraphics[scale=0.4]{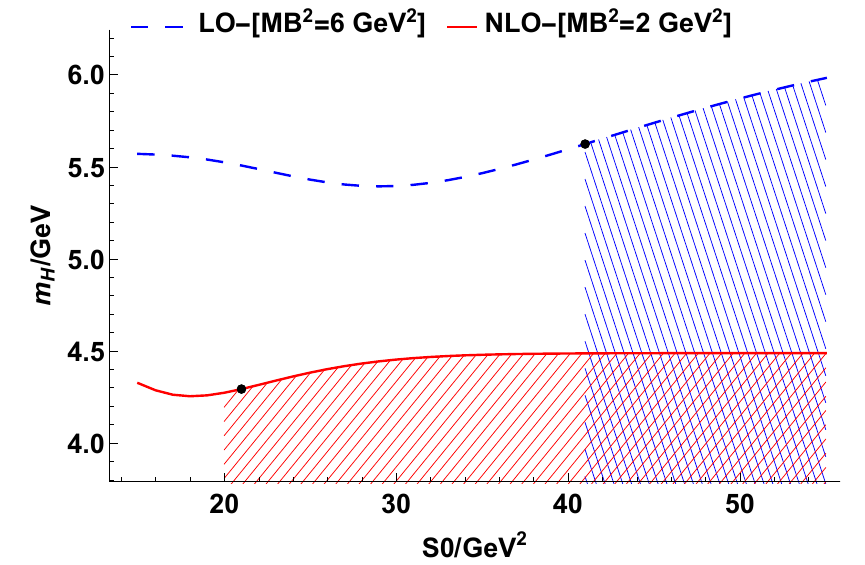}
	}
	\caption{\label{fig:Zc-[1+]-Di-G-3-NLO-MSbar-OS}LO and NLO Result of $J_{3}^{\text{Di-Di}}$ of $Z_c$ system with $\overline{\text{MS}}$ and OS renormalization schemes.}
\end{figure}

\begin{figure}[H]
	\centering
	\subfigure[$\overline{\text{MS}}$]{
		\includegraphics[scale=0.4]{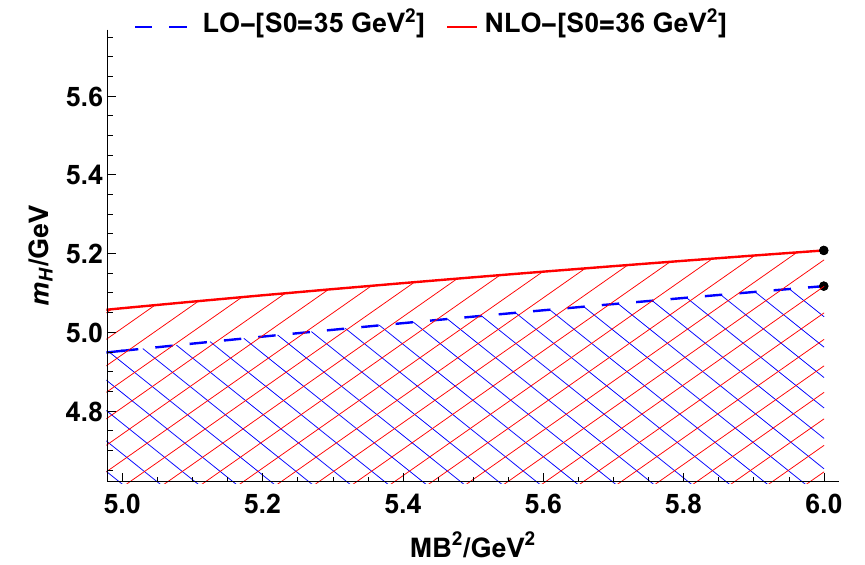}
		\includegraphics[scale=0.4]{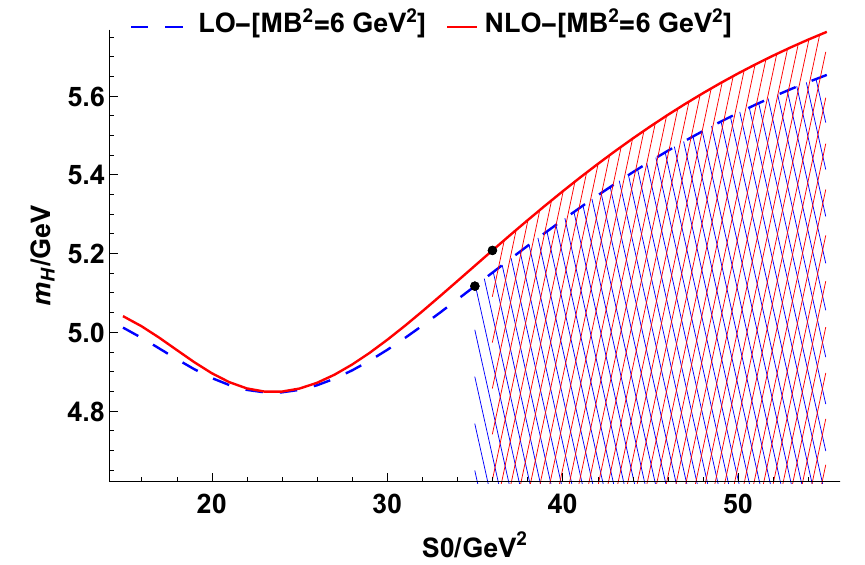}
	}\\
	\subfigure[OS]{
		\includegraphics[scale=0.4]{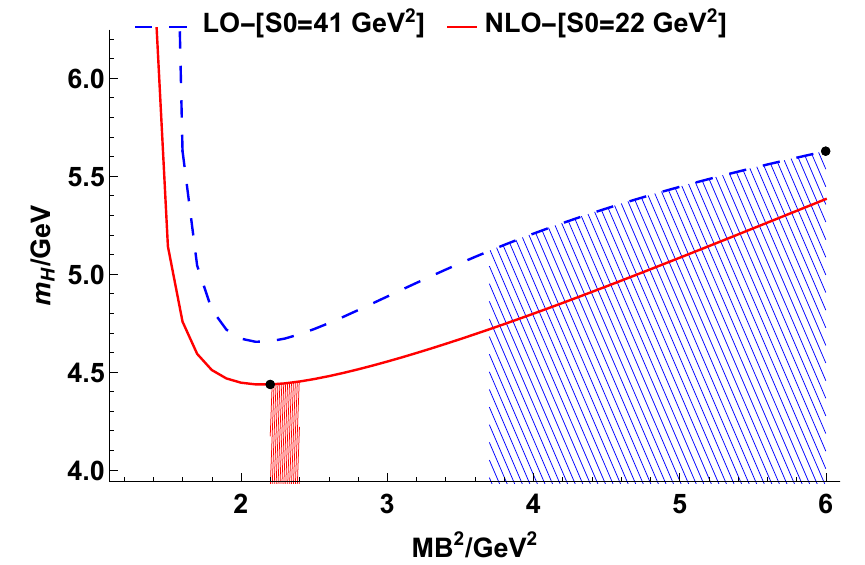}
		\includegraphics[scale=0.4]{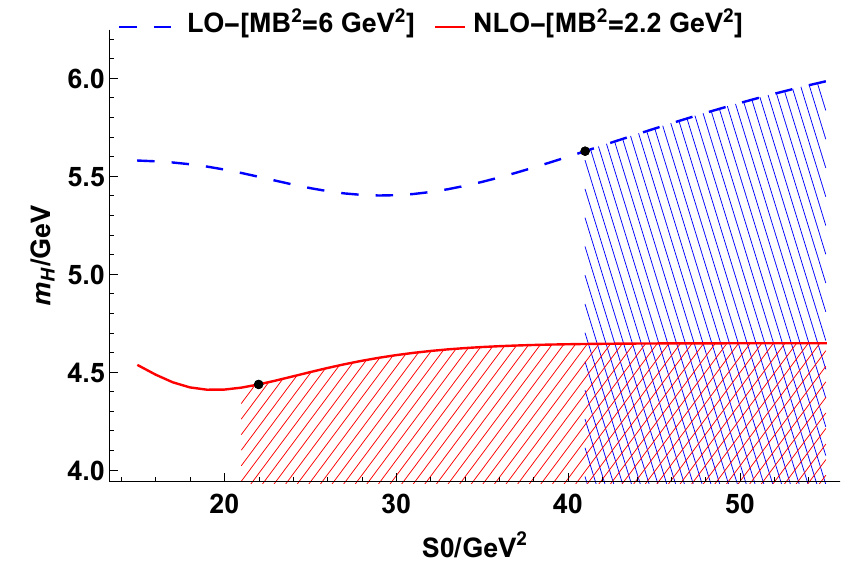}
	}
	\caption{\label{fig:Zc-[1+]-Di-G-4-NLO-MSbar-OS}LO and NLO Result of $J_{4}^{\text{Di-Di}}$ of $Z_c$ system with $\overline{\text{MS}}$ and OS renormalization schemes.}
\end{figure}

\begin{figure}[H]
	\centering
	\subfigure[$\overline{\text{MS}}$]{
		\includegraphics[scale=0.4]{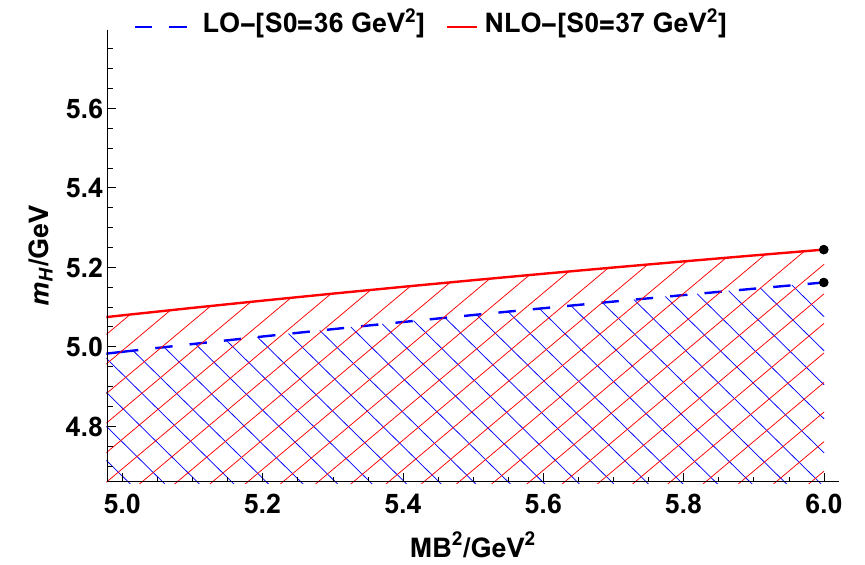}
		\includegraphics[scale=0.4]{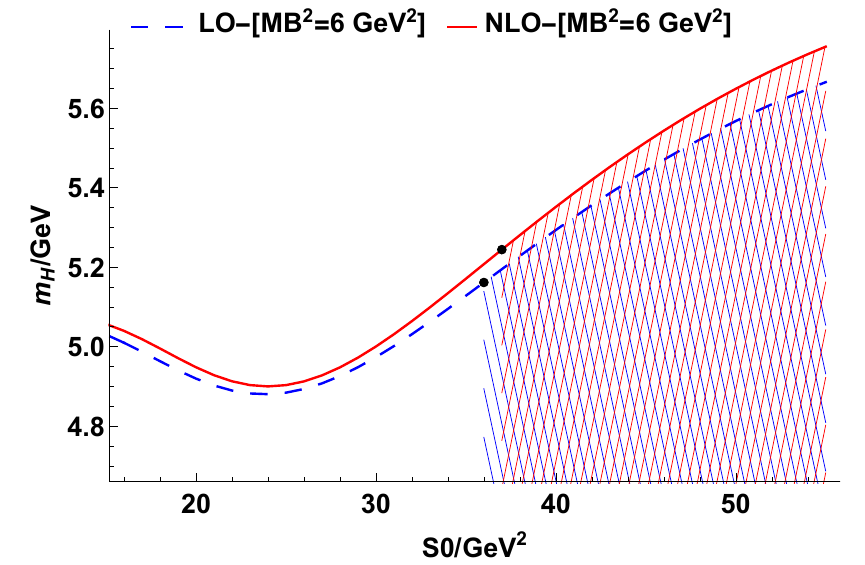}
	}\\
	\subfigure[OS]{
		\includegraphics[scale=0.4]{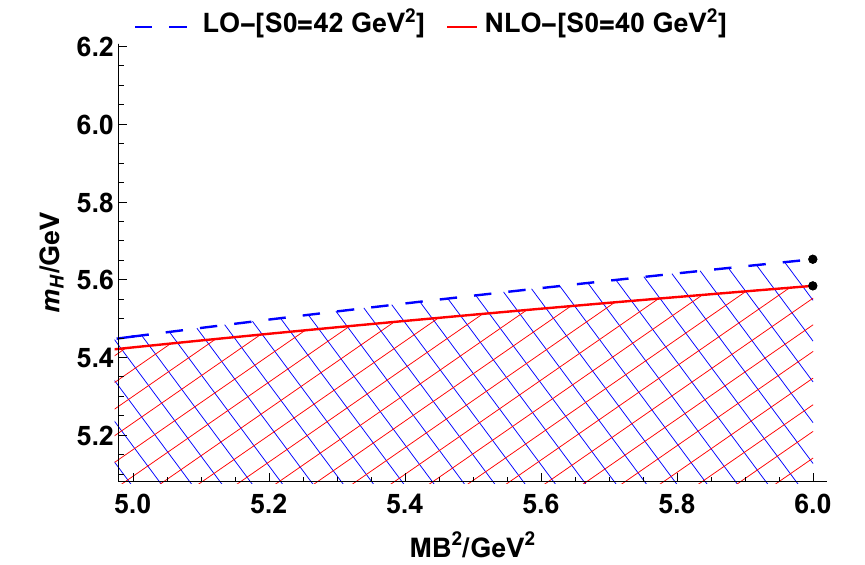}
		\includegraphics[scale=0.4]{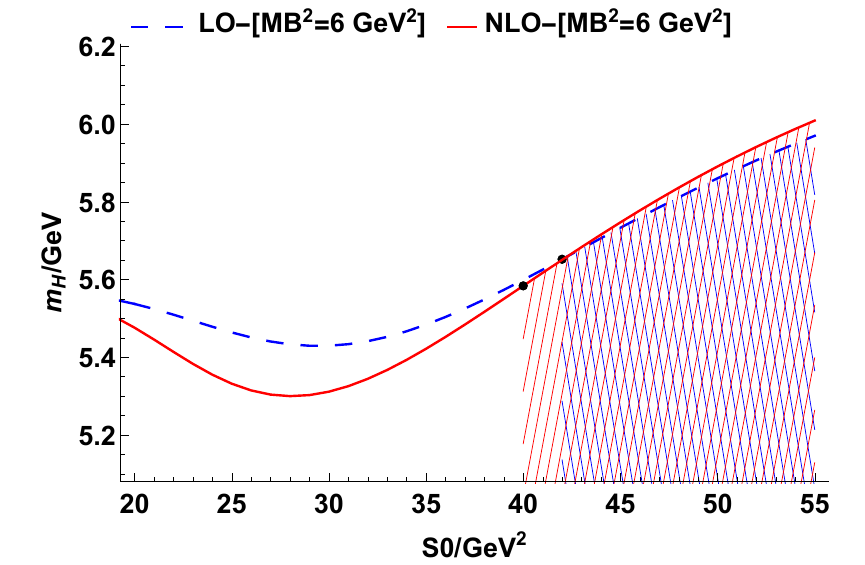}
	}
	\caption{\label{fig:Zc-[1+]-Di-G-5-NLO-MSbar-OS}LO and NLO Result of $J_{5}^{\text{Di-Di}}$ of $Z_c$ system with $\overline{\text{MS}}$ and OS renormalization schemes.}
\end{figure}

\begin{figure}[H]
	\centering
	\subfigure[$\overline{\text{MS}}$]{
		\includegraphics[scale=0.4]{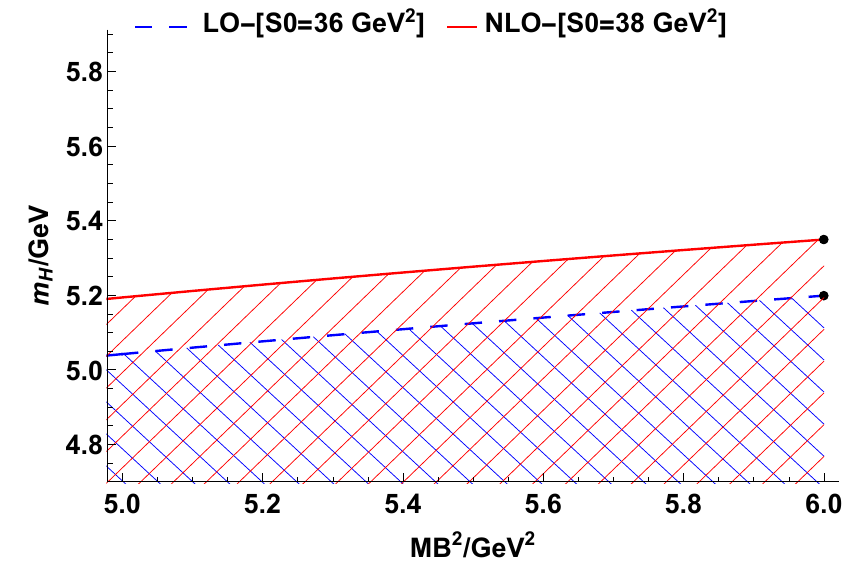}
		\includegraphics[scale=0.4]{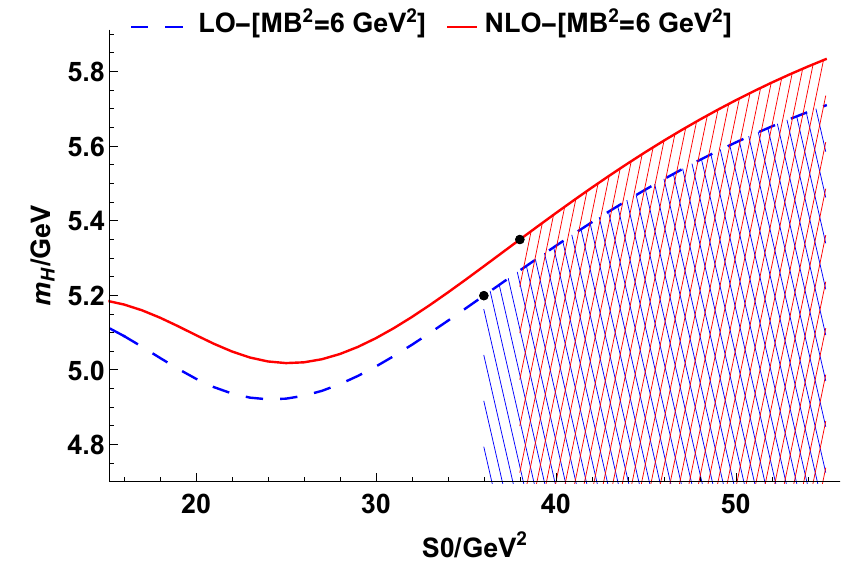}
	}\\
	\subfigure[OS]{
		\includegraphics[scale=0.4]{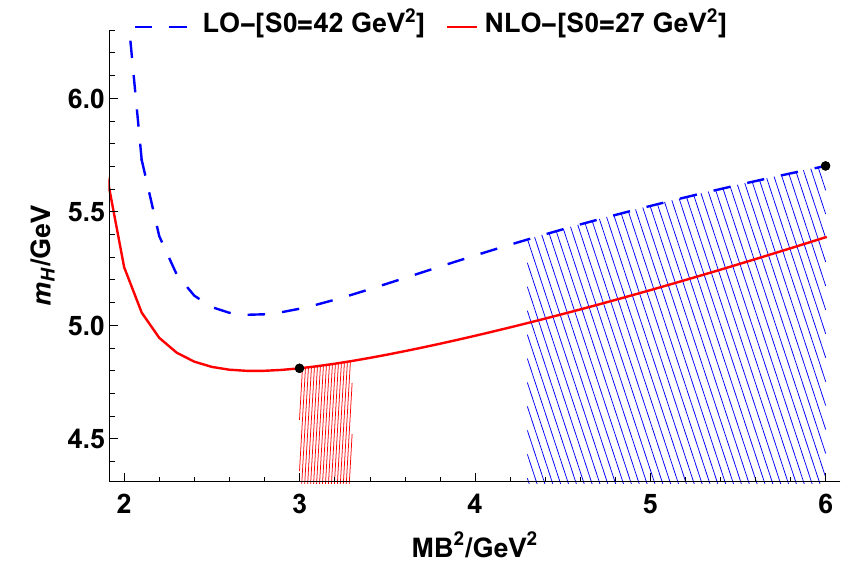}
		\includegraphics[scale=0.4]{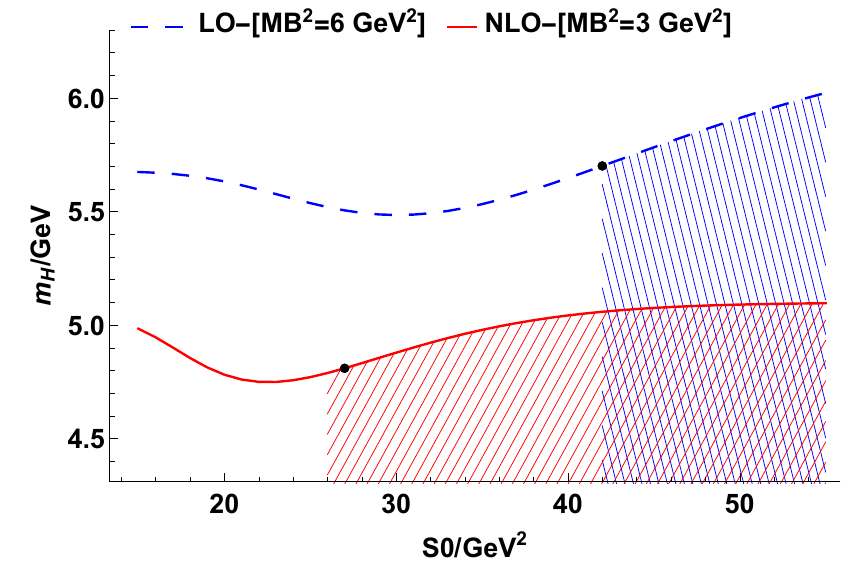}
	}
	\caption{\label{fig:Zc-[1+]-Di-G-6-NLO-MSbar-OS}LO and NLO Result of $J_{6}^{\text{Di-Di}}$ of $Z_c$ system with $\overline{\text{MS}}$ and OS renormalization schemes.}
\end{figure}

\begin{figure}[H]
	\centering
	\subfigure[$\overline{\text{MS}}$]{
		\includegraphics[scale=0.4]{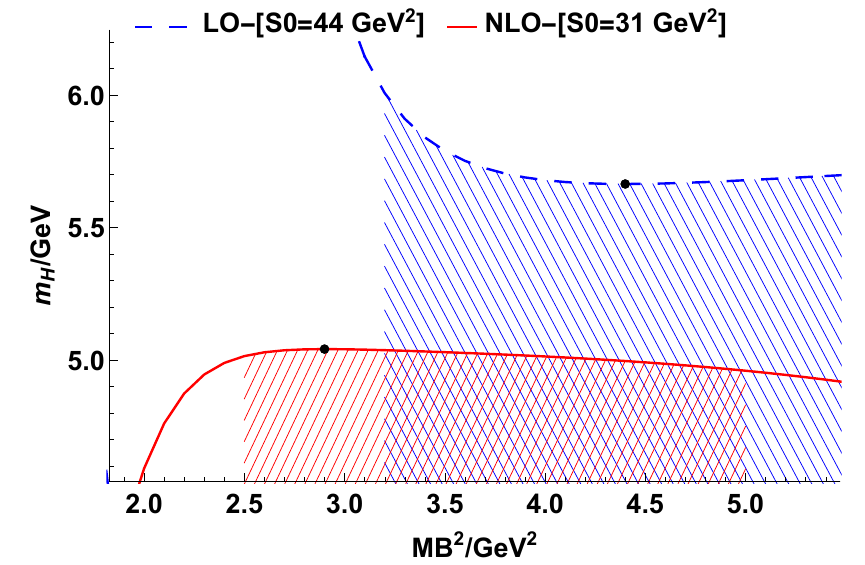}
		\includegraphics[scale=0.4]{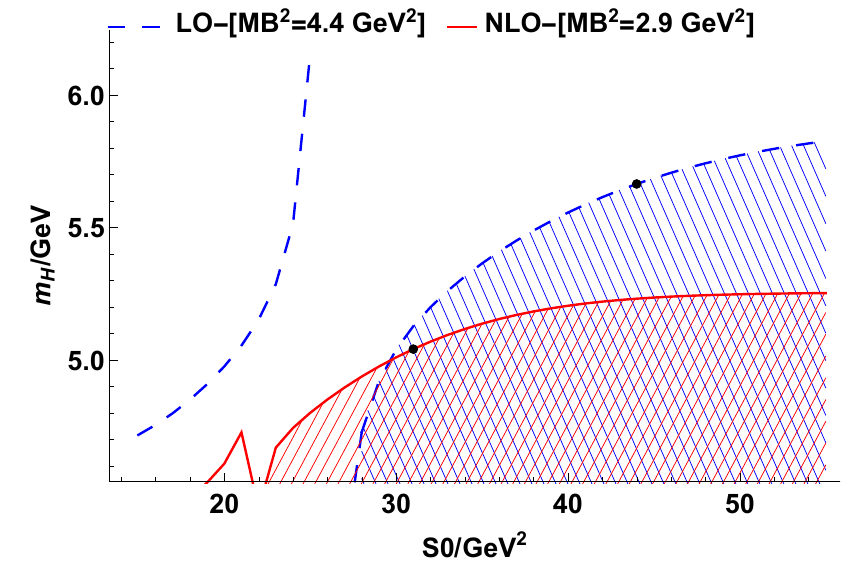}
	}\\
	\subfigure[OS]{
		\includegraphics[scale=0.4]{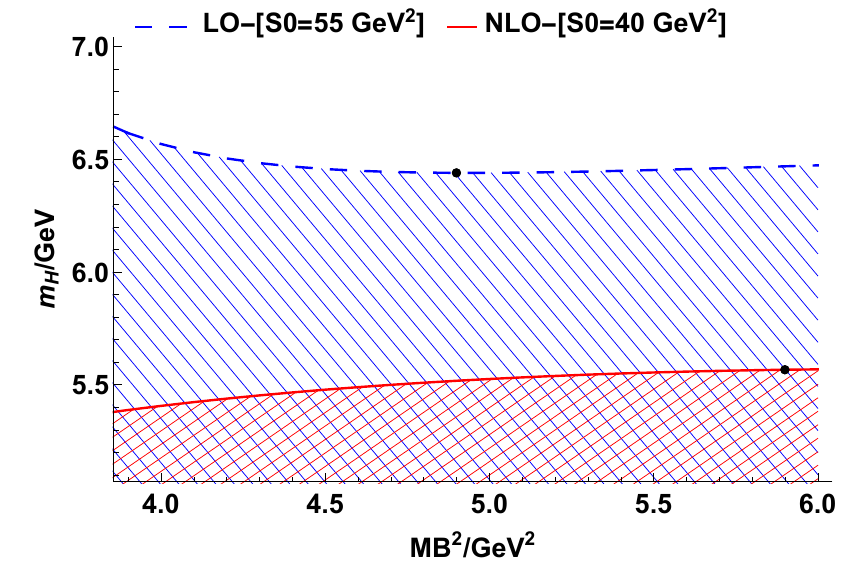}
		\includegraphics[scale=0.4]{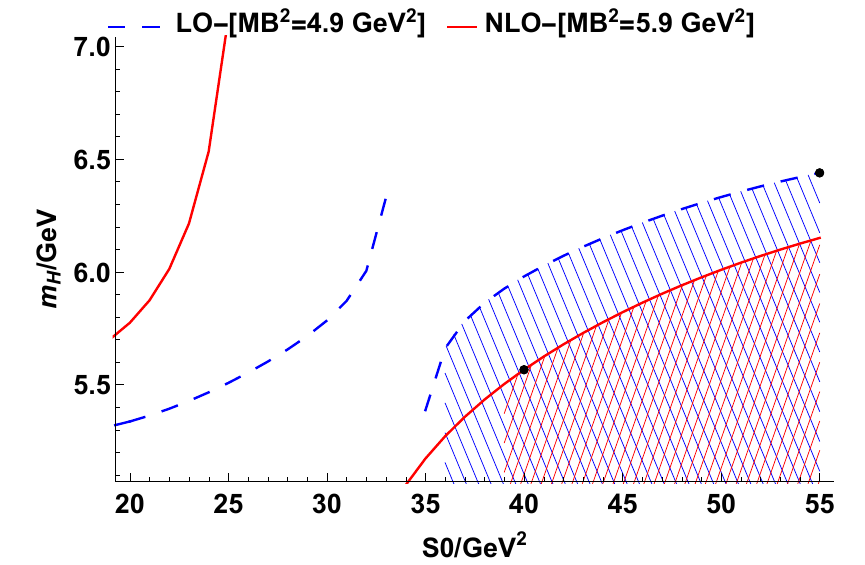}
	}
	\caption{\label{fig:Zc-[1+]-Di-G-7-NLO-MSbar-OS}LO and NLO Result of $J_{7}^{\text{Di-Di}}$ of $Z_c$ system with $\overline{\text{MS}}$ and OS renormalization schemes.}
\end{figure}

\begin{figure}[H]
	\centering
	\subfigure[$\overline{\text{MS}}$]{
		\includegraphics[scale=0.4]{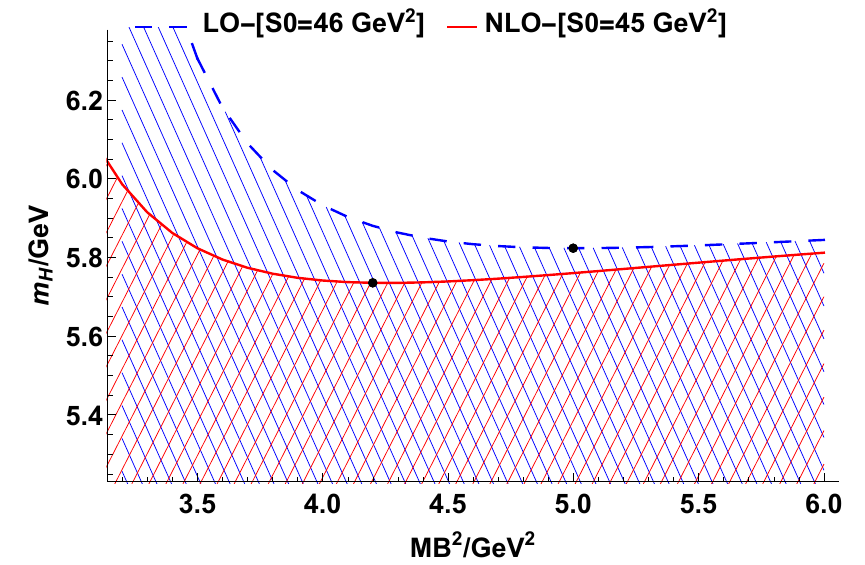}
		\includegraphics[scale=0.4]{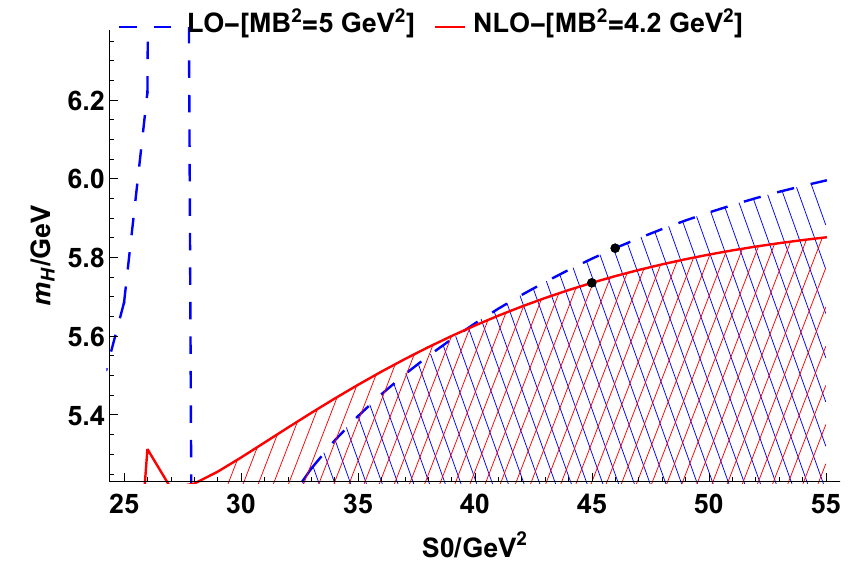}
	}\\
	\subfigure[OS]{
		\includegraphics[scale=0.4]{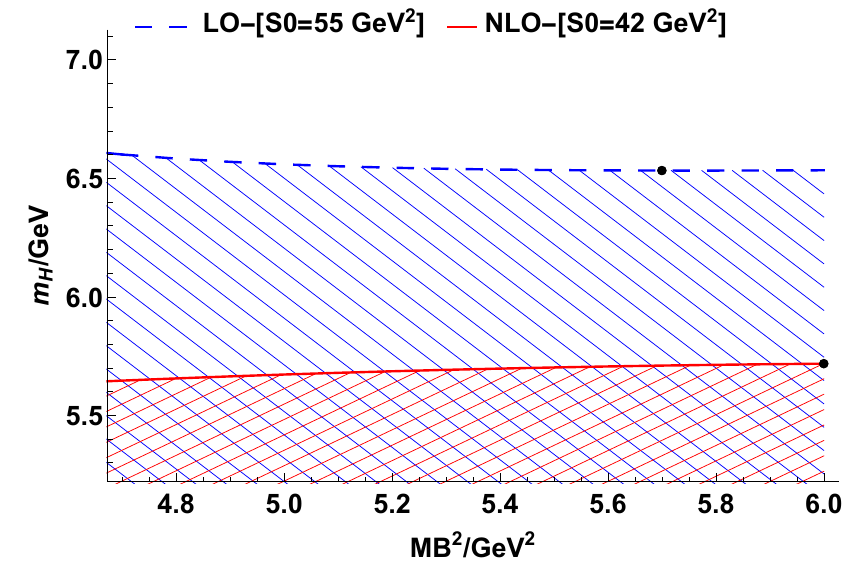}
		\includegraphics[scale=0.4]{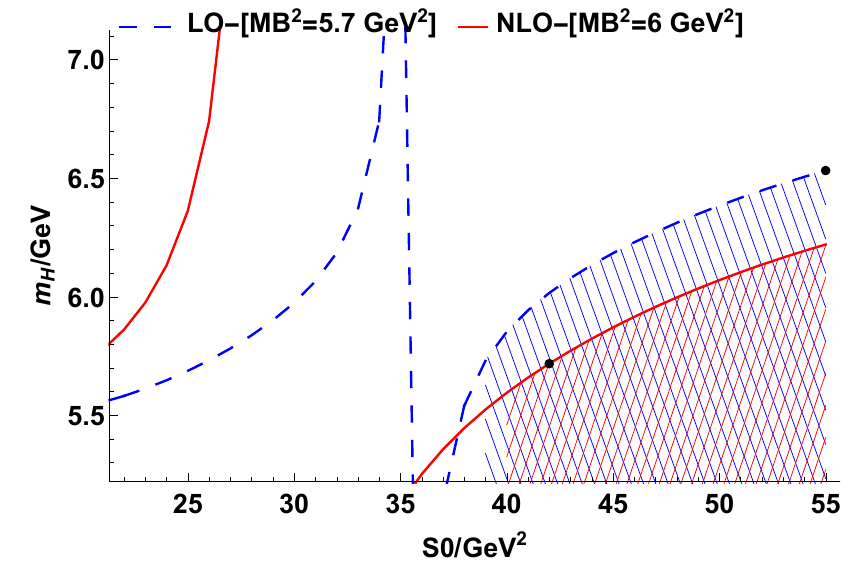}
	}
	\caption{\label{fig:Zc-[1+]-Di-G-8-NLO-MSbar-OS}LO and NLO Result of $J_{8}^{\text{Di-Di}}$ of $Z_c$ system with $\overline{\text{MS}}$ and OS renormalization schemes.}
\end{figure}
\newpage

\subsection{Numerical results for mixed operators of $Z_c$ system}

\begin{table}[H]
	\renewcommand\arraystretch{2}
	\begin{center}
		\setlength{\tabcolsep}{4 mm}
		\begin{tabular}{|c|c|c|c|c|c|c|c|c|c|}
			\hline
			\multirow{3}{*}{Current} &
			\multicolumn{3}{c|}{LO}& \multicolumn{3}{c|}{NLO($\overline{\text{MS}}$)} \\ \cline{2-7} &
			\makecell{$M_H$ \\ (GeV)} & \makecell{$s_0$ \\ ($\text{GeV}^2$)} & \makecell{$M_B^2$ \\ ($\text{GeV}^2$)} &
			\makecell{$M_H$ \\ (GeV)} & \makecell{$s_0$ \\ ($\text{GeV}^2$)} & \makecell{$M_B^2$ \\ ($\text{GeV}^2$)} \\ \hline
			
    $J_{1,5}^\text{{Mixed}}$ &$3.94^{+0.02}_{-0.02}$ &$24.(\pm 10\%)$ &$1.30(\pm 10\%)$    &$3.89^{+0.02}_{-0.02}$ &$22.(\pm 10\%)$ &$1.10(\pm 10\%)$\\ \hline
$J_{2,6}^\text{{Mixed}}$ &$4.01^{+0.06}_{-0.05}$ &$22.(\pm 10\%)$ &$1.50(\pm 10\%)$    &$4.03^{+0.02}_{-0.02}$ &$25.(\pm 10\%)$ &$1.20(\pm 10\%)$\\ \hline
$J_{3,7}^\text{{Mixed}}$ &$4.32^{+0.06}_{-0.08}$ &$26.(\pm 10\%)$ &$1.80(\pm 10\%)$    &$4.45^{+0.01}_{-0.02}$ &$28.(\pm 10\%)$ &$1.40(\pm 10\%)$\\ \hline
$J_{4,8}^\text{{Mixed}}$ &$4.32^{+0.06}_{-0.06}$ &$27.(\pm 10\%)$ &$1.80(\pm 10\%)$    &$4.83^{+0.03}_{-0.02}$ &$34.(\pm 10\%)$ &$1.90(\pm 10\%)$\\ \hline
		\end{tabular}

		\caption{LO and NLO results of mixed operators of $Z_c$ system with $\overline{\text{MS}}$ renormalization scheme. The errors of masses shown in this table just come from the parametric dependence on $s_0$ and $M_B^2$.}
		\label{tab:Zc-Mixed-MSbar}
	\end{center}
\end{table}

\begin{table}[H]
	\renewcommand\arraystretch{2}
	\begin{center}
		\setlength{\tabcolsep}{4 mm}
		\begin{tabular}{|c|c|c|c|c|c|c|c|c|c|}
			\hline
			\multirow{3}{*}{Current} &
			\multicolumn{3}{c|}{LO}& \multicolumn{3}{c|}{NLO($\text{OS}$)} \\ \cline{2-7} &
			\makecell{$M_H$ \\ (GeV)} & \makecell{$s_0$ \\ ($\text{GeV}^2$)} & \makecell{$M_B^2$ \\ ($\text{GeV}^2$)} &
			\makecell{$M_H$ \\ (GeV)} & \makecell{$s_0$ \\ ($\text{GeV}^2$)} & \makecell{$M_B^2$ \\ ($\text{GeV}^2$)} \\ \hline
			
  $J_{1,5}^\text{{Mixed}}$ &$4.18^{+0.04}_{-0.04}$ &$23.(\pm 10\%)$ &$1.30(\pm 10\%)$    &$3.84^{+0.07}_{-0.09}$ &$17.(\pm 10\%)$ &$1.00(\pm 10\%)$\\ \hline
$J_{2,6}^\text{{Mixed}}$ &$4.34^{+0.02}_{-0.02}$ &$28.(\pm 10\%)$ &$1.40(\pm 10\%)$    &$4.00^{+0.02}_{-0.02}$ &$22.(\pm 10\%)$ &$1.00(\pm 10\%)$\\ \hline
$J_{3,7}^\text{{Mixed}}$ &$4.69^{+0.06}_{-0.03}$ &$31.(\pm 10\%)$ &$1.70(\pm 10\%)$    &$3.94^{+0.13}_{-0.20}$ &$18.(\pm 10\%)$ &$1.30(\pm 10\%)$\\ \hline
$J_{4,8}^\text{{Mixed}}$ &$4.61^{+0.07}_{-0.07}$ &$28.(\pm 10\%)$ &$1.80(\pm 10\%)$    &$4.22^{+0.02}_{-0.02}$ &$27.(\pm 10\%)$ &$1.20(\pm 10\%)$\\ \hline
		\end{tabular}

		\caption{LO and NLO results of mixed operators of $Z_c$ system with $\text{OS}$ renormalization scheme. The errors of masses shown in this table just come from the parametric dependence on $s_0$ and $M_B^2$.}
		\label{tab:Zc-Mixed-OS}
	\end{center}
\end{table}

\begin{figure}[H]
	\centering
	\subfigure[$\overline{\text{MS}}$]{
		\includegraphics[scale=0.4]{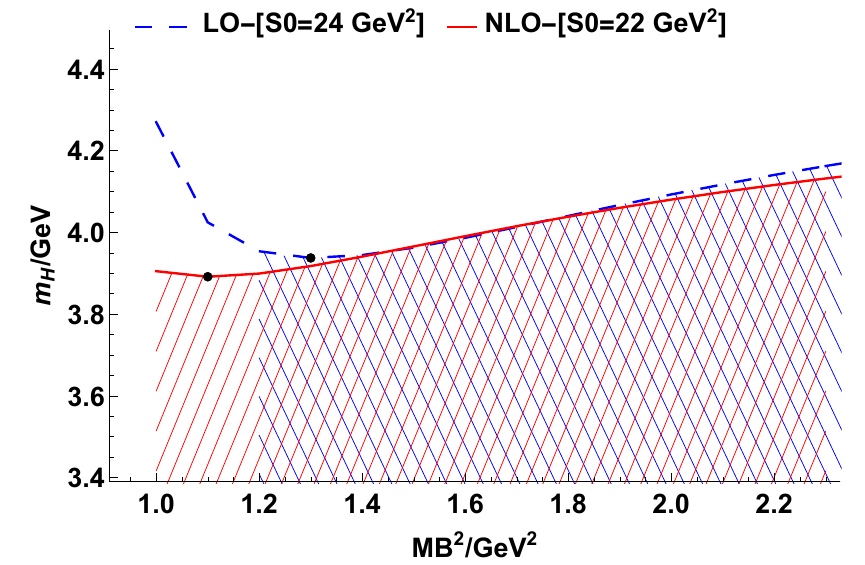}
		\includegraphics[scale=0.4]{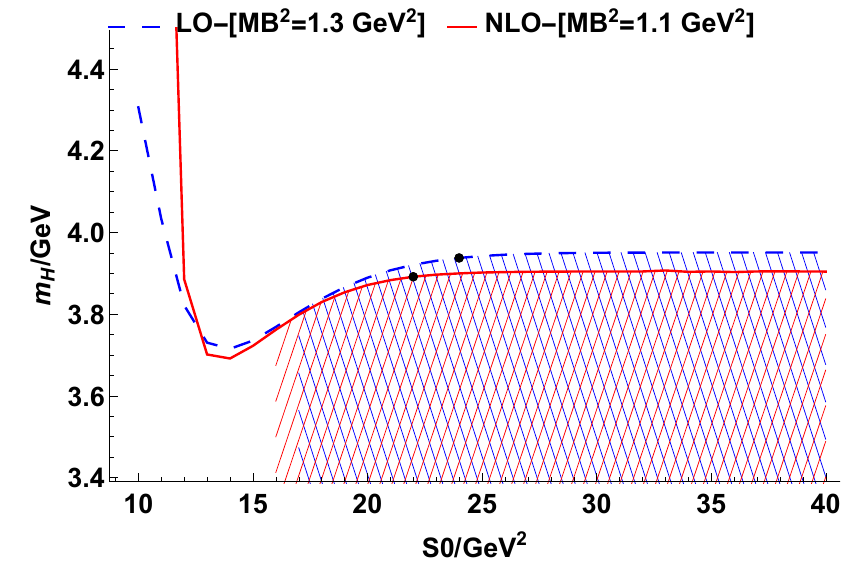}
	}\\
	\subfigure[OS]{
		\includegraphics[scale=0.4]{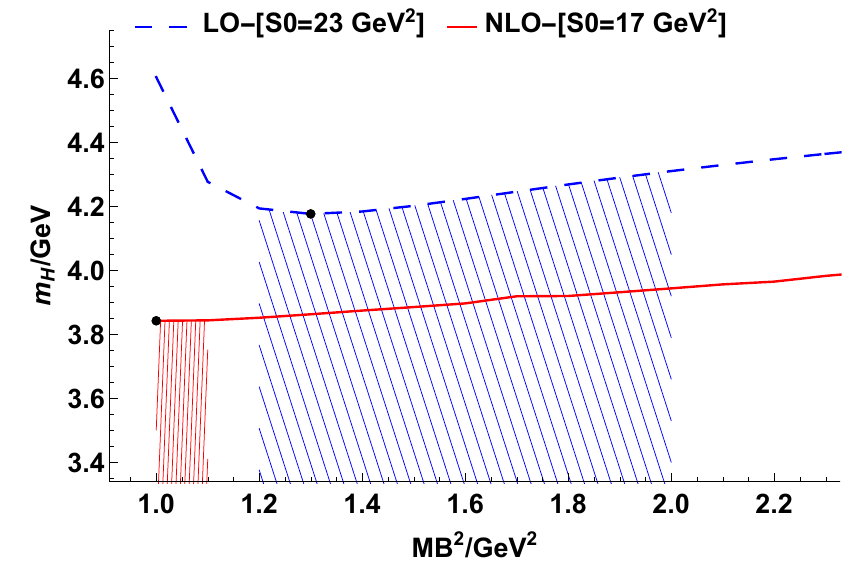}
		\includegraphics[scale=0.4]{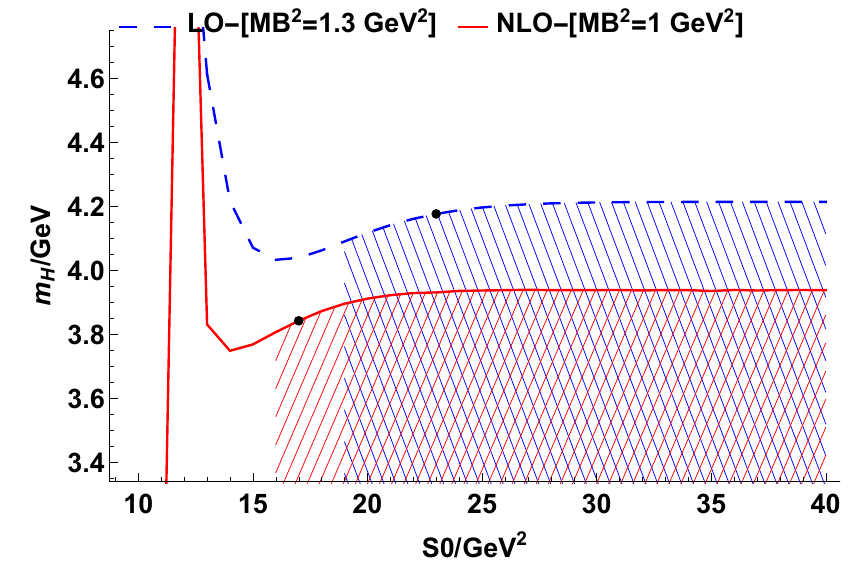}
	}
	\caption{\label{fig:Zc-[1+]-M-Mixed-1-NLO-MSbar-OS}LO and NLO Result of $J_{1,5}^\text{Mixed}$ of $Z_c$ system with $\overline{\text{MS}}$ and OS renormalization schemes.}
\end{figure}

\begin{figure}[H]
	\centering
	\subfigure[$\overline{\text{MS}}$]{
		\includegraphics[scale=0.4]{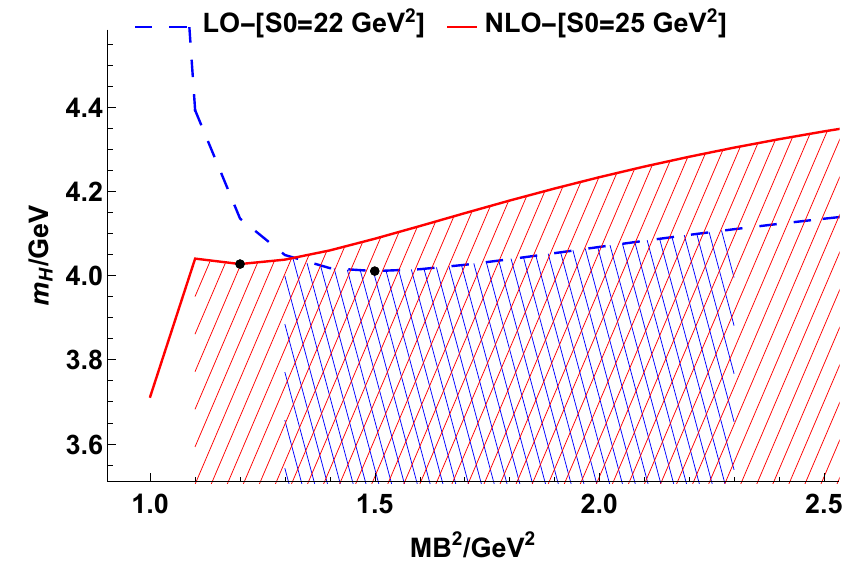}
		\includegraphics[scale=0.4]{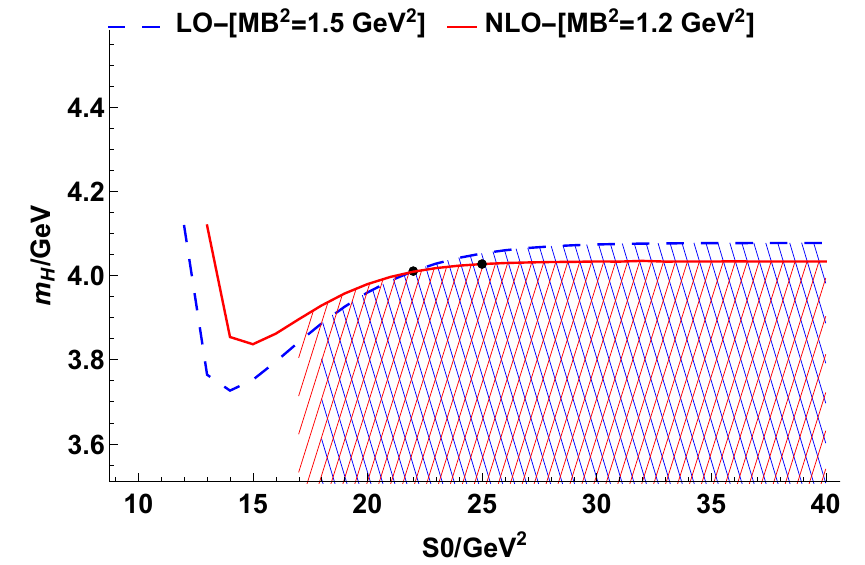}
	}\\
	\subfigure[OS]{
		\includegraphics[scale=0.4]{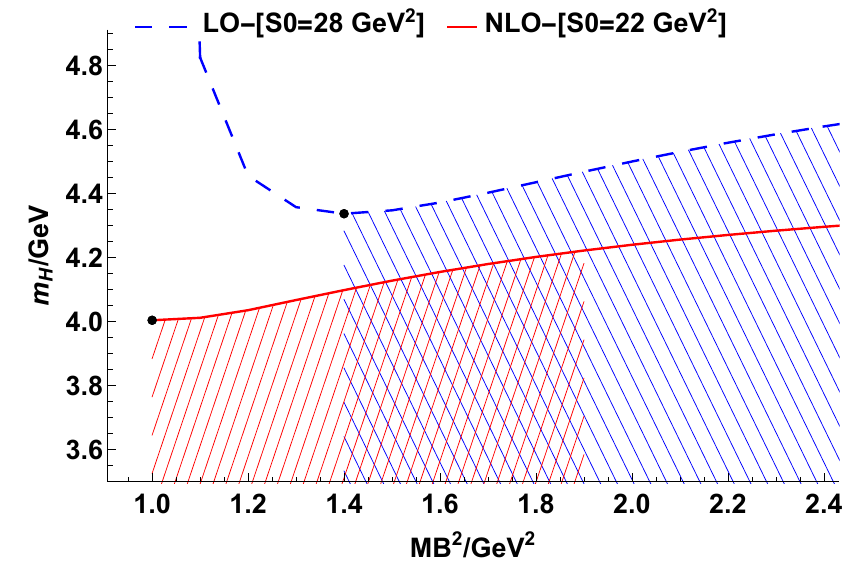}
		\includegraphics[scale=0.4]{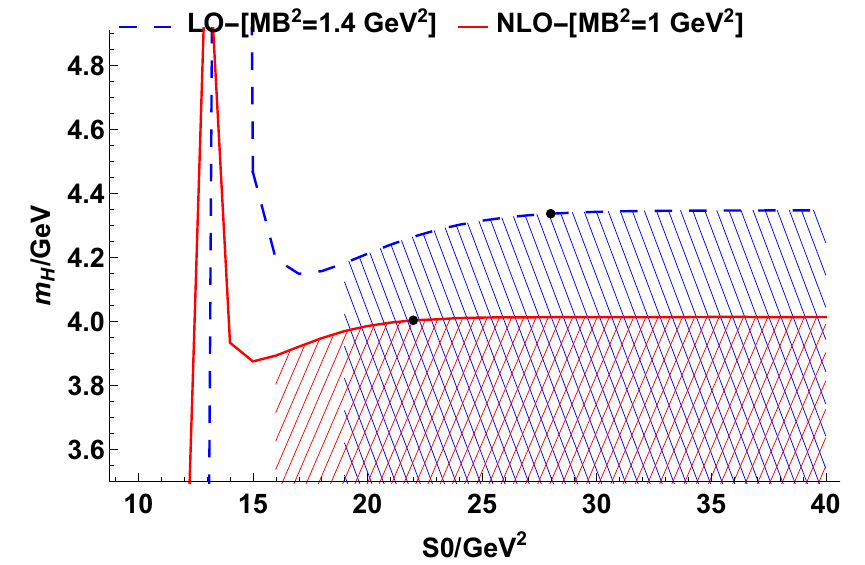}
	}
	\caption{\label{fig:Zc-[1+]-M-Mixed-2-NLO-MSbar-OS}LO and NLO Result of $J_{2,6}^\text{Mixed}$ of $Z_c$ system with $\overline{\text{MS}}$ and OS renormalization schemes.}
\end{figure}

\begin{figure}[H]
	\centering
	\subfigure[$\overline{\text{MS}}$]{
		\includegraphics[scale=0.4]{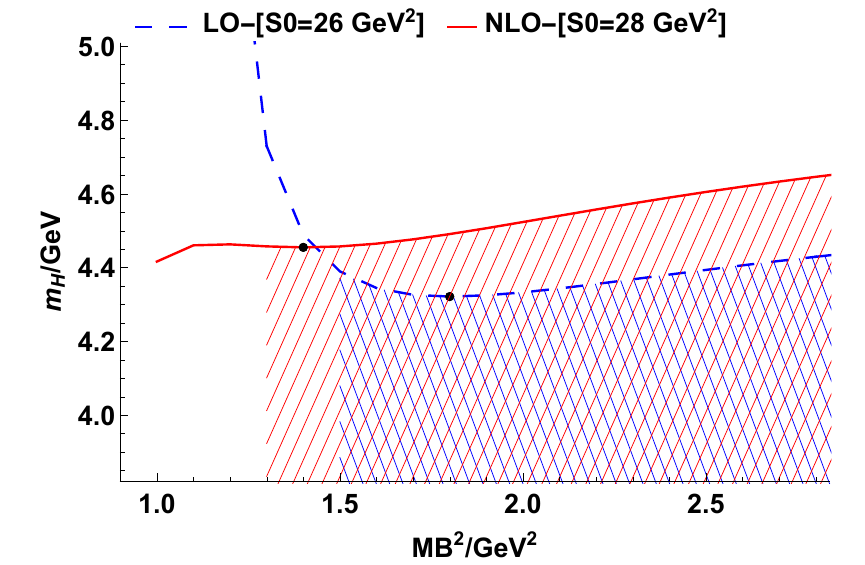}
		\includegraphics[scale=0.4]{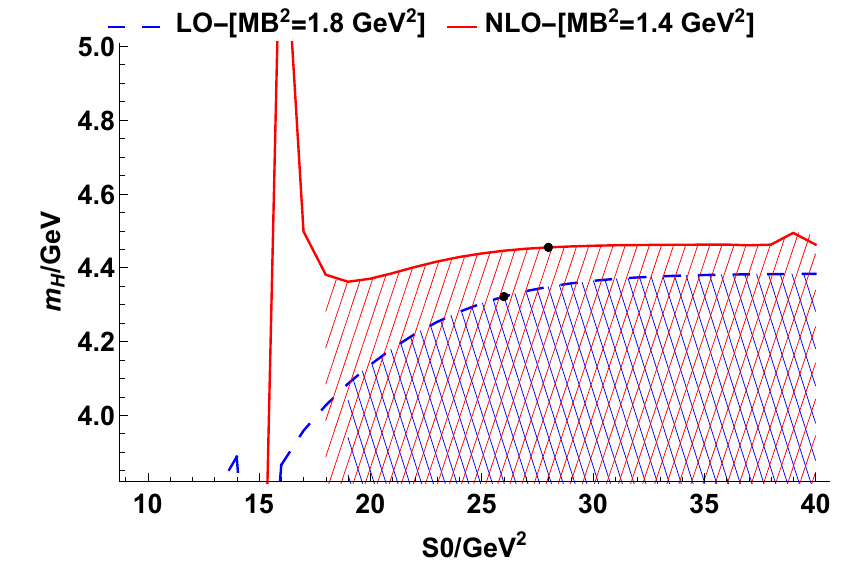}
	}\\
	\subfigure[OS]{
		\includegraphics[scale=0.4]{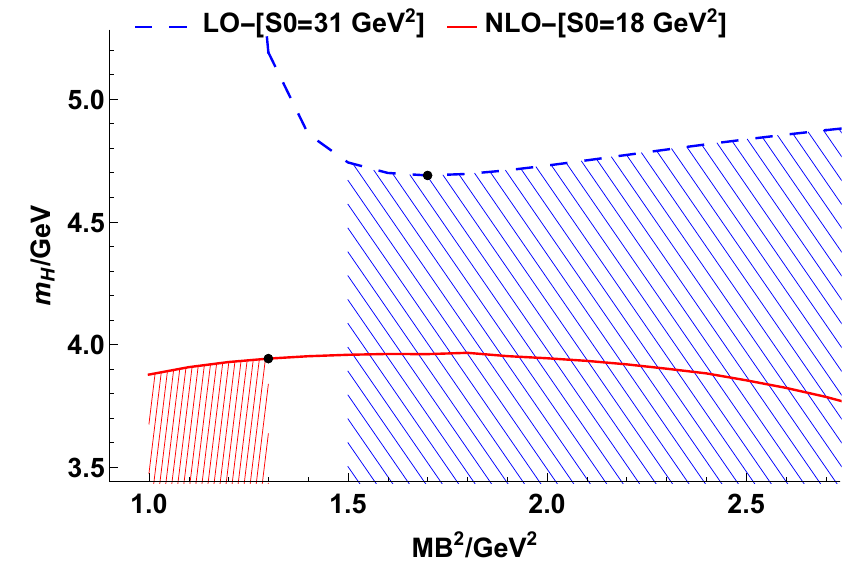}
		\includegraphics[scale=0.4]{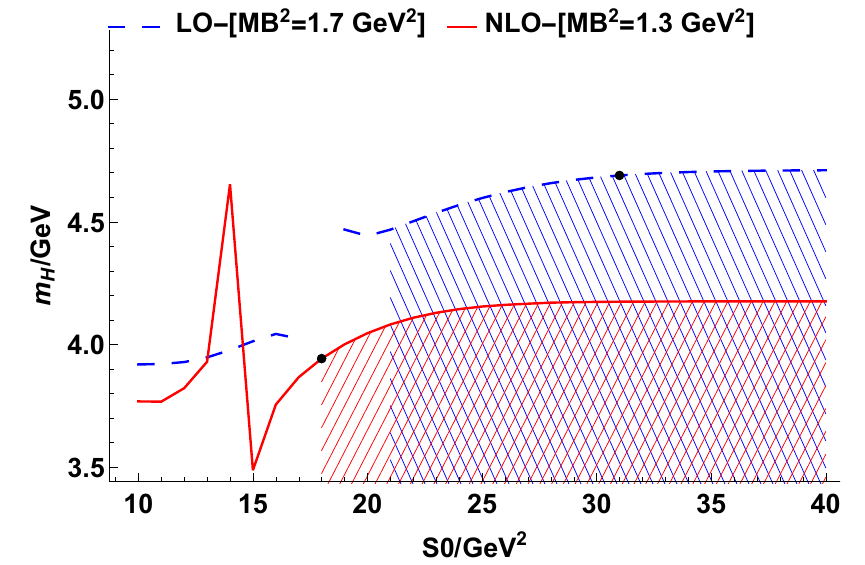}
	}
	\caption{\label{fig:Zc-[1+]-M-Mixed-3-NLO-MSbar-OS}LO and NLO Result of $J_{3,7}^\text{{Mixed}}$ of $Z_c$ system with $\overline{\text{MS}}$ and OS renormalization schemes.}
\end{figure}

\begin{figure}[H]
	\centering
	\subfigure[$\overline{\text{MS}}$]{
		\includegraphics[scale=0.4]{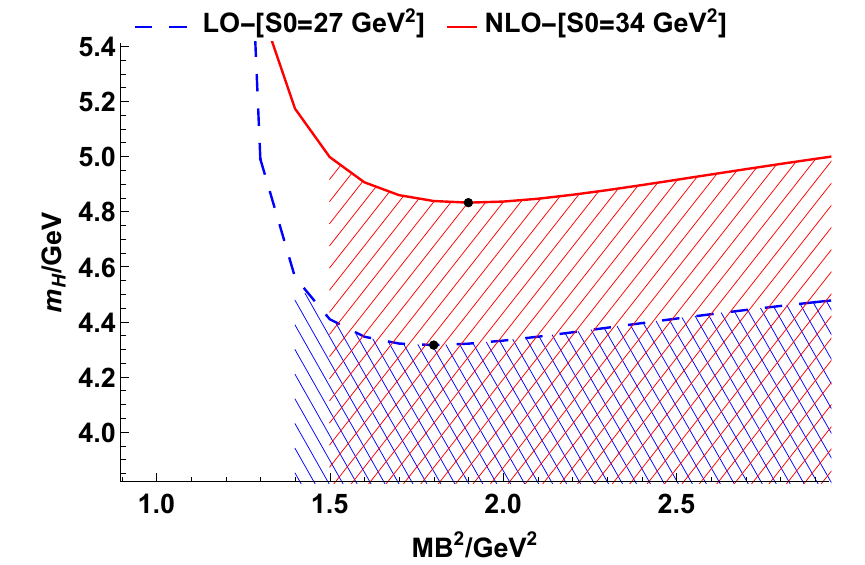}
		\includegraphics[scale=0.4]{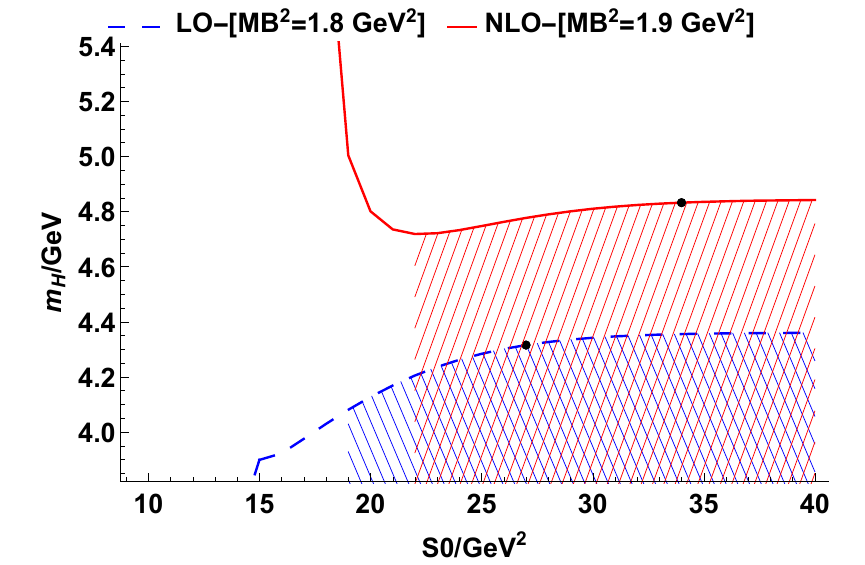}
	}\\
	\subfigure[OS]{
		\includegraphics[scale=0.4]{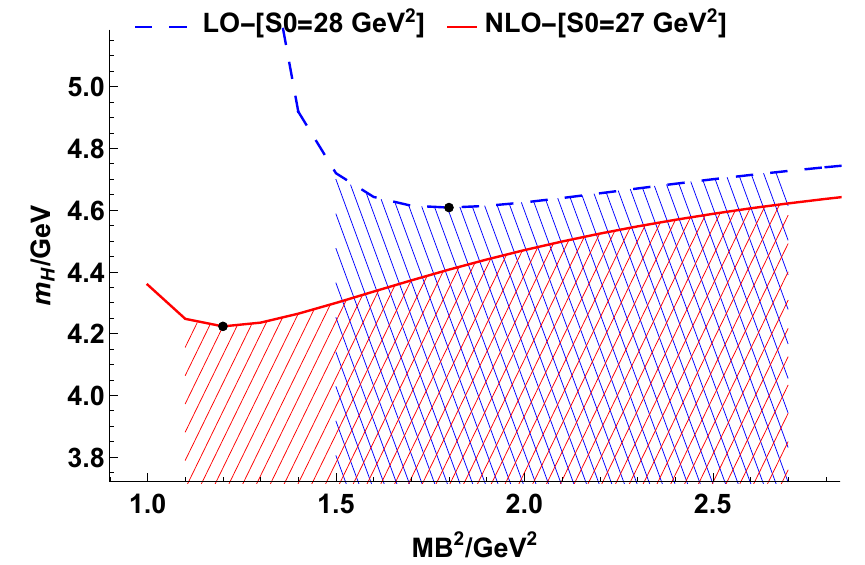}
		\includegraphics[scale=0.4]{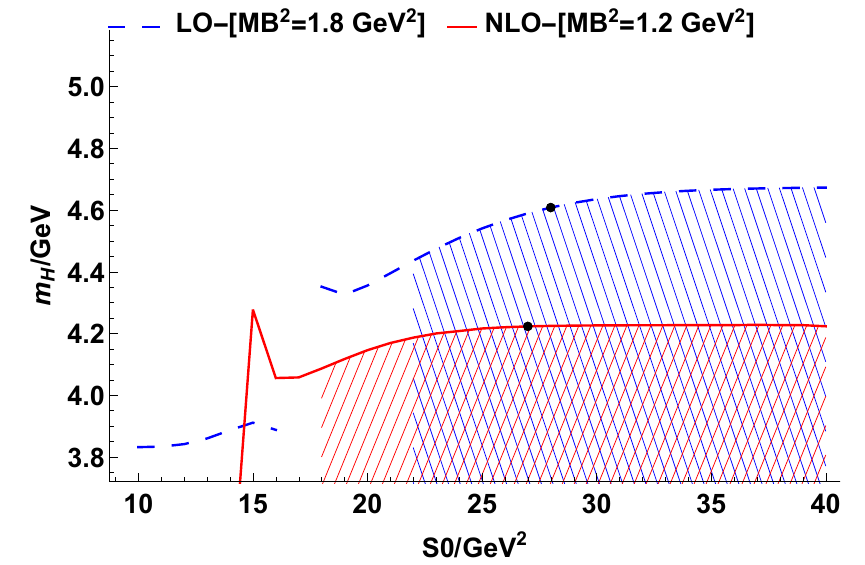}
	}
	\caption{\label{fig:Zc-[1+]-M-Mixed-4-NLO-MSbar-OS}LO and NLO Result of $J_{4,8}^\text{{Mixed}}$ of $Z_c$ system with $\overline{\text{MS}}$ and OS renormalization schemes.}
\end{figure}

\subsection{Numerical results for mixed operators of $Z_{cs}$ system}

\begin{table}[H]
	\renewcommand\arraystretch{2}
	\begin{center}
		\setlength{\tabcolsep}{4 mm}
		\begin{tabular}{|c|c|c|c|c|c|c|c|c|c|}
			\hline
			\multirow{3}{*}{Current} &
			\multicolumn{3}{c|}{LO}& \multicolumn{3}{c|}{NLO($\overline{\text{MS}}$)} \\ \cline{2-7} &
			\makecell{$M_H$ \\ (GeV)} & \makecell{$s_0$ \\ ($\text{GeV}^2$)} & \makecell{$M_B^2$ \\ ($\text{GeV}^2$)} &
			\makecell{$M_H$ \\ (GeV)} & \makecell{$s_0$ \\ ($\text{GeV}^2$)} & \makecell{$M_B^2$ \\ ($\text{GeV}^2$)} \\ \hline
			
			 $J_{1,5}^\text{Mixed}$ &$4.14^{+0.06}_{-0.05}$ &$24.(\pm 10\%)$ &$1.70(\pm 10\%)$    &$4.02^{+0.05}_{-0.06}$ &$21.(\pm 10\%)$ &$1.40(\pm 10\%)$\\ \hline
			$J_{2,6}^\text{Mixed}$ &$4.24^{+0.05}_{-0.08}$ &$25.(\pm 10\%)$ &$1.80(\pm 10\%)$    &$4.21^{+0.05}_{-0.04}$ &$24.(\pm 10\%)$ &$1.50(\pm 10\%)$\\ \hline
			$J_{3,7}^\text{Mixed}$ &$4.53^{+0.06}_{-0.07}$ &$29.(\pm 10\%)$ &$2.20(\pm 10\%)$    &$4.71^{+0.05}_{-0.05}$ &$31.(\pm 10\%)$ &$2.10(\pm 10\%)$\\ \hline
			$J_{4,8}^\text{Mixed}$ &$4.49^{+0.06}_{-0.06}$ &$29.(\pm 10\%)$ &$2.10(\pm 10\%)$    &$5.07^{+0.04}_{-0.06}$ &$35.(\pm 10\%)$ &$2.50(\pm 10\%)$\\ \hline
		\end{tabular}

		\caption{LO and NLO results of $Z_{cs}$ system with $\overline{\text{MS}}$ renormalization scheme. The errors of masses shown in this table just come from the parametric dependence on $s_0$ and $M_B^2$.}
		\label{tab:Zcs-Mixed-MSbar}
	\end{center}
\end{table}

\begin{table}[H]
	\renewcommand\arraystretch{2}
	\begin{center}
		\setlength{\tabcolsep}{4 mm}
		\begin{tabular}{|c|c|c|c|c|c|c|c|c|c|}
			\hline
			\multirow{3}{*}{Current} &
			\multicolumn{3}{c|}{LO}& \multicolumn{3}{c|}{NLO($\text{OS}$)} \\ \cline{2-7} &
			\makecell{$M_H$ \\ (GeV)} & \makecell{$s_0$ \\ ($\text{GeV}^2$)} & \makecell{$M_B^2$ \\ ($\text{GeV}^2$)} &
			\makecell{$M_H$ \\ (GeV)} & \makecell{$s_0$ \\ ($\text{GeV}^2$)} & \makecell{$M_B^2$ \\ ($\text{GeV}^2$)} \\ \hline
			
		 $J_{1,5}^\text{Mixed}$ &$4.53^{+0.06}_{-0.05}$ &$28.(\pm 10\%)$ &$1.70(\pm 10\%)$    &$4.09^{+0.03}_{-0.02}$ &$23.(\pm 10\%)$ &$1.10(\pm 10\%)$\\ \hline
		$J_{2,6}^\text{Mixed}$ &$4.62^{+0.06}_{-0.06}$ &$29.(\pm 10\%)$ &$1.80(\pm 10\%)$    &$4.14^{+0.04}_{-0.04}$ &$22.(\pm 10\%)$ &$1.20(\pm 10\%)$\\ \hline
		$J_{3,7}^\text{Mixed}$ &$5.01^{+0.05}_{-0.05}$ &$36.(\pm 10\%)$ &$2.20(\pm 10\%)$    &$4.32^{+0.05}_{-0.06}$ &$23.(\pm 10\%)$ &$1.40(\pm 10\%)$\\ \hline
		$J_{4,8}^\text{Mixed}$ &$4.90^{+0.05}_{-0.04}$ &$34.(\pm 10\%)$ &$2.10(\pm 10\%)$    &$4.41^{+0.02}_{-0.02}$ &$29.(\pm 10\%)$ &$1.40(\pm 10\%)$\\ \hline
		\end{tabular}

		\caption{LO and NLO results of $Z_{cs}$ system with $\text{OS}$ renormalization scheme. The errors of masses shown in this table just come from the parametric dependence on $s_0$ and $M_B^2$.}
		\label{tab:Zcs-Mixed-OS}
	\end{center}
\end{table}

\begin{figure}[H]
	\centering
	\subfigure[$\overline{\text{MS}}$]{
		\includegraphics[scale=0.4]{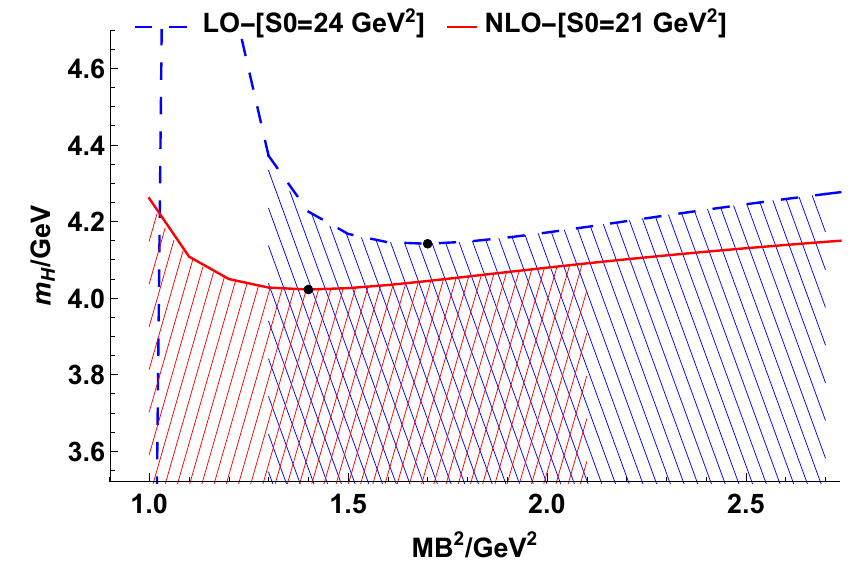}
		\includegraphics[scale=0.4]{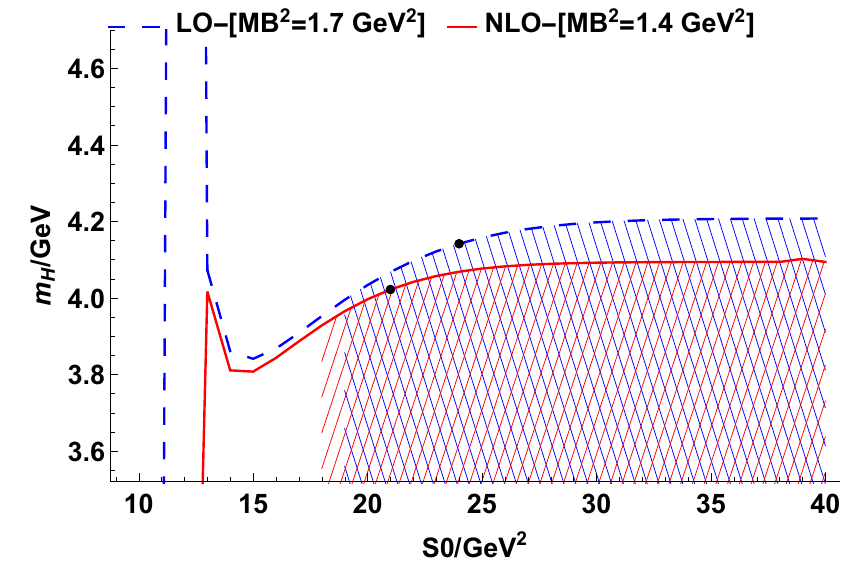}
	}\\
	\subfigure[OS]{
		\includegraphics[scale=0.4]{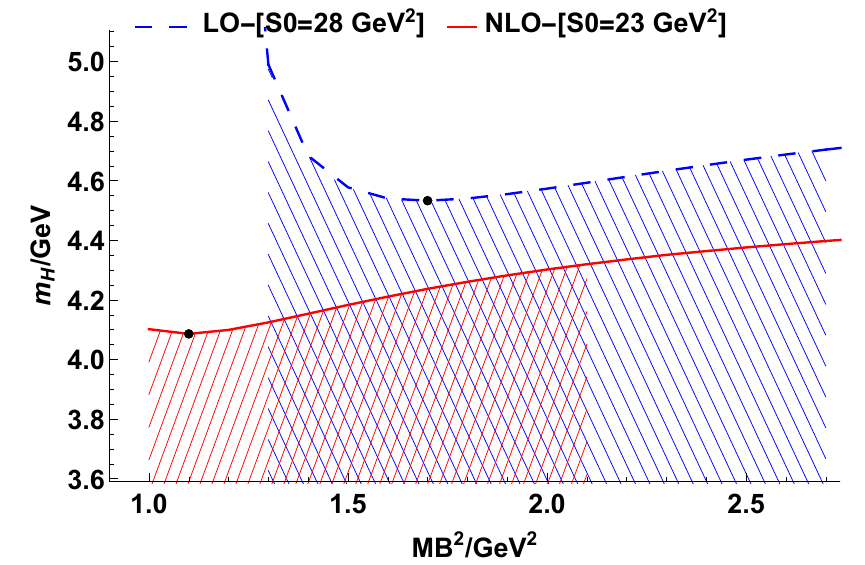}
		\includegraphics[scale=0.4]{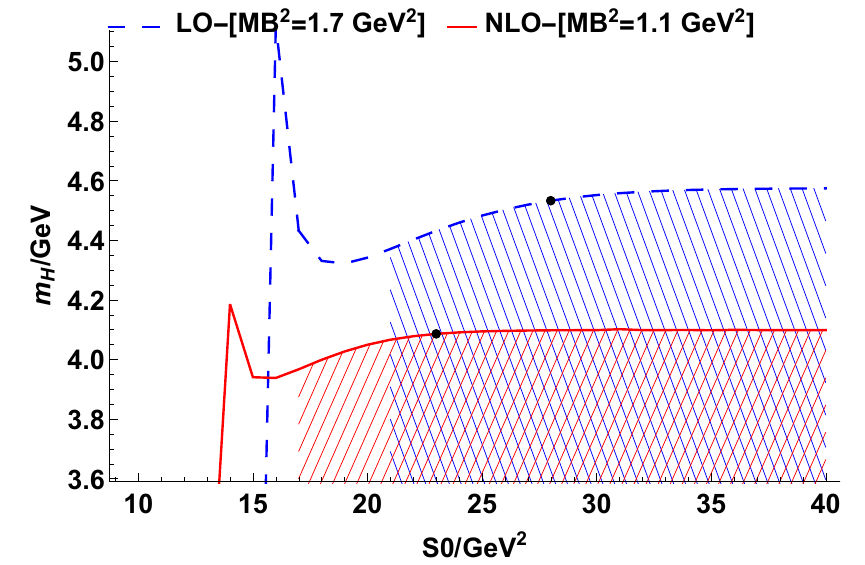}
	}
	\caption{\label{fig:Zcs-[1+]-M-Mixed-1-NLO-MSbar-OS}LO and NLO Result of $J_{1,5}^\text{Mixed}$ with $\overline{\text{MS}}$ and OS renormalization schemes of $Z_{cs}$ system. }
\end{figure}

\begin{figure}[H]
	\centering
	\subfigure[$\overline{\text{MS}}$]{
		\includegraphics[scale=0.4]{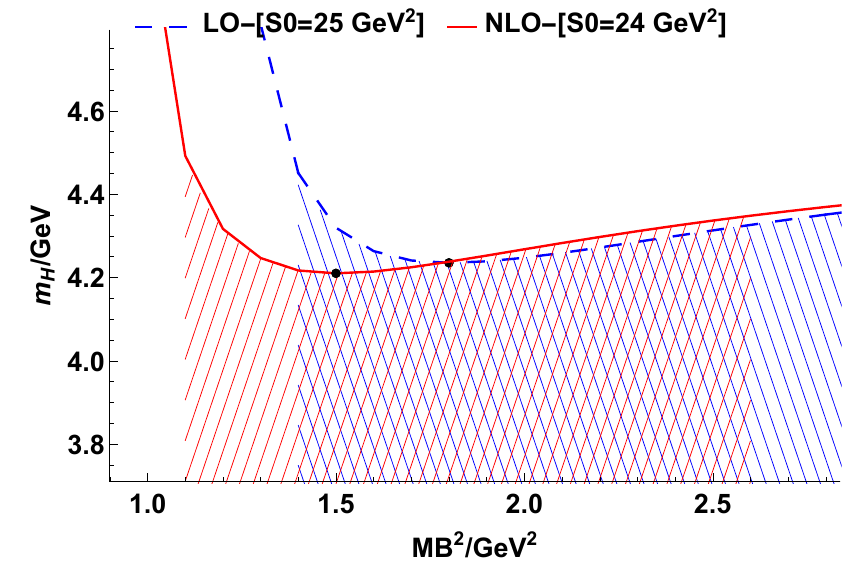}
		\includegraphics[scale=0.4]{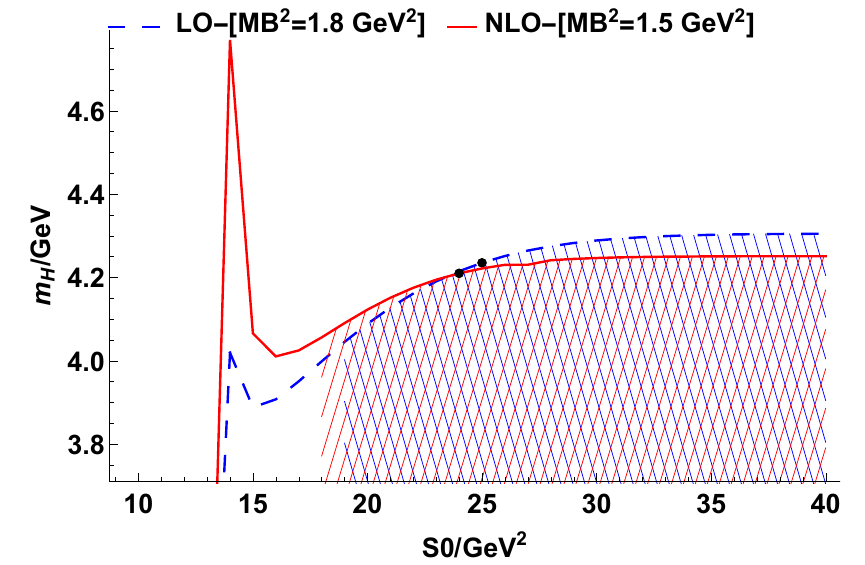}
	}\\
	\subfigure[OS]{
		\includegraphics[scale=0.4]{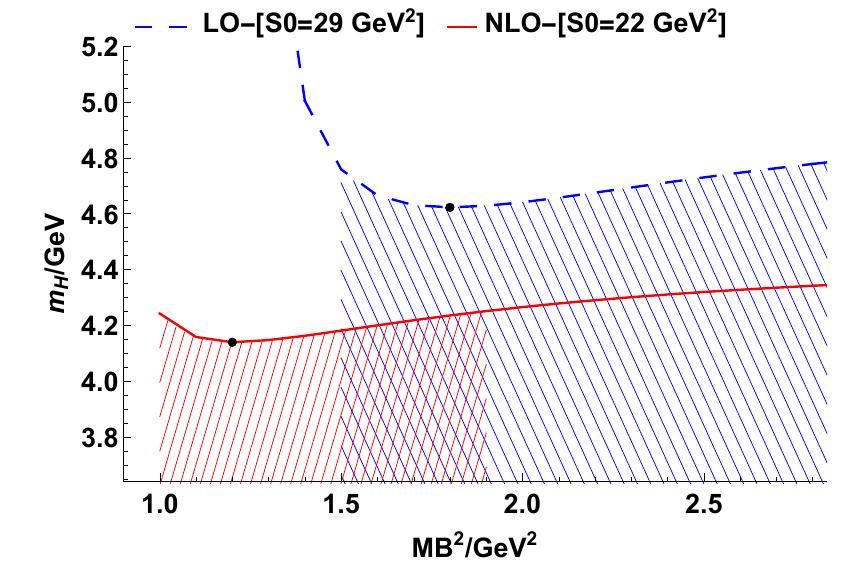}
		\includegraphics[scale=0.4]{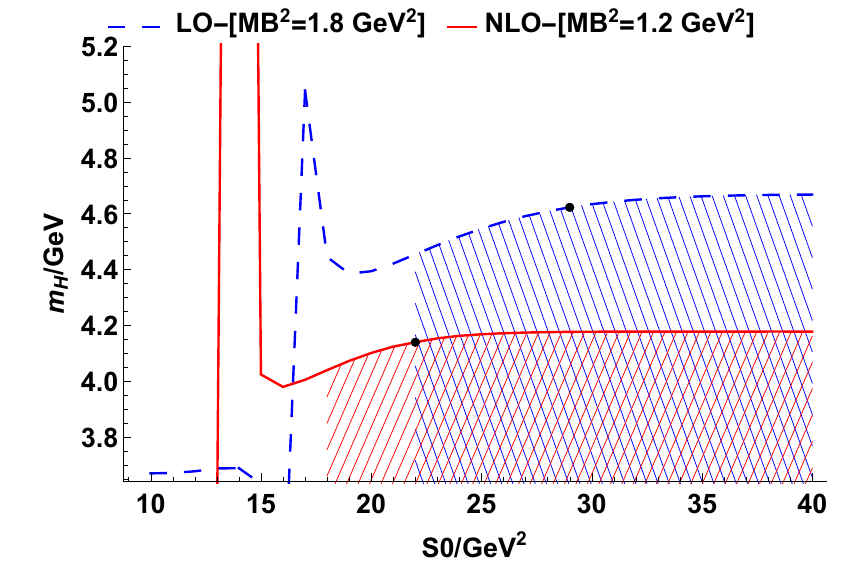}
	}
	\caption{\label{fig:Zcs-[1+]-M-Mixed-2-NLO-MSbar-OS}LO and NLO Result of $J_{2,6}^\text{Mixed}$ with $\overline{\text{MS}}$ and OS renormalization schemes of $Z_{cs}$ system. }
\end{figure}

\begin{figure}[H]
	\centering
	\subfigure[$\overline{\text{MS}}$]{
		\includegraphics[scale=0.4]{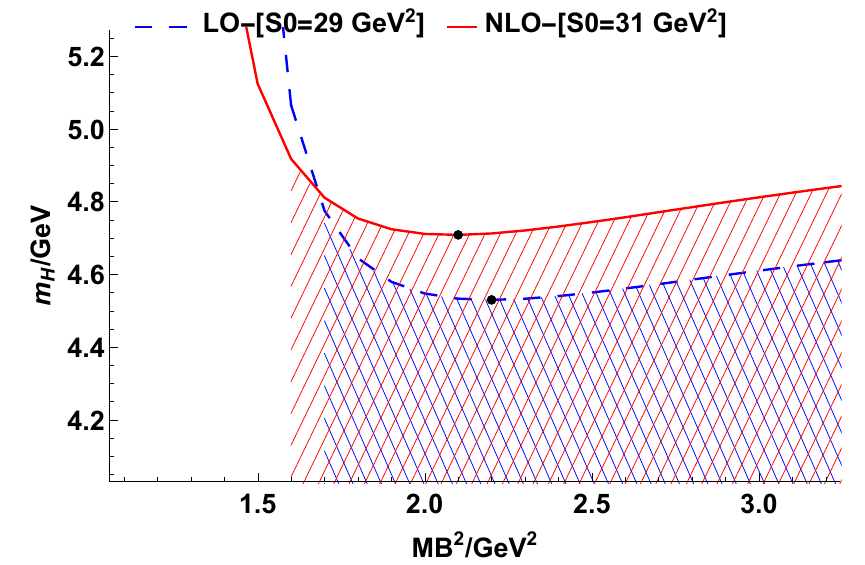}
		\includegraphics[scale=0.4]{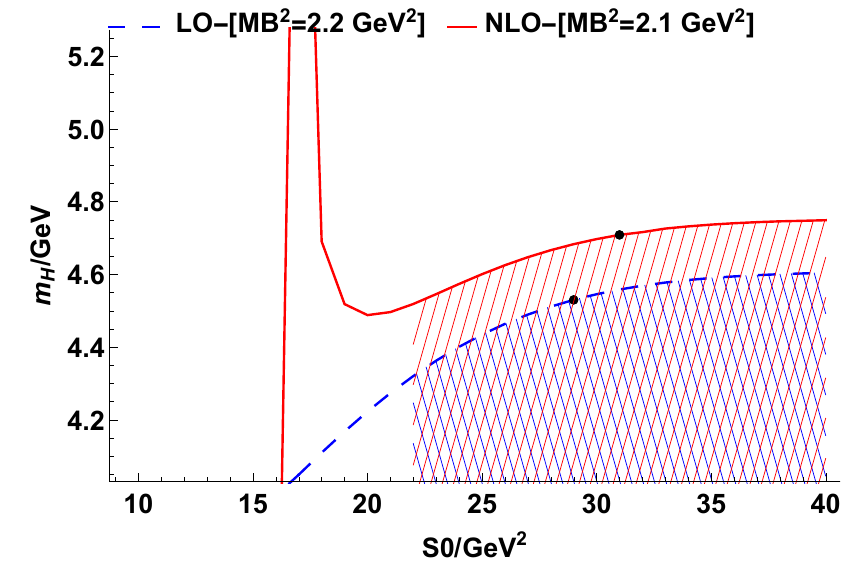}
	}\\
	\subfigure[OS]{
		\includegraphics[scale=0.4]{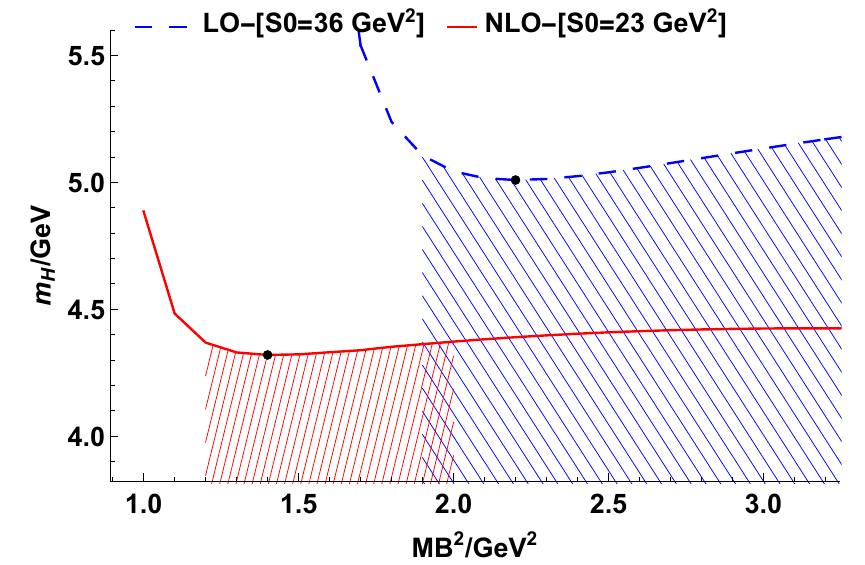}
		\includegraphics[scale=0.4]{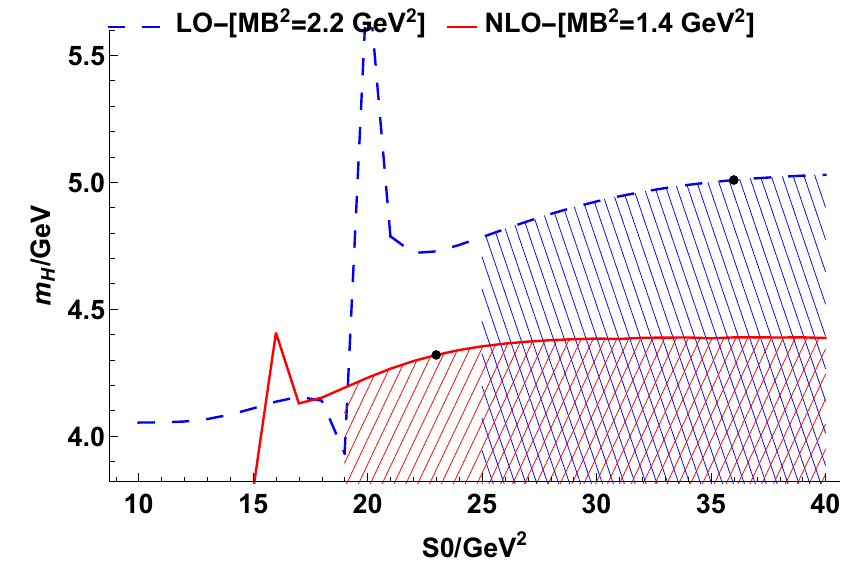}
	}
	\caption{\label{fig:Zcs-[1+]-M-Mixed-3-NLO-MSbar-OS}LO and NLO Result of $J_{3,7}^\text{Mixed}$ with $\overline{\text{MS}}$ and OS renormalization schemes of $Z_{cs}$ system. }
\end{figure}

\begin{figure}[H]
	\centering
	\subfigure[$\overline{\text{MS}}$]{
		\includegraphics[scale=0.4]{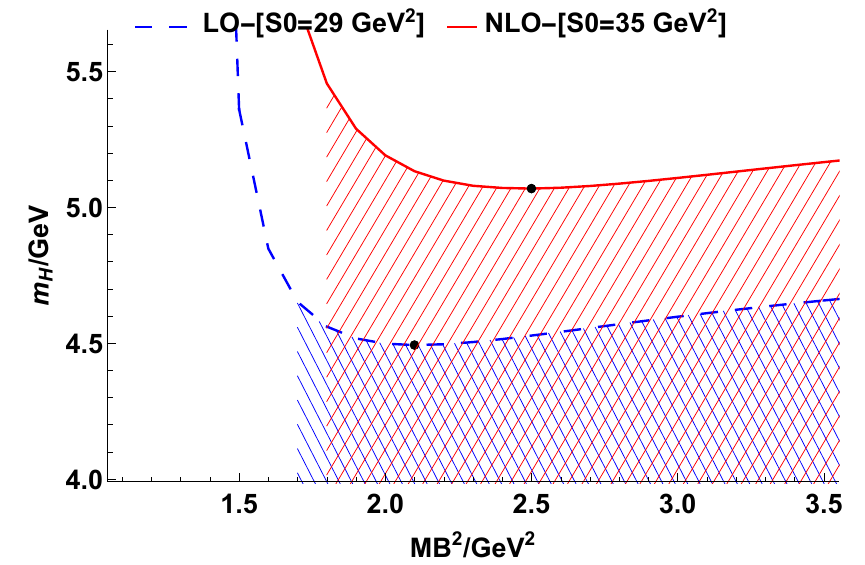}
		\includegraphics[scale=0.4]{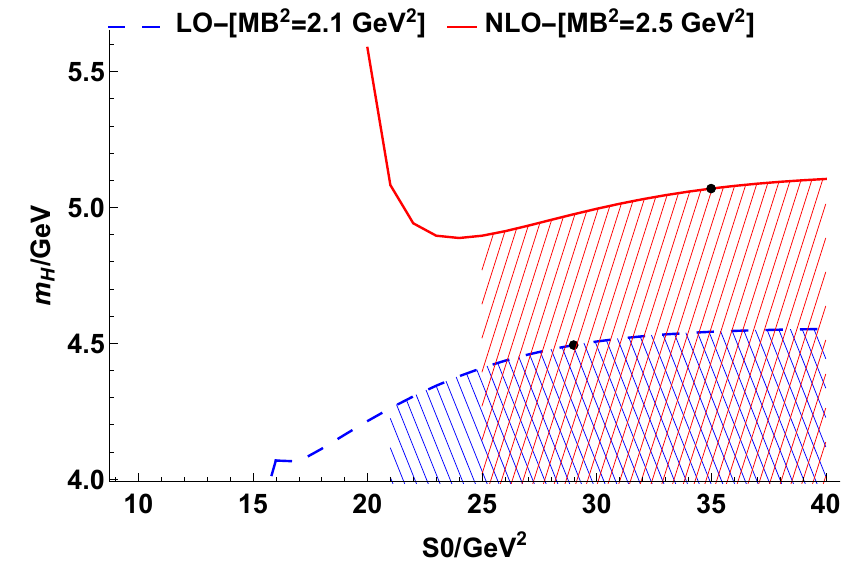}
	}\\
	\subfigure[OS]{
		\includegraphics[scale=0.4]{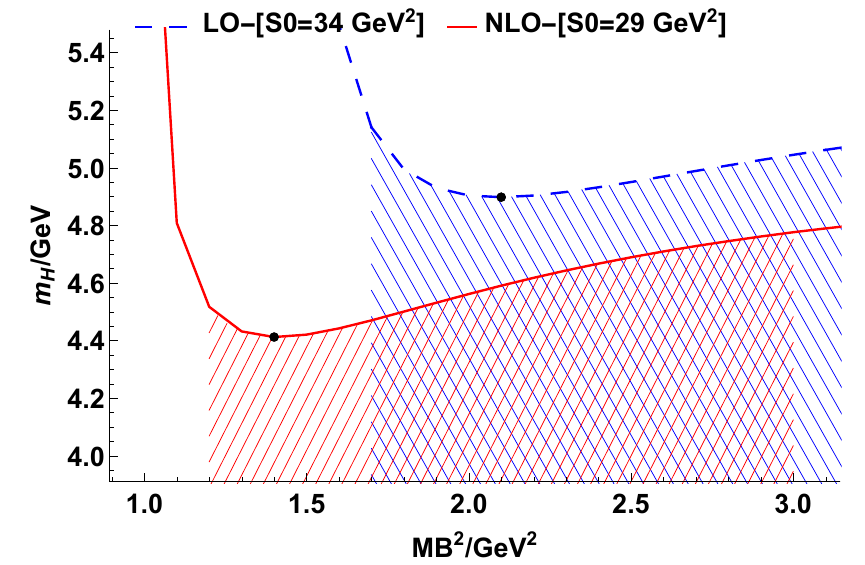}
		\includegraphics[scale=0.4]{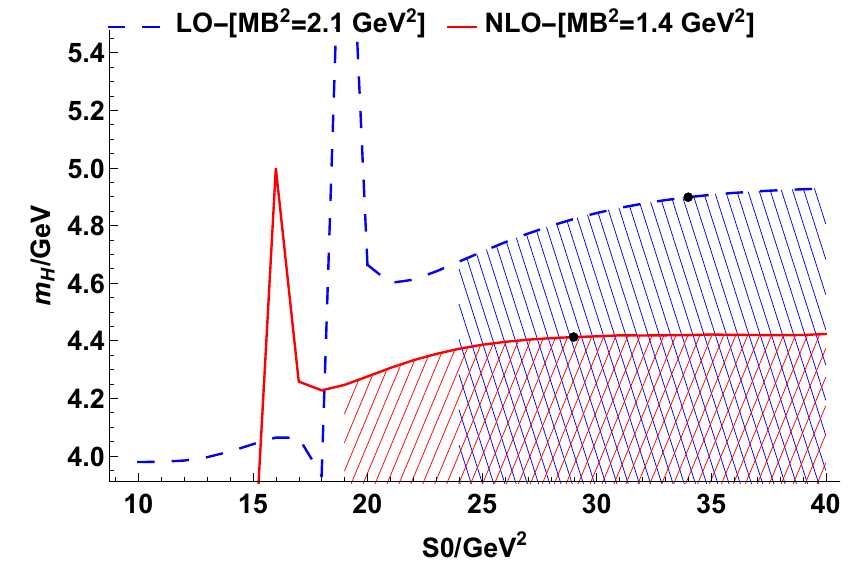}
	}
	\caption{\label{fig:Zcs-[1+]-M-Mixed-4-NLO-MSbar-OS}LO and NLO Result of $J_{4,8}^\text{Mixed}$ with $\overline{\text{MS}}$ and OS renormalization schemes of $Z_{cs}$ system. }
\end{figure}

\newpage
\bibliography{SumRule}

\end{document}